# Machine Learning and Factor-Based Portfolio Optimization[*]

Thomas Conlon[a], John Cotter[b], and Iason Kynigakis[c]

Current Version: July 08, 2021

[*] The authors gratefully acknowledge the support of Science Foundation Ireland under grant number 16/SPP/3347 and 17/SPP/5447.

[a] Smurfit Graduate Business School, University College Dublin, Dublin, Ireland. Email: conlon.thomas@ucd.ie
[b] Smurfit Graduate Business School, University College Dublin, Dublin, Ireland. Email: john.cotter@ucd.ie
[c] Smurfit Graduate Business School, University College Dublin, Dublin, Ireland. Email: iason.kynigakis@ucd.ie

# Machine Learning and Factor-Based Portfolio Optimization

**Abstract**

We examine machine learning and factor-based portfolio optimization. We find that factors based on autoencoder neural networks exhibit a weaker relationship with commonly used characteristic-sorted portfolios than popular dimensionality reduction techniques. Machine learning methods also lead to covariance and portfolio weight structures that diverge from simpler estimators. Minimum-variance portfolios using latent factors derived from autoencoders and sparse methods outperform simpler benchmarks in terms of risk minimization. These effects are amplified for investors with an increased sensitivity to risk-adjusted returns, during high volatility periods or when accounting for tail risk.

*JEL classifications*: C38, C4, C45, C5, C58, G1, G11

*Keywords*: Autoencoder; Covariance matrix; Dimensionality reduction; Factor models; Machine learning; Minimum-variance; Principal component analysis; Partial least squares; Portfolio optimization; Sparse principal component analysis; Sparse partial least squares.

# 1. Introduction

In this paper we examine the characteristics and benefits of latent factors generated from machine learning dimensionality reduction techniques for asset allocation. The analysis is conducted under the framework of factor-based covariance matrices used to construct minimum-variance portfolios.

Machine learning is well suited to many theoretical and empirical problems and has been shown to have the potential to dramatically improve performance in financial applications. In an early study Rapach, Strauss and Zhou (2013) use the elastic net to examine international stock return predictability, while a comprehensive comparison of machine learning methodologies is provided by Gu, Kelly and Xiu (2020), Bianchi, Büchner and Tamoni (2021) and Wu, Chen, Yang and Tindall (2020) for the equity, bond and hedge fund markets, respectively.

Our relation to this literature is twofold. First, we contribute to the studies that construct factors through machine learning. Specifically, we examine the economic value of latent factors generated using a variety of supervised and unsupervised dimensionality reduction methods and their relation to popular characteristics. In addition to classical approaches, such as principal component analysis (PCA) and partial least squares (PLS), their respective regularized versions that induce sparsity through a penalty in the objective function are also considered. We also investigate the performance of factors generated by autoencoders; a type of unsupervised neural network used for dimensionality reduction. The second contribution stems from the framework through which we bridge the gap between machine learning and finance. Particularly, we explore the impact that the proposed latent factors have on the structure of factor-based covariance matrices and to the composition and performance of minimum-variance portfolios.

The assumption that asset returns are driven by a set of observed (Sharpe, 1963; Chen, Roll and Ross, 1986) or latent (Connor and Korajczyk, 1988) factors has a long history in finance. More recently, the advantages of factor construction using machine learning has been shown by Feng, Giglio, and Xiu (2020). They combine the lasso with Fama-MacBeth regressions to systematically evaluate the contribution of an additional factor to a set of preexisting factors, finding that many proposed factors do not add much incremental value. Kelly, Pruitt and Su (2019) propose Instrumental PCA (IPCA), where factors are latent, and the time-varying loadings depend on characteristics. They show that a small number of factors can explain the cross section of average returns more accurately than other leading factor models. Subsequently, Gu, Kelly and Xiu (2021), extend the analysis, without the linearity assumption of IPCA, using machine learning. Feng, Polson, and Xu (2020), build factors that maximize the fit in the cross-section of returns using a deep feed-forward neural network, while Huang, Li and Zhou (2019) find that using reduced-rank regression outperforms popular observed factor models.

PLS applications can also be found in the finance literature. For example, Light, Maslov and Rytchkov (2017) develop a PLS-based approach to aggregate information about expected returns from a large number of firm characteristics into a few composite variables that predict the cross-section of expected stock returns. Lettau and Pelger (2020) introduce risk premia PCA, which identifies factors with small time-series variation that are useful in the cross-section of returns. Kozak, Nagel, and Santosh (2020) find that a stochastic discount factor with a small number of principal components, using Bayesian shrinkage to select a subset of characteristics, provides good out-of-sample explanatory power for average returns.

Machine learning methods have also previously been used in asset allocation. Callot, Caner, Önder, and Ulasan (2019) use nodewise regression to directly estimate the sparse precision matrix. Minimum-variance portfolios based on their proposed approach exhibit lower variances and higher Sharpe ratios compared to commonly used covariance estimators. D'Hondt, DeWinne, Ghysels and Raymond (2020), consider a mean-variance strategy, where the returns are estimated by various machine learning models and the covariance using a dynamic shrinkage estimator, to assess the benefits of robo-investing strategies. They find that machine learning methods lead to portfolios with higher returns that vastly outperform other passive investments during the financial crisis. Lassance, DeMiguel and Vrins (2020), examine the performance of factor-risk-parity portfolios by choosing a set of uncorrelated factors using independent component analysis. They show that portfolios based on independent components provide greater diversification benefits and outperform those using principal components and other benchmarks.

We add to the literature which shows that simpler benchmarks such as the equal-weighted (EW) strategy can be outperformed. In this study we focus on modelling the covariance as a means to improve asset allocation. The improved performance can be attributed to two aspects. First, factor-based covariance matrices tend to significantly reduce the risk of a portfolio consisting of individual stocks. This finding remains robust in an out-of-sample setting, using different risk measures, across covariance and factor specifications, for a varying number of assets, alternative portfolio objective formulations and when transaction costs are taken into account. Second, we demonstrate that using machine learning can lead to significant economic gains. For example, using a factor-implied covariance based on machine learning, can lead to a decrease in out-of-sample portfolio standard deviation of up to 29% and an increase in the Sharpe ratio of over 25%.

The mean-variance criterion of Markowitz (1952) requires knowledge of the mean and covariance matrix of the assets in the portfolio, which in practice are derived from the data and carry considerable estimation risk (Broadie, 1993; Kan and Zhou, 2007). It has been argued (see DeMiguel, Garlappi and Uppal, 2009; Duchin and Levy, 2009) that portfolios selected according to this framework are unlikely to outperform the equal-weighted portfolio, since they are notorious for producing extreme weights that fluctuate considerably over time and perform poorly out-of-sample.[1]

We focus instead on minimum-variance portfolios, which correspond to a risk averse investor who aims to minimize the variance without taking the expected return into consideration. This optimization framework is well motivated by Merton (1980), since it requires only estimates of the covariance matrix, which are often considered to be more accurate than the estimates of the means that have been found to be the principal source of estimation risk (Best and Grauer, 1991; Chopra and Turner, 1993).[2] Although the minimum-variance framework avoids the problem of estimation error associated with expected returns, its performance remains crucially dependent on the quality of the estimated covariance matrix (DeMiguel, Garlappi, Nogales and Uppal, 2009; Ardia, Bolliger, Boudt and Gagnon-Fleury, 2017). To lessen the impact of covariance misspecification on the optimal weights, we impose a factor structure on the covariance matrix, which reduces the number of parameters to be estimated.[3]

---

[1] Extensive effort has been devoted (see e.g., Kritzman, Page and Turkington, 2010; Hjalmarsson and Manchev, 2012; Branger, Lucivjanska and Weissensteiner, 2019; Platanakis, Sutcliffe and Ye, 2021) to the issue of reducing expected return and covariance estimation error with the goal of improving the performance of the Markowitz model relative to simple benchmarks.

[2] The literature (see Chan, Karceski, Lakonishok, 1999; Ledoit and Wolf, 2003; Kempf and Memmel, 2006; Frahm, 2008; Ledoit and Wolf, 2017; Carroll, Conlon, Cotter, Salvador, 2017; Moura, Santos and Ruiz, 2020, among others) has frequently advocated the use of constrained and unconstrained minimum-variance portfolios.

In addition to having observed or latent factors, factor models can be static, such as in the arbitrage pricing theory (APT) of Ross (1976), or dynamic (see Sargent and Sims, 1977; Geweke, 1977). It has been shown that introducing factor structure to the covariance matrix can improve portfolio performance (Green and Hollifield, 1992; Chan, Karceski and Lakonishok, 1999). The performance of factor-implied covariances has also been examined by MacKinlay and Pastor (2000) for tangency portfolios and by Moskowitz (2003) for minimum-variance portfolios. The benefits of using the factor model-based approach to estimate the covariance matrix have also been investigated by Fan, Fan and Lv (2008) and Fan, Liao and Mincheva (2011; 2013) who propose estimators of the covariance for exact and approximate factor models, respectively. More recently, De Nard, Ledoit and Wolf (2019), use a factor framework and evaluate portfolios for different estimates of the error covariance matrix. Our contribution to the literature of factor-based portfolio optimization arises from conducting a comparative analysis of a static and several dynamic specifications of the covariance matrix, based on observed or latent factors. The structure of a dynamic covariance matrix can differ based on whether the factor loadings, the factor covariance matrix or the residual covariance matrix are allowed to vary over time.

To determine the effects of using factors based on machine learning in covariance estimation and portfolio optimization, we first explore the potential links between the latent factors and popular factor proxies, examine the structure of the factor-implied covariance matrices and analyze the properties of portfolio weights. The results show that machine learning yields factors that cause the covariances and portfolio weights to diverge from those based on commonly used estimators. Latent factors produced by PCA and PLS-type methods exhibit a stronger connection with well-known factors (such as those from the Fama and French (2015) five-factor model) throughout the out-of-sample period, compared to factors based on autoencoders. Furthermore, the covariance matrices whose structure deviates most from the sample estimator are based on unsupervised methods or allow the residual covariance matrix to be time-varying. Portfolios based on machine learning also have weights that are smaller, vary less over time and are more diversified, than models based on observed factors. Covariance matrices based on unsupervised methods also lead to portfolios with lower turnover and thus reduced sensitivity to transaction costs.

---

[3] Alternative solutions involve imposing short-selling constraints (Hui, Kwan and Lee, 1993; Jagannathan and Ma, 2003), limiting turnover via norm constraints (DeMiguel, Garlappi, Nogales and Uppal, 2009) or penalizing the objective function, such as in Olivares-Nadal and DeMiguel (2018). Another approach uses either shrinkage estimators of the covariance matrix (Frahm and Memmel 2010; Ledoit and Wolf, 2012), which tend to shrink the covariance matrix towards a specific target covariance or sparse estimators that derive a regularized version of the precision matrix (Friedman, Hastie and Tibshirani, 2008; Callot, Caner, Önder, and Ulasan, 2019). Using higher frequency data can also reduce estimation error (Palczewski and Palczewski, 2014; Fan, Furger and Xiu, 2016).

We evaluate the different factor and covariance specifications by constructing minimum-variance portfolios based on individual stock return data for a sample period spanning 60 years. Overall, our findings suggest that machine learning adds value to factor-based asset allocation. In the baseline case, machine learning leads to portfolios that significantly outperform the equal-weighted benchmark, which DeMiguel, Garlappi and Uppal (2009) show to be a very stringent benchmark, and improves upon portfolios based on the sample estimator or observed factors. The best-performing methods to generate the covariance matrix are autoencoders and sparse principal component analysis, which can lead to portfolios with higher risk-adjusted return and standard deviation that is 3% lower per annum than the equal-weighted one.

In addition, machine learning can improve factor-based portfolio optimization when performance is measured using alternative risk metrics. Covariance matrices based on autoencoders and sparse PCA outperform the equal-weighted portfolio by up to 2.9%, 1.26% and 1.57% per annum, in terms of mean absolute deviation, Value-at-Risk and Conditional Value-at-Risk, respectively. Furthermore, the performance according to certainty equivalent return improves relative to the benchmark as the investor's aversion to risk increases, which falls in line with the optimization goal. In particular, investors with moderate or conservative risk preferences using machine learning factors would realize significant utility gains that are between 2.5% and 4.5% higher than those of the EW portfolio on an annual basis.

Similar to recent studies, the results indicate that shallow learning outperforms deeper learning, which can be attributed to the small size of the data set and the low signal-to-noise ratio. When we consider factors based on neural networks with one to four hidden layers, we find that for the baseline optimization framework the shallowest network outperforms those with deeper architectures. Additionally, unsupervised methods tend to perform better than supervised methods. When we compare PLS and sparse PLS with PCA and sparse PCA, the results show that the PCA-type approaches significantly outperform the equal-weighted benchmark more often. This decline in portfolio performance is potentially associated with the high degree of turnover of strategies based on autoencoders with more hidden layers and supervised methods. Otherwise, the ranking among factors persists across specifications of the covariance matrix. When comparing the results across the alternative specifications of the factor-based covariance matrix, the differences become less pronounced, with approaches that allow the loadings or the residual covariance matrix to vary over time yielding higher risk-adjusted performance.

The performance of portfolios based on machine learning is amplified during periods of high volatility, while the benefits of factor-based optimization become evident during different inflation and credit spread subperiods. All machine learning portfolios generate positive breakeven transaction costs, indicating they outperform the equal-weighted benchmark. The results for machine learning portfolios remain robust when short-selling is allowed, unlike those for observed factors that show a significant increase in portfolio risk. Including a proportional transaction costs penalty to the minimum-variance objective, leads to a decrease in monthly turnover, which benefits primarily the portfolios with high rebalancing requirements such as those based on observed factors, unsupervised methods or deeper autoencoders. The factor-based portfolios continue to outperform the benchmark after increasing the number of assets. Turnover increases considerably with the size of the portfolio, while the breakeven transaction costs decrease. Machine learning models still generate the highest performance across different portfolio sizes. Finally, the results for alternative datasets based on industry or book-to-market sorted portfolios, tend to favor covariances based on observed factors or supervised methods. Overall, the results indicate that machine learning improves factor-based portfolio optimization.

## 2. Methodology

In this section we introduce the machine learning methods for dimensionality reduction used to construct the latent factors, we then describe the different specifications under which the factor-based covariance matrices are estimated and finally, present the optimization framework used to derive the portfolios.

### 2.1 Latent Factors via Machine Learning

The presence of factor structure in asset returns has been widely accepted in the economic literature. The capital asset pricing model (CAPM) of Sharpe (1964) and Lintner (1965) implies a simple factor structure. Other early studies that support the presence of a factor structure in asset returns include the APT of Ross (1976) and the intertemporal CAPM of Merton (1973). Factors can be observable quantities, such as macroeconomic indicators (Chen, Roll and Ross, 1986) or observable proxies, where factors are the returns of portfolios constructed by sorting stocks based on firm characteristics, such as the single index model of Sharpe (1963) or the three-factor model of Fama and French (1993) and its variations.

A factor model for the returns of every asset, $r_{i,t}$, with $i = 1, \dots, N$ assets, $t = 1, \dots, T$ observations and $k = 1, \dots, K$ observed factors takes the form

$$r_{i,t} = a_i + \beta_i F_t + u_{i,t}, \tag{1}$$

where $\beta_i = (\beta_{i,1}, \dots, \beta_{i,K})$ are the time-invariant factor loadings for factors $F_t = (f_{t,1}, \dots, f_{t,K})$, $a_i$ is the time-invariant intercept and $u_{i,t}$ is the error term for asset $i$ at date $t$. The intercepts and factor loadings can be estimated by ordinary least squares (OLS) using the different factor representations.

Factors can also be latent quantities, which are derived from the data using dimensionality reduction techniques. When factors are latent, principal component analysis is a very common approach to reduce dimensionality. The studies of Chamberlain and Rothschild (1983) and Connor and Korajczyk (1988) are among the first to use latent factors in applications of the APT.

The general form of a latent factor model is given by

$$r_{i,t} = a_i + \beta_i(X_t W) + u_{i,t} = a_i + \beta_i F_t + u_{i,t}, \tag{2}$$

where $X_t = (x_{1,t}, \dots, x_{p,t})$ is the $T \times p$ matrix of predictors and $W = (w_1, \dots, w_K)$ is the $p \times K$ matrix of weights, with $K \ll p$. Each $w_k$ is the vector of weights used to construct the $k^{\text{th}}$ latent factor, $f_k$. The $T \times K$ matrix of latent factors is given by $F_t = X_t W$.

In the following sections we describe the classical dimensionality reduction techniques used to generate the latent factors, along with their extensions from the machine learning literature, which rely on regularization and neural networks. The alternative methods we consider are similar in that the dimensionality of the data is reduced by mapping the set of $p$ predictors to a smaller set of $K$ combinations of the original variables.

### 2.1.1 Principal Component Analysis and Partial Least Squares

The two most commonly used dimensionality reduction techniques are principal component analysis and partial least squares. They are both designed to uncover a lower dimensional linear combination of the original predictor set, however, PCA derives the latent factors in an unsupervised way, based only on information from the predictors $X_t$, while in the case of PLS the factors are constructed in a supervised way, by using information from both the predictors $X_t$ and the response $R_t$. The methods differ in the way the latent factor matrix, $F_t$, is extracted, since PCA produces the weight matrix W reflecting the covariance structure between predictors, while PLS computes weights that account for the covariation between the predictors and the response.

Principal component analysis can be viewed as a regression-type problem where the goal is to find the first $K$ principal component weight vectors by minimizing:

$$\underset{W}{\text{argmin}} \|X_t - X_t W W'\|^2, \quad \text{s.t.} \quad W'W = I_K, \tag{3}$$

where $I_K$ is a $K \times K$ identity matrix. The solution to this problem is most often obtained via singular value decomposition: $X_t = UDV'$, by setting $W = V$. The columns of $V = (v_1, \dots, v_K)$ are the principal components weights. Each $v_j$ is used to derive the $k^{\text{th}}$ principal component, $f_k = X_t v_k$, thus, $F_t V$ is the dimension reduced version of the original predictors. The derived variable $f_1$ is the first principal component of $X_t$ and has the largest sample variance amongst all linear combinations of the columns of $X_t$.

Partial least squares, introduced by Wold (1966), identifies the features in a supervised way, by constructing linear combinations based on both $R_t$ and $X_t$. Specifically, PLS decomposes the matrix of predictors $X_t$ and the matrix of asset returns $R_t$ into the form: $X_t = F_t P' + E_t$ and $R_t = F_t Q' + H_t$, where the matrices P and Q are the loadings, while E and H are the residuals. In order to find the PLS component matrix $F_t$, the columns of the weight matrix W need to be obtained through consecutive optimization problems. The criterion to find the $k^{\text{th}}$ estimated weight vector $w_k$ is

$$\underset{w}{\operatorname{argmax}}[w'(X_t' R_t R_t' X_t) w], \quad \text{s.t.} \quad w'w = 1, \quad w' \Sigma_{XX} w_k = 0, \tag{4}$$

where $\Sigma_{XX}$ is the covariance of $X_t$. The latent factor matrix is then given by $F_t = X_t W$. The version of PLS we employ is SIMPLS proposed by de Jong (1993). If $K = p$ then PLS would give a solution equivalent to the OLS estimates.

PCA and PLS have the drawback that for each latent factor the weights are typically non-zero, which leads to difficulties in high dimensional settings. To address this issue, we consider methods that produce modified latent factors with sparse weights, such that each latent factor is a combination of only a few of the original variables.

Sparse principal component analysis (SPCA), developed by Zou, Hastie and Tibshirani (2006), is based on the regression/reconstruction property of PCA and produces modified principal components with sparse weights, such that each principal component is a linear combination of only a few of the original predictors. They show how PCA can be viewed in terms of a ridge regression problem and by adding the $l_1$ penalty, they convert it to an elastic net regression, which allows for the estimation of sparse principal components. The following regression criterion is proposed to derive the sparse principal component weights

$$\underset{W,C}{\operatorname{argmin}}[\|X_t - X_t W C'\|^2 + \lambda_1 \|w\|_1 + \lambda_2 \|w\|^2], \quad \text{s.t.} \quad W'W = I_K, \tag{5}$$

where W and C are both $p \times K$. If $\lambda_1 = \lambda_2 = 0$, $T > p$ and we restrict $C = W$, then the minimizer of the objective function is exactly the first $K$ weight vectors of ordinary PCA. When $p \gg T$, in order to obtain a unique solution, $\lambda_2 > 0$ is required. The $l_1$ penalty on $c_k$ induces sparseness of the weights, with larger values of $\lambda_1$ leading to sparser solutions.

Sparse partial least squares (SPLS) is an extension of PLS that imposes the $l_1$ penalty to promote sparsity onto a surrogate weight vector $c$ instead of the original weight vector $w$, while keeping $w$ and $c$ close to each other (Chun and Keles, 2010). The first SPLS weight vector solves

$$\underset{w,c}{\text{argmin}} \left[ -\frac{1}{2} w' M w + \frac{1}{2} (c - w)' M (c - w) + \lambda_1 \|c\|_1 + \lambda_2 \|c\|^2 \right], \quad \text{s.t.} \quad w'w = 1, \tag{6}$$

where $M = X_t' R_t R_t' X_t$, $\lambda_1$ and $\lambda_2$ are non-negative tuning parameters. To solve SPLS a large $\lambda_2$ value is usually required and setting $\lambda_2 = \infty$ yields a solution that has the form of the soft threshold estimator by Zou and Hastie (2005). This reduces the number of tuning parameters to two, the tuning parameter $\lambda_1$ and the number of latent factors $K$.

### 2.1.2 Autoencoder Neural Networks

Another approach we use to construct the latent factors is based on autoencoders (Bourlard and Kamp, 1988; LeCun, Boser, Denker, Henderson, Howard, Hubbard, Jackel, 1989; Hinton and Zemel, 1994), which are a type of unsupervised neural network that can be used for dimensionality reduction. Autoencoders have a similar structure to feed-forward neural networks, which have been shown to be universal approximators for any continuous function (Hornik, Stinchcombe and White, 1989; Cybenko, 1989). However, an autoencoder differs in that the number of inputs is the same as the number of outputs and that it is used in an unsupervised context. Autoencoders have also been shown to be nonlinear generalizations of PCA. The goal of PCA and autoencoders is to learn a parsimonious representation of the original input data, $X_t$, through a bottleneck structure. The difference is that PCA reduces the dimensionality, by mapping the original $p$ inputs into $K \ll p$ factors in a linear way, while autoencoders use non-linear activation functions to discover non-linear representations of the data.

An autoencoder is trying to learn an approximation to the identity function so as the output $\widehat{X}_t$ is similar to the input $X_t$. The network consists of two parts: an encoder and a decoder. The encoder creates a compressed representation of $X_t$ when the input variables pass through the units in the hidden layers, which are then decompressed to the output layer through the decoder. By placing constraints on the network, such as limiting the number of hidden units, it is forced to learn a compressed representation of the input, potentially uncovering an interesting structure of the data. Most often the encoding and decoding parts of an autoencoder are symmetrical, in that they both feature the same number of hidden layers with the same number of hidden units per layer. The output of the decoder is most commonly used to validate information loss, while the smallest hidden layer of the encoder (or code, at the bottleneck of the network) corresponds to the dimension-reduced data representation.

Let $L$ denote the number of hidden layers and $K^{(l)}$ denote the number of hidden units in each layer, for $l = 1, \dots, L$, while the output of unit $k$ in layer $l$ is defined as the vector $z_k^{(l)}$ and the output of layer $l$ as the matrix $Z^{(l)} = \left(z_1^{(l)}, \dots, z_{K^{(l)}}^{(l)}\right)$. The original data, $X_t$, enters the network through the input layer ($l = 0$), while in each hidden layer inputs from the previous layer are transformed through nonlinear activation functions $h(\cdot)$ before being passed as inputs onto the next layer. The output of each hidden unit $k$ in layer $l$ is based on the function

$$z_k^{(l)} = h\big(Z^{(l-1)}W^{(l-1)} + b^{(l-1)}\big), \tag{7}$$

where $W^{(l-1)}$ is a $K^{(l-1)} \times K^{(l)}$ weight matrix and $b^{(l-1)}$ is a $1 \times K^{(l)}$ bias vector. For the first hidden layer the matrix of predictors is used as input, $Z^{(0)} = X_t$, such that $z_k^{(1)} = h\big(X_t W^{(0)} + b^{(0)}\big)$. We use the hyperbolic tangent (tanh) activation function defined as $h(x) = 2/(1 + e^{-2x}) - 1$, which is a zero-centered function whose range lies between -1 to 1. The results from each hidden layer are aggregated in the output layer

$$\widehat{X}_t = h\big(Z^{(L-1)}W^{(L-1)} + b^{(L-1)}\big). \tag{8}$$

Since an autoencoder tries to approximate $X_t$ the dimensions of the input and the output layer are identical, $K^{(0)} = p = K^{(L)}$. We consider four different network architectures based on the depth of the network. First, we construct a shallow autoencoder (AEN1) with a single hidden layer (the code). The other three models include additional hidden layers to the encoder and decoder representation, up to a maximum of three layers (AEN2, AEN3 and AEN4). The number of hidden nodes in each layer is selected according to the geometric pyramid rule by Masters (1993).

It can also be shown that linear autoencoders are equivalent to PCA (Baldi and Hornik, 1989; Karhunen and Joutsensalo 1995; Japkowicz, Hanson and Gluck, 2000). Specifically, when the autoencoder has a single hidden layer, so the network representation becomes $X_t \to Z^{(1)} \to \hat{X}_t$ and all activation functions are linear, it can be shown that the $K^{(1)}$ latent variables at the bottleneck correspond to the first $K$ principal components of the data. Hinton and Salakhutdinov (2006) show that deep autoencoders outperform shallow or linear autoencoders in image recognition tasks, however, recent applications in finance (see Gu, Kelly and Xiu, 2020; Bianchi, Büchner and Tamoni, 2021) find that shallower networks generate better performance.

The weights and biases of the neural network are estimated by minimizing the square loss of the form

$$\underset{b,W}{\mathrm{argmin}}\, \mathcal{L}(X_t, \hat{X}_t) = \underset{b,W}{\mathrm{argmin}} \|X_t - \hat{X}_t\|^2, \quad \text{s.t.} \quad \|W^{(l)}\|^2 = i_{K^{(l)}}. \qquad (9)$$

The estimates of the parameters of a neural network are solutions of a non-convex optimization problem. The neural network is trained using stochastic gradient descent (SGD). Unlike standard gradient descent that uses the entire training sample to evaluate the gradient at each iteration of the optimization, SGD evaluates the gradient from a random subset of the data and iteratively minimizes the objective function through back propagation. The version of SGD we implement is the adaptive moment estimation algorithm (Adam), introduced by Kingma and Ba (2015). Adam computes individual adaptive learning rates for the model parameters using estimates of first and second moments of the gradients.

Training a neural network can be challenging due to the large number of parameters to be estimated and the nonconvexity of the objective function. To alleviate those concerns we modify the loss function by adding a penalty on the output of the layers (activations), encouraging the activations of the nodes to be sparse (Goodfellow, Bengio and Courville, 2016). We consider activity regularization based on the elastic net penalty that shrinks the output of the bottleneck layer. Following papers such as Gu, Kelly and Xiu (2020), we implement early stopping. In each iteration of the optimization algorithm the parameter estimates are updated to reduce prediction errors in the training sample and then the predictive performance of the model for that iteration is evaluated using data from the validation sample. Early stopping is implemented by stopping the training process prematurely when the validation error no longer decreases. Specifically, the optimization process halts when the maximum number of iterations is reached or if the validation error has not improved for a certain number of consecutive iterations. In both cases the parameter estimates of the best performing model are retrieved. The effects of early stopping are that it prevents overfitting and significantly speeds up the training process.

## 2.2 Factor-based Covariance Estimation

After the factor model is estimated from equation (1) or (2) the covariance matrix of returns, $\Sigma_r$, is obtained by its decomposition into two components: the first is based on the factor loadings and the factor covariance matrix, while the second is the covariance matrix of the errors. The time-invariant covariance matrix of the returns $R = (r_1, \ldots, r_N)$ is given by:

$$\Sigma_r = B'\Sigma_f B + \Sigma_u, \tag{10}$$

where B is a $K \times N$ matrix with the $i$th column containing the vector of time-invariant factor loadings $\beta_i$ and $\Sigma_f$ and $\Sigma_u$ denote the time-invariant covariance matrices of the factors and the errors respectively. We focus on exact factor models (Fan, Fan and Lv, 2008), where the covariance matrix of the residuals $u_t$ is diagonal, $\Sigma_u \equiv diag(\Sigma_u)$, by assuming cross-sectional independence.

The models presented so far rely on a static specification. However, Sargent and Sims (1977) and Geweke (1977) introduce dynamic factor models (DFM) as an extension.[4] There are various definitions of DFMs (see Stock and Watson, 2011), the one we follow in this study is a model that allows the factor loadings to be time varying (Avramov and Chordia, 2006; Engle, 2016 and Bali, Engle and Tang, 2017) or models in which either the factor or residual covariance matrix varies over time (Engle, Ng and Rothchild, 1990). In the description below, the factors can be observed quantities or latent factors.[5]

---

[4] DFMs have been further developed by Forni, Hallin, Lippi and Reichlin (2000; 2005), while Boivin and Ng (2005) compare the forecasting performance of models that use static or dynamic factors.

A dynamic factor model is one in which at least one of the following three generalizations holds true: (i) the intercept and factor loadings are time-varying, (ii) the covariance matrix of the factors is time-varying or (iii) the covariance matrix of the errors is time-varying.

When the intercepts $a_i$ and factor loadings $\beta_i$ are allowed to be time-varying the conditional dynamic factor model takes the following form

$$r_{i,t} = a_{i,t} + \beta_{i,t} F_t + u_{i,t}. \tag{11}$$

The estimates of the time-varying regression coefficients are then obtained by $\hat{\beta}_{i,t} = \Sigma_{f,t}^{-1} \sigma_{fr_i,t}$. The coefficients, $\hat{\beta}_{i,t}$, of this expression are the dynamic conditional betas and are based on time-varying estimates of the factor covariance matrix $\Sigma_{f,t}$ and the vector of covariances, $\sigma_{fr_i,t}$ between the returns of asset $i$, $r_i$ and factor $f_k$, with $k = 1, \ldots, K$. The intercept can be obtained by $\hat{a}_{i,t} = \bar{r}_i - \hat{\beta}_{i,t} \bar{F}$. The time-varying covariance matrix of $R_t$ is given by:

$$\Sigma_{r,t} = B_t' \Sigma_f B_t + \Sigma_u, \tag{12}$$

where $B_t$ is a $K \times N$ matrix with the $i$th column containing the vector of time-varying factor loadings $\beta_{i,t}$.

The unconditional dynamic factor model under generalization (ii) and (iii) takes a form similar to equations (1) or (2), but with time-varying conditional covariance matrices for $f_t$ and $u_t$ respectively. If $\Sigma_f$ is assumed to be time-varying, then the covariance matrix of $R_t$ is given by

$$\Sigma_{r,t} = B' \Sigma_{f,t} B + \Sigma_u. \tag{13}$$

Otherwise, if $\Sigma_u$ is assumed to be time-varying, then

$$\Sigma_{r,t} = B' \Sigma_f B + \Sigma_{u,t}. \tag{14}$$

---

[5] The econometric properties of PCA in a static (Bai and Ng, 2002) or a dynamic (Stock and Watson, 2002a; 2002b) setting, have been examined when factor models are used for economic forecasting with many predictors.

The factor covariance matrix, $\Sigma_{f,t}$ is estimated by the dynamic conditional correlation (DCC) model (Engle, 2002) and the diagonal elements of $\Sigma_{u,t}$ are estimated by univariate GARCH models.

## 2.3. Minimum-Variance Portfolios

The sensitivity of portfolio weights to estimates of asset means is well documented (Best and Grauer, 1991; Chopra and Turner, 1993). To this end, we focus on minimum-variance portfolios, which have frequently been used in the portfolio optimization literature (see e.g. Carroll, Conlon, Cotter, Salvador, 2017; Moura, Santos and Ruiz, 2020), thus avoiding the issue of estimation error in expected returns.

Specifically, the different estimates of the covariance matrix, $\hat{\Sigma}_r$, from the factor models are evaluated through the minimum-variance framework with short-selling constraints, where the goal is to minimize portfolio risk. Assuming there are $N$ assets in the investment universe and $r_t = (r_{1,t}, \ldots, r_{N,t})$ is a vector of asset returns, the objective of the minimum-variance optimization problem is

$$\underset{\omega}{\operatorname{argmin}} \, \omega' \hat{\Sigma}_r \omega, \quad \text{s.t.} \quad \omega' i_N = 1, \quad \omega_i \geq 0, \text{for } i = 1, \ldots, N, \tag{15}$$

where $\omega = (\omega_1, \ldots, \omega_N)$ is the portfolio weight vector and $i_N$ is a $N \times 1$ unit vector. The return of the portfolio can then be calculated as $r_{p,t+1} = \hat{\omega}' r_{t+1}$. All portfolios include short-selling and leverage constraints to avoid implausible positions, by imposing a lower bound of zero on all weights and that the sum of the weights does not exceed one. The additional non-negativity constraint on minimum variance portfolios has been shown (Jagannathan and Ma, 2003) to be equivalent to shrinking the elements of the covariance matrix.

## 3. Data, Sample Splitting and Hyperparameter Tuning

The data set consists of monthly total individual stock returns from the Center for Research in Security Prices (CRSP) starting on January 1960 to December 2019, for a period of 60 years (or $T = 720$ monthly observations). Our approach regarding the backtest and the restrictions we impose on the data set is similar to that of Engle, Ledoit and Wolf (2019) and De Nard, Ledoit and Wolf (2019), but adapted to a monthly frequency. We restrict our data set to stocks listed on the NYSE, AMEX, and NASDAQ stock exchanges (exchange codes 1, 2 or 3) and to ordinary common shares (share codes 10 or 11).

We adopt a rolling window approach to examine the out-of-sample (OOS) performance of our models. The size of the rolling window is set to $T_0 = 240$ monthly observations (or 20 years), with the initial window spanning the period from January 1960 to December 1979. The rolling window moves across the full sample by one monthly observation at a time, leading to an out-of-sample size of $T_{OOS} = T - T_0 = 480$ monthly observations (or 40 years), from January 1980 to December 2019. The portfolios are constructed in each iteration of the rolling window, based on stocks that have at least 97.5% history of returns available over the past $T_0 = 240$ months and are also not missing the return observation for the following month after the end of the rolling window. This forward-looking restriction is commonly applied to allow for the out-of-sample evaluation of portfolios, which are based on in-sample estimates of the covariance matrix. Finally, we consider stocks whose price is greater than $5 within each iteration. In the baseline case, the latent factors, covariance matrices and portfolio weights are estimated based on the $N = 100$ stocks with the highest market capitalization within each iteration of the rolling window, before we expand the analysis to larger portfolios.[6]

In each iteration of the rolling window, we cross-sectionally transform the predictors, $X_t$, before estimating the latent factors in a similar way to Freyberger, Neuhierl, and Weber (2020). Specifically, we calculate the rank of a stock based on the return and then divide the ranks by the number of observations. We follow Kelly, Pruitt, and Su (2019) and subtract 0.5, to map the features into the $[-0.5, 0.5]$ interval. This transformation focuses on the ordering of the data and is insensitive to outliers.

The machine learning models used to derive the latent factors rely on hyperparameter tuning. The choice of hyperparameters controls the amount of model complexity and is critical for the performance of the model. Specifically, we adopt the validation sample approach, in which the optimal set of values for the tuning parameters is selected in the validation sample. One of the advantages of using this approach over $k$-fold cross validation is that we maintain the temporal ordering of the data. For a detailed description of cross-validation see Friedman, Hastie and Tibshirani (2009). Specifically, in each iteration of the rolling window, the in-sample ($T_0 = 240$) is split into two disjointed periods, the training subsample, $T_0^{\mathcal{T}}$, consisting of 80% of the observations, with the remaining observations belonging to the validation subsample, $T_0^{\mathcal{V}}$. In the training subsample the model is estimated for several sets of values of the tuning parameters. The second subsample is used to select the optimal set of tuning parameters, by using the latent factor weight and loading estimates for each set of hyperparameters from the training sample, forecasts are constructed for the observations in the validation sample.

---

[6] Figure A1 in the Appendix displays the number of stocks in the sample for each month for the full sample period, while the number of stocks in each iteration of the rolling window for the out-of-sample period is displayed in Figure A2.

Specifically, the factor model is first estimated using information only from the training sample, from the following regression equation

$$r_{i,t} = a_i^{\mathcal{T}} + \beta_i^{\mathcal{T}}(X_t W^{\mathcal{T}}) + u_{i,t}, \text{ for } t = 1, \dots, T_0^{\mathcal{T}}, \tag{16}$$

where $W^{\mathcal{T}}$ is the factor weight matrix estimated by one of the dimensionality reduction methods using a specific set of hyperparameters, $\beta_i^{\mathcal{T}}$ is one of the $N$ columns of the matrix of factor loadings, $B^{\mathcal{T}}$, estimated by OLS using data only from $\mathcal{T}$ and $a_i^{\mathcal{T}}$ is the intercept. The covariance over the training subsample is then calculated by

$$\Sigma_r^{\mathcal{T}} = B^{\mathcal{T}\prime} \Sigma_f^{\mathcal{T}} B^{\mathcal{T}} + \Sigma_u^{\mathcal{T}}, \tag{17}$$

where $\Sigma_f^{\mathcal{T}}$ is the covariance of the factors $X_t W^{\mathcal{T}}$ and $\Sigma_u^{\mathcal{T}}$ is the covariance of the errors $u_t$ for $t = 1, \dots, T_0^{\mathcal{T}}$. The covariance over the training set is compared to that of the validation set, which is derived using the estimated matrices $W^{\mathcal{T}}$ and $B^{\mathcal{T}}$. The covariance using data from the validation subsample is estimated as

$$\Sigma_r^{\mathcal{V}} = B^{\mathcal{T}\prime} \Sigma_f^{\mathcal{V}} B^{\mathcal{T}} + \Sigma_u^{\mathcal{V}}, \tag{18}$$

where $\Sigma_f^{\mathcal{V}}$ is the covariance of the factors $X_t W^{\mathcal{T}}$, for $t = 1, \dots, T_0^{\mathcal{V}}$ and $\Sigma_u^{\mathcal{V}}$ is the covariance of the errors, derived by $u_{i,t} = r_{i,t} - \left(a_i^{\mathcal{T}} + \beta_i^{\mathcal{T}}(X_t W^{\mathcal{T}})\right)$, for $t = 1, \dots, T_0^{\mathcal{V}}$.

The covariance matrices based on different sets of hyperparameters are evaluated by employing a measure of economic performance related to the given portfolio application. Following Engle and Colacito (2006) and Becker, Clements, Doolan and Hurn (2015), the optimal set of hyperparameters is then chosen as to minimize the portfolio variance. The loss function is

$$\mathcal{L}_{\text{MVP}}(\widehat{\omega}, \widehat{\Sigma}_r) = \widehat{\omega}' \widehat{\Sigma}_r \widehat{\omega}. \tag{19}$$

The vector of portfolio weights, $\omega$, is obtained by solving the minimum-variance portfolio problem from equation (15), based on the covariance over the training set, $\Sigma_r^{\mathcal{T}}$. The portfolio variance, $\mathcal{L}_{\text{MVP}}$, is calculated using the weight estimates and setting $\widehat{\Sigma}_r = \Sigma_r^{\mathcal{V}}$, for each set of hyperparameters. The set of optimal tuning parameters is the one that yields the lowest portfolio variance over the validation sample. The weights used to construct the latent factors and the factor loadings are re-estimated using all observations in the rolling window for the optimal set of tuning parameters. Finally, the true out-of-sample performance is evaluated by constructing the return of the portfolio using the asset returns one month ahead from the end of each rolling window, which are not included in the validation procedure or parameter estimation.

## 4. Characteristics of a Machine Learning Portfolio

In this section we explore the links between the latent factors and popular factor models, examine the structure of the alternative covariance matrix specifications and analyze the properties of the portfolio weight vectors. Specifically, we investigate how the factors based on several dimensionality reduction approaches relate to those of the Fama and French (2015) five-factor model and the long-short anomaly portfolios constructed by Chen and Zimmermann (2020).[7] Furthermore, we compare the covariance matrices for different factor model specifications to the sample covariance estimator (Sample) and factor models based on observed factors such as the single index (Market) model (Sharpe, 1963) or the three-factor model (FF3) by Fama and French (1993).[8] The portfolio weights based on factor models are also compared to the equal-weighted (EW) portfolio, a scheme which requires no parameter estimation, since the weights are $\omega_i = 1/N$, for $i = 1, ..., N$.

---

[7] Data on the long-short anomaly portfolios are obtained from Andrew Y. Chen's website. We consider portfolios that have a full history of returns from January 1960 to December 2019, which reduces the initial number of available series from 205 to 110.
[8] Data on the Fama-French factors were downloaded from Kenneth French's Data Library.

Latent factors produced by PCA, PLS and their regularized versions better relate to various factor proxies, compared to their neural network counterparts from an analysis of the Fama-French five-factor model. This is illustrated by higher explanatory power for these models' factors compared to those from autoencoder models. The structure of covariance matrices based on unsupervised methods exhibit the greatest differences from the sample estimator. When comparing across covariance specifications, the differences become more apparent in the covariance structure when $\Sigma_u$ varies over time. Finally, portfolios based on machine learning tend to produce weights which are smaller, less volatile throughout time and more diversified, than models based on observed factors. Covariance matrices based on unsupervised methods also lead to portfolios that require less frequent rebalancing relative to their supervised counterparts.

## 4.1 Links to Popular Factors

In this section we examine the links that the estimated factors have with popular characteristic-based factors from the literature. We investigate the links of the latent factors, first with the five-factors from Fama and French (2015) and then with a larger dataset which consists of the long-short anomaly portfolios constructed by Chen and Zimmermann (2020).

For each dimensionality reduction method, we regress each of the estimated latent factors $f_{t,k}$, for $k \in [1, 5]$, on the factors from the Fama and French (2015) five-factor model using OLS.[9] To compare how well the Fama and French model explains the latent factors across dimensionality reduction approaches, we report the adjusted $R^2$, averaged over the 438 estimation windows, in Figure 1. A higher average adjusted $R^2$ indicates that the Fama and French five-factor model is well suited to explain the variability of that specific latent factor throughout the out-of-sample period.

---

[9] Information on the five-factors by Fama and French (2015) is available from July 1963. Given a rolling window size of 240 observations, we have an out-of-sample size of 438 observations, from July 1983 to December 2019. The OLS regressions are re-estimated in each iteration of the rolling window for all combinations of dimensionality reduction techniques and fixed number of factors $K$. The OLS regressions include an intercept, but since we focus on the potential relationship between the latent factors with the observed factors, we omit the intercept from the Figures for the sake of brevity.

[Insert Figure 1 about here]

The factors based on supervised methods tend to have higher $R^2$ than those of unsupervised methods, while the $R^2$ of factors based on autoencoders varies less with $K$ compared to the remaining approaches. Specifically, the results for PCA show that the Fama and French five-factor model is better at explaining the first and third principal components, with $R^2$ between 34% and 36%, while the $R^2$ for the remaining components is between 8% and 15%, decreasing as $K$ increases. The pattern for sparse principal components is similar, but the average $R^2$ is smaller for the first and third factor, decreasing to about 33% and 27%, respectively, while for the remaining components it increases to between 14% and 22%. The Fama and French model explains almost 50% of the variability of the first factor by PLS or SPLS. For the remaining four factors the average $R^2$ varies between approximately 13% and 23%. The average $R^2$ of latent factors based on autoencoders remains relatively stable across different values of $K$ and number of hidden layers, with values between 18% and 25%.

Following Gu, Kelly and Xiu (2020), we quantify the influence of the Fama-French factors as the change in $R^2$ from setting the observations of a factor proxy to zero within each estimation window. The values are averaged to obtain a single variable importance measure for each of the five Fama and French factors and then scaled to sum to 100. The variable importance is presented in Figure 2.

[Insert Figure 2 about here]

According to the change in $R^2$, the results for PCA indicate that the market factor (MKTRF) is most influential for the first principal component, $K = 1$, while the value factor (HML) relates mainly to the third latent factor, $K = 3$. The profitability (RMW) and investment (CMA) factors are better at explaining the $K = 4$ and $K = 5$ components, respectively. In contrast, none of the proxies particularly dominates in terms of explaining the second component, $K = 2$. The characteristics that are influential to the corresponding sparse principal components are similar to those for PCA. However, their importance is decreased with the remaining proxies contributing more to the explanation of the latent factors. In particular, the fourth sparse principal component poses an exception since it relates more to the size factor (SMB) than RMW.

The two PLS methods exhibit a very similar pattern in their relation to the Fama-French five factor model. The first latent factor of both approaches exhibits a connection to the market, which is much stronger than the one observed for the respective principal component. The value characteristic can explain the second, third and fourth latent factors of PLS and SPLS, while RMW and CMA equally explain the fifth factor. The results for the autoencoders remain relatively unchanged when comparing across different values of $K$ and number of hidden layers. The value and, to a lesser extent, the market factors are those which exhibit the strongest relation to the latent factors.

We further examine the connection of the latent factors to the Fama and French five-factor model by aggregating the distributional properties of the $t$-statistics, estimated throughout the out-of-sample period, using boxplots. Boxplots of the $t$-statistics provide a simple five-number summary of their distribution, which consists of the median (marked by the line within the box), first and third quartiles (the edges of the box, with its length representing the interquartile range, IQR), and the minimum and maximum individual $t$-statistics (depicted by the two lines or whiskers, with the distance from the end of each line representing the range).

We consider that a latent factor of a specific dimensionality reduction technique is linked to one of the observed factor proxies, when the median $t$-statistic is non-zero and when the size of the box (IQR) or the distance between the ends of the two whiskers (range) is large, indicating that the majority of the $t$-statistics throughout the rolling window iterations are significantly different from zero. On the contrary, the relationship of a latent factor with a particular Fama-French factor will tend to be weak if the box falls within the lines depicting the Student's $t$ critical values at the 5% level.

[Insert Figure 3 about here]

The results based on the five factors by Fama and French (2015) are presented in Figure 3 and corroborate those from the variable importance analysis. Overall, there is substantial significance for the latent factors based on PCA and PLS methods with some of the Fama-French factors, and especially the market factor. When comparing PCA with SPCA, the relationship of the latent factors with the observed factors is diminished for the latter. In contrast, both PLS approaches generate latent factors that exhibit a similar pattern. Autoencoders do not display any strong links to a particular factor, indicating that latent factors based on neural networks cannot be adequately explained by the Fama-French factors.

Focusing on the results for PCA, the first principal component, $K = 1$, is primarily related to the market factor (MKTRF), with the middle 50% of the $t$-statistics exceeding 5 in absolute value and to the size factor (SMB) where most of the $t$-statistics are around three standard errors from zero. For both factors the distribution of $t$-statistics is right skewed since the median is located towards the left and the left whisker is shorter, indicating a negative relationship. The second factor, $K = 2$, based on PCA does not exhibit any strong links to any of the observed proxies, since the $t$-statistics are less dispersed, as evidenced by the interquartile range and values that are clustered within the 5% critical values bands. The third principal component, $K = 3$, has a strong relation to the value (HML) factor, with $t$-statistics in the range of -11 to 11. The fourth latent factor, $K = 4$, relates to the size and profitability (RMW) characteristics, while the fifth factor, $K = 5$, is primarily linked to the investment (CMA) factor. The pattern of the $t$-statistics for SPCA is similar to that of PCA, however the connection of the sparse principal components to the five factors is overall weaker, as evidenced by the smaller dispersion of the $t$-statistics.

Turning to the results for supervised methods, the patterns observed when comparing the boxplots of both PLS-type approaches appear similar. The first latent factor of PLS and SPLS is positively related to the market factor and, to a lesser extent, the size factor according to the median, with the middle 50% of the $t$-statistics lying approximately between -9 and 8 for MKTRF and from approximately -2 to 5 for SMB. The remaining latent factors are all positively related to the market, albeit to a lesser degree than the first factor. The second latent factor of both methods is also related to HML, according to the negatively skewed boxplot, with the lower quartile being approximately equal to -1 and the third quartile to 5. The third factor is also linked to the value factor, although the distribution of the $t$-statistics is more symmetrical. The fourth factor displays a weak relationship with the size and value factors, while the fifth factor is weakly related to RMW and CMA. Finally, the results for the latent factors based on autoencoders do not reveal any significantly strong links with a particular proxy, when compared to the other dimensionality reduction approaches. For the majority of the autoencoder factors the boxes are symmetrical and smaller, with the middle 50% of the $t$-statistics being within the 5% critical value bands and concentrated around the median, however, the whiskers are longer indicating a greater range of potential values over the out-of-sample period.

We also examine the relationship of the latent factors with the Hou, Xue and Zhang (2015) q-factor model augmented by the expected growth factor (Hou, Mo, Xue and Zhang, 2021).[10] For PCA and PLS methods the q-factor model is better at explaining the first latent factor, with an average $R^2$ between 35% and 50%, which decreases significantly for the remaining four factors (8% to 16%). The $R^2$ does not vary as much when comparing across factors based on autoencoders, with values in the range of 14% to 21%. The market factor is better at explaining the first factor of PCA and PLS approaches, while the investment, return-on-equity and expected growth factors exert greater influence on the remaining factors. For supervised methods the market also relates strongly to the fourth latent factor. The factors based on autoencoders are primarily related to the market and to a lesser extent the investment and return-on-equity characteristics.

We extend the variable importance analysis to a larger dataset, by examining the relationship between the latent factors and long-short anomaly portfolios constructed by Chen and Zimmermann (2020). The equity portfolios are based on a comprehensive reconstruction of firm-level characteristics, which have been replicated using the same data and methods as the original papers. Due to the large number of variables, the influence of each feature to the respective latent factor is derived based on the lasso (Tibshirani, 1996). The penalized model is estimated using a rolling window approach and the results are averaged throughout the out-of-sample period, from January 1980 to December 2019. The variable importance is estimated as the change in $R^2$ from setting the observations of a feature to zero within each iteration of the rolling window. The results are further aggregated by summing the variable importance of the characteristics-based portfolios belonging in the same group. We use the same categorization as Chen and Zimmermann (2020) to assign a variable to a group. Details on which anomaly portfolios belong in each of the six groups can be found in Table A1 in the Appendix. Figure 4 presents the results of the variable importance.

---

[10] The results based on OLS regressions of each of the five latent factors on factors from the augmented q-factor model are reported in Figures A3, A4 and A5 in the Appendix.

[Insert Figure 4 about here]

Across all different models the group comprised of portfolios sorted by accounting characteristics is the most influential according to the lasso. Important variables within this group include portfolios sorted on gross profits/total assets, dividend yield, book-to-market and leverage. The second most influential group is that of portfolios sorted on price characteristics. Individual portfolios with high variable importance values are based on characteristics such as CAPM beta, earnings-to-price ratio, tail risk beta, coskewness and idiosyncratic risk. Portfolios based on trading characteristics, such as liquidity and volume indicators, also relate significantly to the majority of the latent factors. The portfolios from the remaining categories contribute a relatively small part to the total variable importance. The variable importance pattern remains similar across autoencoders with different number of hidden layers, while the results for PCA and PLS methods are less homogenous.

As an alternative we consider the macroeconomic dataset by McCracken and Ng (2015).[11] The first latent factor of PCA and PLS methods is strongly related to interest rates and exchange rates. For the remaining factors the results vary based on the dimensionality reduction approach and the value of $K$. Factors constructed using autoencoders are better explained by interest and exchange rate variables, followed by economic indicators in the labor market or money and credit categories.

## 4.2 Structure of the Covariance Matrices

Given the dependence of the minimum-variance portfolio on the estimates of the covariance, it is informative to further investigate how machine learning factors affect the structure of the covariance matrix. The differences between factor-implied covariance matrices are examined by comparing the structure of the alternative covariance matrix, Σ, to that of the sample estimator, S.

---

[11] The results based on the McCracken and Ng (2015) macroeconomic dataset can be found in Figure A6 in the Appendix, while details on the eight groups and the variables within each group are reported in Table A2.

Following the analysis of Moskowitz (2003) we consider three measures. The first measure examines the similarity between two matrices and is given by $Eig_t = \sqrt{\text{tr}(\Sigma'_t \Sigma_t)/\text{tr}(S'_t S_t)}$, where tr(·) denotes the trace of a matrix. This metric represents the sum of the eigenvalues of the factor-implied covariance matrix as a fraction of the sum of the eigenvalues of the sample covariance matrix. The matrices are squared to capture the absolute amount of covariation. The sum of the eigenvalues provides a measure of total covariation represented by the matrix. The second measure compares the factor-implied covariance matrix with the sample covariance in terms of magnitude and is calculated as $Mag_t = (i'|S_t - \Sigma_t|i)/(i'|S_t|i)$. This measure sums the absolute value of all the elements of the difference between the two matrices and scales this sum with the sum of the absolute value of all the elements of the sample covariance matrix. Finally, we compare the alternative covariance estimates with the benchmark in terms of the direction of the covariances, by determining the fraction of the covariances from the alternative models that have the same sign with those of the sample covariance matrix and is given by $Dir_t = (i'\text{sign}(S_t \circ \Sigma_t)i)/\text{rank}(S_t)^2$.

We report the average value of the measures across the out-of-sample period in Table 1. A high $Eig$ ratio indicates that the alternative covariance matrix is close to the sample estimator. Additionally, if the factor-implied covariance matrix captures similar information to the sample covariance matrix then the magnitude, $Mag$, should be close to zero and the direction, $Dir$, should be close to unity. We also examine whether the difference from the sample estimator of a factor-implied covariance is the same as that of a covariance matrix based on the market factor. The two-sided bootstrapped $p$-value is adjusted for autocorrelation up to 12-month lags.

[Insert Table 1 about here]

Overall, covariance matrices based on latent factors and especially those where the factors are derived from unsupervised methods, exhibit the greatest differences from the sample estimator, for all three measures. The structure of the covariances based on SPCA and shallow autoencoders are the ones that deviate most from the benchmark. In contrast, the structure of covariances using observed factors is closest to the sample estimator. When comparing across covariance specifications, the differences become more pronounced in the covariance structure when the residual covariance matrix is dynamic. Furthermore, the information captured by the covariance matrices based on latent factors relative to the sample estimator is significantly different at the 1% level to that of the corresponding matrix based on the market factor for all measures.

Specifically, according to the $Eig$ measure, the covariance matrices based on latent factors differ considerably from the sample estimator, by a range of 40% to 52%. In contrast, covariance matrices based on the three Fama-French factors or the market factor, differ to the sample covariance matrix by only 1% to 7% or 7% to 9%, respectively. The results under the magnitude and direction measures, remain consistent, with covariances based on latent factors deviating considerably from the sample covariance, while those of observed factors remaining very close to the benchmark. In particular, the values of the latent factors for the $Mag$ measure are closer to unity (0.60 to 0.77) further highlighting the difference from the sample estimator, while the results for the observed factors are much closer to zero (0.01 to 0.1). For $Dir$ the results show that on average, 36% to 65% of the covariances have a different sign from those of the benchmark covariance, compared to observed factors where there is only a 2.5% to 5% difference. It is interesting to note that the values for the direction measure do not significantly change across covariance specifications.

### 4.3 Properties of Portfolio Weights

We now explore how the weighing structure of the portfolios based on machine learning differs compared to the naïve allocation, which assigns equal weight to all assets, and to that of the sample estimator and the observed factors.

We start by analyzing the properties of the portfolio weights, $\widehat{\omega}$, using the maximum weight (MAX), the standard deviation of the portfolio weights ($SD_\omega$) and the mean absolute deviation from the equal-weighted portfolio ($MAD_{EW}$) calculated as $1/N \sum_{i=1}^{N} |\widehat{\omega}_i - 1/N|$. In line with DeMiguel, Garlappi and Uppal (2009), we report the average monthly portfolio turnover (TO) computed as the average absolute

change of the portfolio weights over the $T_{OOS}$ rebalancing periods across the $N$ assets. The turnover at time $t + 1$ is given by $\|\omega_{t+1} - \omega_t\|_1$, where $\omega_{t+1}$ is the vector of portfolio weights at time $t + 1$ and $\omega_t$ are the portfolio weights at the time before rebalancing. Table 2 reports the average value of each weight characteristic over the 480 out-of-sample periods.

[Insert Table 2 about here]

Overall, the portfolios based on machine learning methods tend to produce weights which are smaller, less volatile and closer to those of EW portfolio, than models based on observed factors. The maximum weight attained for a single asset over the out-of-sample period by the sample estimator is 38%. Portfolios based on observed factors also have high maximum weights, from 30% to 44.5% for the market and from 31% to 45% for FF3. On the contrary, the maximum weight for portfolios based on latent factor models is considerably lower, with values between 6.6% and 16.5%. Comparing across covariance specifications, when the error covariance is allowed to vary over time the portfolios produce higher maximum weights than the other specifications. Allocations based on latent factors and static covariance matrices tend to have smaller maximum weights than those based on observed factors or dynamic covariances, which is an indicator of less extreme portfolio positions.

Portfolios with the least volatile weights are those based on unsupervised methods, with a standard deviation that is lower by approximately a factor of three from that of the sample or observed factors. For latent factors, $SD_\omega$ is between 3.5% and 6% per annum, while observed factors have the most dispersed weights, with annualized values between 8.5% and 12% (monthly $SD_\omega$ is annualized by scaling with $\sqrt{12}$). The weights of allocations based on dynamic error covariance vary more over time, with an increase in standard deviation by over 20% to that of static covariance. Portfolios based on unsupervised learning methods tend to produce weights with the smallest deviation from the $1/N$ portfolio, with $MAD_{EW}$ values between 2.7% to 4% per annum, indicating that these portfolios are better diversified. In contrast, observed factor models exhibit the highest deviation from the benchmark, having annualized $MAD_{EW}$ in the range of 5.3% to 5.9%, potentially concentrating on a smaller number of assets. The weights of portfolios based on the dynamic error covariance deviate more from the benchmark than the remaining specifications. Having weights that vary less throughout the out-of-sample period and closer to the $1/N$ strategy, is a positive indicator regarding the effects of transaction costs on portfolio performance. This is due to portfolios of unsupervised methods being more likely to exhibit lower turnover (and thus transaction costs) than those of observed factors or the sample estimator.

The results show that monthly turnover is considerably lower for portfolios using latent factors. Portfolios based on unsupervised methods for the static specification have low rebalancing requirements, producing monthly turnover that is up to 30% lower compared to the sample or observed factor portfolios. Lower turnover indicates that these portfolios would be less affected by transaction costs. Turnover for the equal-weighted portfolio is naturally very low, since the only cause for rebalancing are changes in the stock universe from one period to the next. The portfolio based on the sample estimator generates a monthly turnover of 41.269%, which is higher than most of the portfolios, except those based on observed factors or those that allow $\Sigma_{u,t}$ to vary over time. Comparing across different factor model specifications, portfolios based on dynamic covariance matrices exhibit higher turnover than those using the static factor covariance. Regardless of the specification, factor models with observed factors generate higher turnover than those with latent factor models. The lowest turnover is produced by portfolios with static and dynamic factor covariances based on PCA (29.737% and 30.017%) and SPCA (29.931% and 29.973%) factors, followed by portfolios using autoencoder factors with turnover varying from 31.060% to 49.029%, depending on factor specification. Neural networks with more hidden layers have consistently higher rebalancing requirements than shallower networks.

## 5. Asset Allocation

We now proceed to analyze the out-of-sample performance of the portfolios based on the factor and covariance specifications explored in Section 4 using a variety of performance measures. The buy-and-hold portfolio returns are calculated for the period of one month and the portfolio is rebalanced monthly until the end of the evaluation period (January 1980 to December 2019). In addition, we analyze the behavior of the portfolios during different subperiods and investigate the effects of transaction costs on portfolio performance. We further evaluate the performance of portfolios for allocations without short-selling constraints or for a penalized minimum-variance objective function. Finally, we examine the effects on portfolio performance for a different number of assets or alternative data sets.

The results show that portfolios based on machine learning significantly outperform the equal-weighted benchmark and improve upon allocations based on observed factors or the sample covariance estimator. These economic benefits persist when performance is evaluated using alternative risk metrics. Furthermore, the risk-adjusted performance of the machine learning portfolios significantly improves relative to the benchmark as the investor's aversion to risk increases. The best performing models are those based on shallow autoencoders or sparse PCA and covariance specifications that allow B or $\Sigma_u$ to be time-varying. The benefits of machine learning are amplified during periods of high volatility, while the economic value of factor-based optimization is also evident during different inflation and credit spread regimes.

All portfolios generate positive breakeven transaction costs, indicating they outperform the benchmark. However, an investor would realize greater economic gains over the EW portfolio by choosing machine learning portfolios. When the short-selling restriction is relaxed, performance of latent factor portfolios remains stable relative to those using observed factors, indicating a greater inherent error in covariance estimation by the latter. Likewise, penalizing portfolio turnover, is beneficial primarily to portfolios that were more sensitive to transaction costs, such as those of observed factors or latent factors based on supervised methods. The factor-based portfolios continue to outperform the benchmark even when the number of assets is increased. The turnover of the portfolios increases dramatically with the number of assets, while the breakeven transaction costs decrease, with machine learning models still generating the highest performance across different portfolio sizes. Finally, considering alternative datasets based on industry or book-to-market sorted portfolios, the results tend to favor observed factors or supervised methods.

## 5.1. Portfolio Performance

We focus on evaluating the performance of the portfolios based on measures of risk and risk-adjusted returns, since equation (15) is designed to minimize variance rather than maximize the expected return. Therefore, similar to Ledoit and Wolf (2017) and De Nard, Ledoit and Wolf (2019), we primarily compare the economic value of the alternative covariance matrix estimates using the standard deviation, followed by the Sharpe ratio. In Table 3 we report the monthly performance of the portfolios over the out-of-sample period, $T_{OOS}$, based on the standard deviation (SD) of the 480 out-of-sample portfolio returns in excess of the risk-free rate and the Sharpe ratio (SR) of the portfolio calculated as $(\bar{r}_p - \bar{r}_f)/\text{SD}$, where $\bar{r}_p$ is the average value of the 480 out-of-sample portfolio returns and $\bar{r}_f$ is the average value of the risk-free rate.

We also consider the question whether one portfolio delivers improved out-of-sample performance relative to another portfolio at a level that is statistically significant. Since we consider 42 portfolios, there would be 861 pairwise comparisons. DeMiguel, Garlappi, and Uppal (2009) provide persuasive evidence that the simple equal-weighted portfolio should serve as a natural benchmark to assess the performance of more sophisticated asset allocation strategies. To avoid a multiple-testing problem and since one of the major goals of this study is to outperform the $1/N$ rule, we restrict our focus to the comparison of the equal-weighted with the alternative portfolios. For each case, a two-sided $p$-value is obtained by the prewhitened $\text{HAC}_{PW}$ test proposed by Ledoit and Wolf (2011) for the null hypothesis of equal standard deviations and by Ledoit and Wolf (2008) for the null hypothesis of equal Sharpe ratios.

[Insert Table 3 about here]

The results from Table 3 indicate that portfolios using machine learning consistently outperform the equal-weighted benchmark in terms of standard deviation and Sharpe ratio by a wide margin. Using a covariance matrix based on machine learning factors, can lead to a statistically significant decrease in out-of-sample standard deviation of up to 29% and a significant increase in Sharpe ratio of over 25% relative to the $1/N$ portfolio. Factors based on sparse principal component analysis and autoencoders are found to yield the best performance, with shallower neural networks outperforming those with more hidden layers. Additionally, estimators that allow the loadings or error covariance to be time-varying outperform the static or dynamic covariance specifications. It is interesting to note that the factor models whose covariance structure diverges most from the sample covariance correspond to those that generate the highest portfolio performance. The latent factor models that tend to diverge the most from the sample estimator are based on PCA, sparse PCA and an autoencoder with one hidden layer.

Unsupervised learning approaches used to derive the latent factors tend to produce portfolios with lower standard deviation. The best performing model is AEN1, with monthly standard deviation of 3.216% and 3.318% depending on covariance specification, except for dynamic beta covariance where SPLS yields the lowest standard deviation with a value of 3.261%. The outperformance of the shallow autoencoder (AEN1) from the EW strategy is between 2.9% and 3.3% per annum, varying based on the covariance specification. The shallow autoencoder and SPCA have the highest Sharpe ratios, with values between 0.82 and 0.85 in annual terms. AEN1 has the highest ratio in the dynamic beta covariance specification, while SPCA is the best performing model for the remaining specifications. Overall, the best performing portfolios are based on the dynamic error covariance or dynamic beta covariance specifications, while portfolios based on static or dynamic factor covariance generate comparable performance.

The outperformance over $1/N$ portfolio across factors and model specifications in terms of standard deviation is statistically significant at the 1% level. The outperformance in terms of Sharpe ratio is, however, statistically significant only for latent factor models. For the Sharpe ratio, latent factor models generate statistically significant outperformance, compared to observed factors and the sample estimator that yield insignificant results. Specifically, SPCA consistently outperforms the $1/N$ portfolio at the 1% level, followed by AEN1 that for dynamic error covariance is statistically significant at the 5% level and at the 1% level for the remaining specifications. All other models are statistically significant at the 5% or 10% levels. The exception are portfolios based on PLS, for static or dynamic factor specifications, which are the only latent factor models that are not found to be statistically different from the $1/N$ portfolio.

The latent factor models consistently outperform the sample estimator across all covariance specifications, for both performance measures. For example, a portfolio based on autoencoders would result in a standard deviation and Sharpe ratio which improve by up to 8% and 17%, respectively, compared to the portfolio formed using the sample estimator. On the other hand, the observed factor models improve upon the sample estimator primarily in terms of Sharpe ratio, with the Fama-French three-factor model generating better results than the market. Latent factor models generate lower standard deviation and higher Sharpe ratio than models using the market or FF3 factors. The decrease in annualized standard deviation compared to portfolios based on the worst performing factor model (Market) can be up to 1.5%. Furthermore, choosing the machine learning factors can lead to an increase in Sharpe ratio of 10% to over 16% relative to the market factor, depending on the covariance specification. Latent factors tend to favor a time-varying error covariance, $\Sigma_{u,t}$, while the observed factors yield better results when the matrix $B_t$ is time-varying.

The results throughout the paper are based on models using a validation window consisting of the last 20% of the observations in the rolling window, as described in Section 3. We also examine the performance of the portfolios when the size of the validation window is reduced to 10% or increased to 30% of the observations in the rolling window. The results are reported in Table A3 in the Appendix and are qualitatively similar to the baseline case, with methods such as SPCA and autoencoders favoring the longer validation window.

Furthermore, we examine the out-of-sample monthly portfolio performance using several alternative risk measures, including the mean absolute deviation (MAD), value-at-risk (VaR) and conditional value-at-risk (CVaR) of the 480 out-of-sample portfolio returns in excess of the risk-free rate. Tail risk is of particular importance for portfolios during periods of financial distress. The historical VaR at the $100(1-a)\%$ level is computed as $-z_a\bar{\sigma}_p - \bar{r}_p$, where $z_a = \Phi^{-1}(1-a)$ is the $a\%$ quantile of a standard normal distribution and $\Phi$ is the cumulative standard normal distribution function. The historical CVaR at the $100(1-a)\%$ level calculated as $(1-a)^{-1}\varphi(z_a)\bar{\sigma}_p - \bar{r}_p$, where $\varphi$ is the probability density function of the standard normal distribution. The results for the MAD along with the VaR and CVaR calculated at the 95% confidence level are reported in Table 4.

[Insert Table 4 about here]

Results for the alternative risk criteria also point towards the benefits of using machine learning latent factors for covariance estimation, with all models outperforming the equal-weighted benchmark. Using a factor-implied covariance matrix based on machine learning can lead to an improvement of up to 27% in terms of MAD and between 23% to 26% in terms of VaR and CVaR compared to the $1/N$ portfolio, offering a measure of protection to investors concerned with tail risk.

The best performing models are based on autoencoders or SPCA, outperforming the EW benchmark by up to 2.9%, 1.26% and 1.57% per annum, in terms of MAD, VaR and CVaR respectively. Latent factor models outperform portfolios based on the sample covariance or observed factor models. In comparison to the market factor the best performing factors based on machine learning can lead to an over 17% reduction in terms of MAD and an improvement of over 13% and 12% in terms of VaR and CVaR respectively. Observed factors tend to favor covariance specifications that allow $B_t$ to vary over time. For latent factors, the greatest improvement from the benchmark, according to MAD, occurs when the residual covariance is time-varying, while the results for VaR and CVaR appear more favorable in the static specification.

We further investigate the economic value of the covariance matrix estimates using the certainty equivalent return (CER), defined as: $\text{CER} = \bar{r}_P - 0.5\gamma\bar{\sigma}_P^2$, where $\bar{r}_P$ and $\bar{\sigma}_P^2$ are the mean and variance of the portfolio returns over the out-of-sample period. The CER can be interpreted as the risk-free return that a mean-variance investor with a coefficient of relative risk aversion $\gamma$ is willing to accept instead of investing in the risky portfolio. We consider values of the risk aversion parameter $\gamma$, that roughly correspond to an aggressive ($\gamma = 2$), moderate ($\gamma = 5$) or conservative ($\gamma = 10$) investor.[12] Given that the optimization objective is to minimize variance, the portfolios would be of interest to investors more sensitive to risk, which makes CER with higher values of $\gamma$ a more representative performance measure.

Following Neely, Rapach, Tu and Zhou (2014), we report the difference in monthly CER ($\Delta$CER), which is equivalent to the percentage CER generated by the portfolio based on alternative covariance matrices minus that of the EW benchmark. ΔCER can be interpreted as the performance fee that the investor would be willing to pay to use the information of each alternative covariance estimator instead of the benchmark. To test the statistical significance of the CER gains, relative to the EW allocation approach, we use the two-sided $p$-value obtained by the test developed by DeMiguel, Garlappi and Uppal (2009), for the null hypothesis of equal CER.

---

[12] Similar values for the parameter of risk aversion have been used in DeMiguel, Garlappi and Uppal (2009).

[Insert Table 5 about here]

The results reported in Table 5 show that portfolios based on machine learning outperform simpler benchmarks, leading to statistically significant utility gains that exceed those of the EW by 2.5% to 4.5% on an annual basis, for investors with moderate or conservative risk preferences, respectively. The CER for the $1/N$ strategy decreases as the parameter of risk aversion increases, turning negative when $\gamma = 10$. All portfolios generate positive ΔCER, with the outperformance from the EW portfolio increasing as the parameter of risk aversion, $\gamma$, increases in value. The best performing model in the case of the static factor covariance is SPCA with monthly ΔCER between 0.109% and 0.346% depending on the degree of risk aversion. In the case of dynamic estimators, autoencoders yield the highest ΔCER for $\gamma = 10$. For $\gamma = 5$, AEN1 outperforms the other models for dynamic beta covariance, while SPCA performs best when $\Sigma_{f,t}$ and $\Sigma_{u,t}$ are time varying.

Latent factor models generate statistically significant results at the 1% level across all covariance specifications for a conservative investor. In the case of a moderate investor, the portfolios that offer significant CER gains over the EW allocation at the levels of 5% or 10%, are those based on unsupervised methods, except for SPLS which generates statistically significant results when $B_t$ varies over time. Compared to latent factor models, portfolios using the sample estimator or observed factors do not generate statistically significant CER relative to the naïve allocation when $\gamma = 5$. For $\gamma = 10$, the outperformance from the EW portfolio tends to be insignificant, with only a few exceptions. The sample covariance has statistically significant ΔCER at the 10% level. Portfolios based on observed factors yield statistically significant results only in the case of the dynamic beta covariance specification or at the 10% level for the Fama-French three-factor model based on the dynamic error covariance.

However, in the case of an aggressive investor ($\gamma = 2$) the outperformance of the alternative estimators of the covariance matrix is not statistically significant.

To summarize, we find that the performance of the models significantly improves relative to the benchmark as the coefficient of risk aversion increases. The comparison of the ΔCER across models confirms the conclusions reached from the analysis of the Sharpe ratios, with unsupervised methods that induce sparsity or introduce non-linearities to the factors yielding the highest significant performance. The factor models can earn a consistently higher out-of-sample CER return than the EW portfolio, especially when latent factors are derived by machine learning methods, with performance that remains statistically significant for investors with moderate or conservative risk appetite. Ideally the results would remain statistically significant across all values of $\gamma$. However, since the objective of this asset allocation exercise is to minimize variance, the portfolios would primarily be of interest to investors that are not willing to undertake a high degree of risk, while more aggressive investors would opt for an objective function that seeks to maximize return or Sharpe ratio.

## 5.2 Subperiod Analysis

In this section we examine portfolio performance during different subperiods as defined by market volatility, inflation and credit spread. The impact of different market regimes on asset allocation has been well documented in the literature (see Ang and Bekaert, 2002a; Guidolin and Timmermann, 2008; Pettenuzzo and Timmermann, 2011). To analyze the results under different market regimes, we derive the periods of high and low market volatility using a Markov Switching model.[13]

[Insert Table 6 about here]

During high volatility periods (Panel A) all portfolios outperform the equal-weighted benchmark in terms of both standard deviation and Sharpe ratio. Strategies based on machine learning models, especially unsupervised methods, generate the highest statistically significant results. Specifically, the best performing strategy is that of a shallow autoencoder with monthly Sharpe ratios ranging from 0.197 to 0.205 and significant at the 1% level. An exception is the case of covariances with loadings allowed to vary over time, which leads to the Fama-French three-factor model having the highest performance with a ratio of 0.213 that is significant at the 5% level. For low volatility periods (Panel B) machine learning portfolios are less risky by up to 6%, albeit statistically indistinguishable from the $1/N$ portfolio. Unsupervised methods outperform supervised methods, producing a standard deviation that is lower by 13% to 28% than portfolios based on observed factors. Furthermore, portfolios based on machine learning are statistically indistinguishable from the EW portfolio also in terms of Sharpe ratio. In contrast, portfolios based on observed factors are more negatively affected than those based on latent factors and exhibit significantly reduced performance from the benchmark.

---

[13] The regimes are determined based on the filtered probabilities of the following two-state Markov Switching model: $r_{m,t} = \mu_s + e_{t,s}$, with $e_{t,s} \sim N(0, \sigma_{t,s}^2)$, where $r_m$ is the market factor return, $s$ represents the latent state and $\mu_s$ and $\sigma_s^2$ denote the state dependent mean and variance. When the filtered probability of the low volatility state is lower than 0.5 the market is in a high-volatility period, while observations where the filtered probability of the low volatility state is higher than 0.5 are low-volatility periods.

Asset allocation is also affected by shifts in the economic environment due to inflation and credit spread, where the presence of separate regimes has been detected (see Ang and Bekaert, 2002b; Kritzman, Page and Turkington, 2010 for example). To this end, we also investigate changes in portfolio performance during subperiods of high and low inflation or credit spread.[14]

[Insert Table 7 about here]

Table 7 presents the results for different inflation subperiods. During periods of high inflation (Panel A) all strategies dramatically outperform the EW portfolio. Additionally, an investor interested more in minimizing volatility would prefer latent factor models. Specifically, portfolios based on SPCA or AEN have statistically significant standard deviations that are up to 24% lower than the benchmark. On the other hand, an investor opting to maximize the Sharpe ratio would choose observed factor models, which offer slightly higher ratios than latent factors, but are significant at a lower level. The results for observed factors tend to vary in terms of significance, while for latent factors they are consistently significant at the 1% level across specifications for both measures. In contrast, during low inflation periods, the Sharpe ratio of the EW allocation is high, with strategies based on unsupervised methods outperforming it, albeit insignificantly. In terms of standard deviation all strategies yield improved and significant performance, with unsupervised methods generating the lowest risk. An investor would realize a reduction of over 20% in standard deviation and an increase of up to 7% in Sharpe ratio by choosing strategies based on machine learning. Comparing across covariance specifications, those that have time-varying B and $\Sigma_u$ tend to have higher performance.

---

[14] The series used are based on the year-on-year change of Consumer Price Index for all urban consumers, for inflation and the difference between Moody's Baa corporate bond yield and the yield on a 10-year Treasury, for credit spread. Both series are retrieved from FRED. Periods of high (low) inflation or credit spread are defined by whether inflation or credit spread for a specific month is higher (lower) than the median over the out-of-sample period.

[Insert Table 8 about here]

Turning to the results for the credit spread in Table 8, we observe that during the high credit spread regime (Panel A), all models outperform the EW. The standard deviation is significantly lower than the $1/N$ strategy at the 1% level for all portfolios, however, the Sharpe ratios tend to be insignificant even though they are higher than the benchmark. Some outliers include the latent factor models for strategies based on dynamic error covariance, the Fama-French factors for time-varying loadings and the shallow autoencoder for static factor models, which produce statistically significant values. In the low credit spread regime the differences across models become more apparent. Compared to periods with a high spread, the covariance specification that benefits most asset allocation is based on static factor models or a covariance with dynamic loadings or factors. The standard deviation of portfolios based on latent factors is lower by over 17% from the EW and statistically significant at the 1% level across specifications, as opposed to the observed factors that are significant only for dynamic loadings. Finally, the factor-based portfolios of unsupervised methods have significant Sharpe ratios, which can be 28% higher than those of the benchmark portfolio.

## 5.3 Effects of Transaction Costs

Following Han (2006), we consider the breakeven transaction costs in basis points that would cause an investor to be indifferent between a certain strategy and the benchmark strategy. The breakeven transaction costs in terms of Sharpe ratio of a portfolio relative to the equal-weighted portfolio, $c_{ew}$, are calculated as $\Delta SR/\Delta TO$, where $\Delta SR = SR_p - SR_{ew}$ is the difference in Sharpe ratios of the alternative portfolio from the equal-weighted benchmark and $\Delta TO = TO_p - TO_{ew}$ is the difference in the respective average turnover. Breakeven transaction costs become important in the absence of reliable estimates of transaction costs, with positive values of $c_{ew}$ indicating that the alternative model outperforms the benchmark.

[Insert Table 9 about here]

Table 9 shows that all models generate positive $c_{ew}$, indicating outperformance from the benchmark. However, an investor would realize greater economic benefits over the EW portfolio by choosing machine learning portfolios, since they exhibit higher breakeven transaction costs by a factor of three from portfolios based on observed factors. The breakeven transaction costs are higher for unsupervised methods than the sample estimator or the remaining factor models. Furthermore, static and dynamic factor covariance strategies produce the highest breakeven transaction costs across all specifications. Portfolios with factors estimated by SPCA have annualized breakeven transaction costs between 48 and 69 bps, while a shallow autoencoder would realize $c_{ew}$ from 46 to 64 bps per annum. On the contrary, the worst performing factor-based portfolios are those of the market factor, which yield the lowest breakeven transaction costs (14 to 22 basis points) followed by that of the sample estimator with breakeven transaction costs of approximately 22 basis points. This concurs with the earlier evidence from portfolio turnover, which shows that these strategies have lower rebalancing requirements. The results for average turnover and breakeven transaction costs suggest that rebalancing frequency is a key contributor to portfolio performance. This is in keeping with the literature (see Han, 2006), that finds increasing breakeven transaction costs for reduced rebalancing frequency.

We also examine the performance of the portfolio for specific levels of transaction costs. In this setting transaction costs would arise from changes to the stock universe from one month to the next and from the change in weights of stocks that remain in the stock universe for multiple iterations. The portfolio's return is modified to account for transaction costs based on portfolio turnover. Given a transaction cost of $c$, the trading cost of the entire portfolio is $c\|\omega_{t+1} - \omega_t\|_1$. The return of the portfolio after transaction costs becomes $r_{p,t+1}^{TC} = (1 + r_{p,t+1})(1 - c\|\omega_{t+1} - \omega_t\|_1) - 1$. We consider values for transaction costs of $c \in \{5, 20\}$ basis points.[15]

The results are reported in Table A4 in the Appendix, for transaction costs of $c = 5$ basis points (Panel A) and for $c = 20$ basis points (Panel B). Overall, the results remain consistent with those of Table 9. Models that exhibit higher average monthly turnover and lower breakeven transaction costs are more affected by transaction costs. In terms of standard deviation, the results remain virtually unchanged compared to the baseline case, regardless of the level of transaction costs. Unsurprisingly, the $1/N$ portfolio suffers the least when transaction costs are introduced. For transaction costs of 5 bps the values of the Sharpe ratios also remain relatively unaffected, but there is a change in their levels of significance. Specifically, sparse PCA and the shallow autoencoder are now significant at the 5% instead at the 1% level, while PLS-type methods and AEN4 are no longer statistically significant in most covariance specifications. When transaction costs are increased to 20 bps, portfolios based on the sample covariance and market factor either marginally or fail to outperform the EW benchmark. The latent factors that significantly outperform the benchmark at the 10% level are those of SPCA and AEN1, for all specifications except when the error covariance is dynamic.

## 5.4 Portfolios Allowing Short-Selling

In this section we consider the case of the minimum-variance portfolio in the absence of short-selling constraints. Imposing a restriction to short positions is known to limit extreme portfolio weights. It is interesting to investigate how removing short-selling constraints, thus potentially allowing for large short positions, affects portfolio performance. The objective becomes

$$\underset{\omega}{\operatorname{argmin}} \, \omega'\hat{\Sigma}_r\omega, \quad \text{s.t.} \quad \omega'i_N = 1. \tag{20}$$

---

[15] The value of 5 bps may be low by academic standards, where values as high as 50 bps (Kirby and Ostdiek, 2012) have been used. Other studies are less conservative and use a range of transaction costs. Ledoit and Wolf (2017) consider transaction costs of $c \in \{3, 50\}$ bps, pointing out that 3 bps is representative for liquid stocks, while Moura, Santos and Ruiz (2020), use $c \in \{5, 10\}$ bps.

Compared to the $1/N$ portfolio or the problem in (15), the asset weights, $\omega$, can be negative. The results are reported in Table 10.

[Insert Table 10 about here]

Removing the short-selling constraint leads to an overall decrease in portfolio performance. Portfolios based on latent factors remain relatively unaffected, in terms of magnitude and significance of the various measures. On the contrary, portfolios based on the sample covariance and observed factors show a considerable increase in standard deviation. However, an increase in return results in Sharpe ratios of the same level as portfolios that allow short-sales. Furthermore, when loadings are dynamic, allocations based on observed factors generate standard deviations at a level closer to the baseline results. The increase in volatility for the observed factors is between 6% and 22% depending on the covariance specification. Allowing the portfolio weights to be negative has also led to a dramatic increase in turnover and a decrease in breakeven transaction costs for observed factors and the sample estimator. This indicates that imposing a short-selling constraint to portfolios based on the sample covariance or observed factors helps to reduce estimation error due to covariance misspecification. In contrast, even without restricting short positions, the risk of portfolios based on latent factors remains at a similar level to the portfolios with $\omega_i \geq 0$.

The results on the maximum weight, standard deviation of portfolio weights and deviation of the weights from those of the EW portfolio are reported in Table A5 in the Appendix. When short-selling is allowed, portfolios based on the sample estimator, followed by those of observed factors, have weights that are larger, more volatile and further away from those of the $1/N$ portfolio than portfolios based on latent factors. Equivalently, minimum-variance portfolios based on a time-varying residual covariance matrix exhibit higher values across the three measures than the remaining specifications.

## 5.5 Turnover-Constrained Portfolios

We follow Olivares-Nadal and DeMiguel (2018) who show that incorporating a $l_p$ transaction cost term in the portfolio problem may help to reduce the impact of estimation error. Here we consider the case of proportional transaction costs and modify the minimum-variance optimization problem by adding a $l_1$ transaction cost term, which is equivalent to assuming that transaction costs are proportional to the amount traded. The new constrained optimization problem becomes

$$\underset{\omega}{\mathrm{argmin}}\, \omega' \hat{\Sigma}_r \omega + \kappa \|\omega - \omega_0\|_1, \quad \text{s.t.} \quad \omega' i_N = 1, \quad \omega_i \geq 0, \text{for } i = 1, \dots, N, \tag{21}$$

where $\kappa$ is the transaction cost parameter that controls for the degree to which portfolio turnover is penalized and $\omega_0$ are the weights of the portfolio from the previous period before rebalancing. For the first period of the expanding window the weights $\omega_0$ are initialized based on the original minimum-variance allocation. The transaction costs are set to $\kappa = 5$ bps for each asset. When $\kappa = 0$ the above optimization problem becomes equivalent to the one in (15). Table 11 presents the results for the penalized portfolios.

[Insert Table 11 about here]

Overall, including a turnover penalty has an adverse effect for portfolios based on unsupervised methods that exhibit low turnover, but is beneficial to portfolios with high turnover such as supervised methods and observed factors. This improved performance is due to an increase in portfolio return since standard deviation remains either at the same or at a higher level. The effects of regularization become more evident when looking at the results in Panel B. Turnover for observed factors and supervised methods is reduced by approximately 50%, while the breakeven transaction costs more than double compared to unpenalized portfolios.

The results for the remaining weight characteristics for the penalized portfolios can be found in Table A6 in the Appendix. The positive effects of regularization are apparent in the cases of the sample estimator and the observed factors, where the standard deviation of the weights is reduced by up to 7.5% and $\mathrm{MAD}_{EW}$ by over 6% compared to the baseline case.

## 5.6 Increased Number of Assets

Here we examine how portfolio performance is affected when the number of assets increases. Along with the baseline case for $N = 100$, the results for portfolios for $N = \{200, 300, 400, 500\}$ largest stocks by market capitalization are presented in Figure 5 for the case of the static factor covariance specification.[16]

[Insert Figure 5 about here]

When the number of assets is increased, the $1/N$ portfolio is consistently outperformed in terms of standard deviation and Sharpe ratio by all portfolios. The standard deviation of the EW portfolio increases with the size of the portfolio, while the pattern for observed factors is mixed, first increasing for $N = \{200, 300\}$ and then decreasing slightly for $N = \{400, 500\}$. The volatility of latent factor models slightly increases for $N = 200$ and then remains relatively stable across different portfolio sizes. The Sharpe ratio for all factor models is highest for $N = 100$, decreases when $N = \{200, 300\}$ and then increases again for $N = \{400, 500\}$, with latent factor models producing higher ratios than observed factors. Average monthly turnover steadily increases with the number of stocks in the portfolio, with observed factor models generating higher turnover than latent factor models consistently across all portfolio sizes, indicating that portfolios based on the Market and FF3 factors are more sensitive to transaction costs. The results show that increasing the size of the portfolios has a considerable negative impact on breakeven transaction costs, with all models experiencing a sharp decrease in $c_{ew}$, indicating larger portfolios are more sensitive to transaction costs. Comparing across factor specifications, unsupervised methods generate the highest breakeven transaction costs for different values of $N$. In contrast, portfolios based on the market factor generate negative breakeven transaction costs for $N = \{200, 300\}$, failing to outperform the benchmark, and then positive but small values for larger portfolios.

## 5.7 Different Data Sets

In our analysis up to this point we have focused on individual stocks as assets, since this case tends to be of greater interest to investors and fund managers. On the other hand, a great number of academic studies have been devoted to asset portfolios. To this end we consider two alternative data sets. The first is comprised of 49 portfolios based on industry classification and the other of 21 assets, including 20 portfolios sorted based on size and book-to-market (BM) in addition to the S&P 500.[17] The size and splitting of the sample are the same as for individual stock data.

---

[16] The results for the remaining covariance specifications exhibit a similar pattern to that of the static case and are presented in Figures A7, A8 and A9 in the Appendix, for the cases when $B$, $\Sigma_f$ and $\Sigma_u$ are dynamic, respectively. Additionally, the results for autoencoders with three and four hidden layers were similar to those of an autoencoder with 2 hidden layers and are not presented for the sake of brevity.

Looking at the results for the 49 industry portfolios (Table 12), all minimum-variance portfolios outperform the EW allocation, in terms of both standard deviation and Sharpe ratio. The best performing portfolios are those based on PLS methods and the market factor. The standard deviation is statistically significant at the 1% level for all models; however, latent factor methods generate significant outperformance at the 1% level in terms of Sharpe ratios, compared to the sample estimator or observed factors, which are significant at the 5% or 10% levels. The factors that generate the best performing portfolios in terms of breakeven transaction costs and average turnover depend on the covariance specification. Observed factors tend to favor the static covariance specification, while machine learning factors yield lower turnover and higher breakeven transaction costs relative to observed factors, for dynamic covariance specifications.[18]

[Insert Table 12 about here]

Specifically, using a covariance matrix based on SPLS or the market factor leads to Sharpe ratios between 0.78 and 0.83 per annum, which is an increase of 50% compared to the EW portfolio. The annual decrease in standard deviation from the $1/N$ portfolio is between 3.11% to 4.40% depending on the factor and covariance specification, with latent factors favoring allocations based on covariance matrices where $\Sigma_u$ is dynamic. Turnover varies largely depending on the model. Observed factors generate monthly turnover between 2% (Market) and 2.5% (FF3) when the covariance is static, while in a dynamic specification average monthly turnover significantly increases to the range of 6.2% to 25.8% for the market and from 8.6% to 32.1% for the Fama-French three-factor model. In contrast, latent factor models generate lower turnover in a dynamic specification. Supervised approaches tend to require less frequent rebalancing than unsupervised methods, with the exception of the dynamic beta specification that leads to PCA and AEN portfolios with lower turnover (between 16% and 17.8%). The breakeven transaction costs indicate that observed factors yield higher monthly $c_{ew}$ when the factor-based covariance is static (392 bps for the Market and 275 bps for FF3). In a dynamic specification PLS and SPLS yield the highest breakeven transaction costs ranging from 38.9 to 157.5 basis points per month, compared to observed factors (20.3 to 122 bps).

---

[17] Following Wang (2005) and DeMiguel, Garlappi and Uppal (2009), we exclude the five portfolios containing the largest firms because the market, SMB, and HML are almost a linear combination of the Fama and French portfolios.

[18] Return observations are not available from the beginning of the sample period for all industry portfolios. This causes a small increase to monthly turnover as the portfolios that were missing observations in the beginning of the backtest enter the asset universe.

Turning to the results for the 20 book-to-market portfolios and the S&P 500 (Table 13) the standard deviation is significantly lower at the 1% level from the benchmark for all alternative allocations. Fama and French factors and the sample estimator yield standard deviations that are lower by approximately 2.7% from the $1/N$ portfolio per annum compared to a decrease from 1.4% to 1.9% for the latent factor models. The variation in Sharpe ratios across models is small, with the Market factor producing the highest ratio. Sharpe ratios are statistically higher than the benchmark at the 5% or 10% level. However, portfolios based on the sample estimator or observed factors, exhibit increased turnover compared to latent factor models, which has a considerable effect on breakeven transaction costs. For example, the monthly $c_{ew}$ for PLS, SPLS and shallow autoencoders for the static covariance specification, is between 154.5 and 171.7 compared to 94.2 to 97.8 bps for the observed factors or 102.3 bps for the sample covariance.

[Insert Table 13 about here]

## 6. Conclusion

In this paper we explore the properties of machine learning factor-based portfolios. We then examine whether factor-implied covariance matrices based on machine learning dimensionality reduction techniques can benefit minimum-variance portfolios comprised of individual stocks. Overall, our findings indicate that machine learning can help improve factor-based portfolio optimization.

When exploring the characteristics of machine learning portfolios, we find that factors based on PCA and PLS exhibit a stronger relationship with commonly used factor proxies than autoencoders. Furthermore, the structure of covariance matrices that are dynamic or based on unsupervised methods diverges the most from that of the sample estimator. Unsupervised learning methods yield portfolios that require less frequent rebalancing, with weights that are less volatile and more diversified relative to their supervised counterparts or observed factors.

Our analysis also shows that methods which induce sparsity and autoencoder neural networks tend to be the best performing models. Specifically, we find that the proposed models can lead to a statistically significant reduction in portfolio volatility of up to 3.3% per annum and a significant increase in the Sharpe ratios of approximately 25%, relative to the $1/N$ portfolio. Furthermore, using a factor-implied covariance matrix based on machine learning can lead to an improvement of up to 27% in terms of MAD and between 23% to 26% in terms of tail risk measures compared to simpler benchmarks. These findings become more acute as an investor's sensitivity to risk increases. Investors with moderate or conservative risk preferences would realize statistically significant utility gains that exceed those of the equal-weighted allocation by 2.5% to 4.5% on an annual basis. The benefits of machine learning to factor-based allocations are increased during periods of high volatility and are also evident in different inflation and credit spread regimes. Furthermore, the results show that shallower neural networks outperform deeper architectures, which falls in line with conclusions reached by recent applications of machine learning in finance. Finally, when comparing various static and dynamic factor model specifications, the results indicate that models which allow the error component of the covariance matrix to vary over time can deliver increased performance but at the cost of higher portfolio turnover.

# Tables

**Table 1**
**Comparison of factor-based covariance matrices**

|  | Static Factor Covariance | | | Dynamic Beta Covariance | | | Dynamic Factor Covariance | | | Dynamic Error Covariance | | |
|---|---|---|---|---|---|---|---|---|---|---|---|---|
|  | Eig | Mag | Dir | Eig | Mag | Dir | Eig | Mag | Dir | Eig | Mag | Dir |
| Market | 0.913 | 0.076 | 0.973 | 0.933 | 0.105 | 0.954 | 0.926 | 0.070 | 0.973 | 0.908 | 0.082 | 0.973 |
| FF3 | 0.932 | 0.041 | 0.976 | 0.990 | 0.062 | 0.951 | 0.963 | 0.015 | 0.973 | 0.926 | 0.047 | 0.976 |
| PCA | 0.501*** | 0.742*** | 0.393*** | 0.517*** | 0.754*** | 0.384*** | 0.523*** | 0.725*** | 0.394*** | 0.485*** | 0.750*** | 0.393*** |
| PLS | 0.584*** | 0.616*** | 0.619*** | 0.603*** | 0.634*** | 0.580*** | 0.597*** | 0.601*** | 0.620*** | 0.573*** | 0.624*** | 0.619*** |
| SPCA | 0.487*** | 0.763*** | 0.479*** | 0.513*** | 0.757*** | 0.462*** | 0.489*** | 0.764*** | 0.474*** | 0.469*** | 0.771*** | 0.479*** |
| SPLS | 0.574*** | 0.626*** | 0.648*** | 0.595*** | 0.637*** | 0.610*** | 0.577*** | 0.619*** | 0.648*** | 0.563*** | 0.634*** | 0.648*** |
| AEN1 | 0.480*** | 0.761*** | 0.370*** | 0.501*** | 0.770*** | 0.363*** | 0.501*** | 0.745*** | 0.368*** | 0.461*** | 0.769*** | 0.370*** |
| AEN2 | 0.498*** | 0.744*** | 0.389*** | 0.509*** | 0.76*** | 0.380*** | 0.532*** | 0.720*** | 0.388*** | 0.479*** | 0.753*** | 0.389*** |
| AEN3 | 0.505*** | 0.735*** | 0.404*** | 0.506*** | 0.759*** | 0.397*** | 0.551*** | 0.701*** | 0.404*** | 0.487*** | 0.743*** | 0.404*** |
| AEN4 | 0.507*** | 0.734*** | 0.405*** | 0.511*** | 0.757*** | 0.396*** | 0.551*** | 0.702*** | 0.406*** | 0.489*** | 0.742*** | 0.405*** |

This table reports monthly measures that compare the factor-implied covariance matrices to the sample estimator, based on total covariation (Eig), Magnitude (Mag) and direction (Dir), over the out-of-sample period from January 1980 to December 2019. The results are presented for four factor-implied covariance specifications: static factor covariance, dynamic beta covariance, dynamic factor covariance and dynamic error covariance. The factor specifications are based on the single factor model (Market), the Fama-French 3-factor model (FF3), principal component analysis (PCA), partial least squares (PLS), sparse principal component analysis (SPCA), sparse partial least squares (SPLS) and autoencoders with 1, 2, 3 and 4 hidden layers (AEN). The significant deviation from the sample estimator of the alternative factor-implied covariance matrices from a covariance matrix based on the market factor is denoted by: *, **, and *** for significance at the 10%, 5%, and 1% level, respectively.

**Table 2**
**Characteristics of the portfolio weight vectors**

*A. Average turnover and maximum weight*

|  | TO | MAX |
|---|---|---|
| EW | 1.081 | 0.010 |
| Sample | 41.269 | 0.383 |

|  | Static Factor Covariance | | Dynamic Beta Covariance | | Dynamic Factor Covariance | | Dynamic Error Covariance | |
|---|---|---|---|---|---|---|---|---|
|  | TO | MAX | TO | MAX | TO | MAX | TO | MAX |
| Market | 41.297 | 0.302 | 60.745 | 0.296 | 43.957 | 0.433 | 56.045 | 0.445 |
| FF3 | 42.785 | 0.311 | 71.744 | 0.312 | 47.423 | 0.450 | 57.039 | 0.421 |
| PCA | 29.737 | 0.070 | 35.064 | 0.078 | 30.017 | 0.073 | 46.098 | 0.139 |
| PLS | 33.408 | 0.095 | 39.693 | 0.092 | 33.496 | 0.094 | 49.269 | 0.157 |
| SPCA | 29.931 | 0.071 | 36.287 | 0.087 | 29.973 | 0.071 | 46.137 | 0.134 |
| SPLS | 33.372 | 0.100 | 40.552 | 0.108 | 33.332 | 0.098 | 49.203 | 0.194 |
| AEN1 | 31.196 | 0.066 | 36.925 | 0.075 | 31.396 | 0.067 | 46.615 | 0.119 |
| AEN2 | 31.060 | 0.066 | 35.694 | 0.079 | 31.368 | 0.068 | 46.855 | 0.123 |
| AEN3 | 33.136 | 0.074 | 37.462 | 0.076 | 33.577 | 0.073 | 48.857 | 0.166 |
| AEN4 | 33.346 | 0.072 | 37.276 | 0.076 | 33.885 | 0.078 | 49.029 | 0.132 |

*B. Standard deviation of the weights and mean absolute deviation from the equal-weighted portfolio*

|  | $SD_\omega$ | $MAD_{EW}$ |
|---|---|---|
| EW | 0.000 | 0.000 |
| Sample | 3.146 | 1.621 |

|  | Static Factor Covariance | | Dynamic Beta Covariance | | Dynamic Factor Covariance | | Dynamic Error Covariance | |
|---|---|---|---|---|---|---|---|---|
|  | $SD_\omega$ | $MAD_{EW}$ | $SD_\omega$ | $MAD_{EW}$ | $SD_\omega$ | $MAD_{EW}$ | $SD_\omega$ | $MAD_{EW}$ |
| Market | 2.824 | 1.616 | 2.452 | 1.535 | 2.769 | 1.582 | 3.433 | 1.711 |
| FF3 | 2.738 | 1.582 | 2.473 | 1.520 | 2.669 | 1.546 | 3.342 | 1.669 |
| PCA | 1.091 | 0.820 | 1.047 | 0.791 | 1.098 | 0.824 | 1.323 | 0.928 |
| PLS | 1.419 | 1.050 | 1.334 | 0.997 | 1.423 | 1.052 | 1.700 | 1.165 |
| SPCA | 1.067 | 0.772 | 1.059 | 0.777 | 1.064 | 0.771 | 1.288 | 0.882 |
| SPLS | 1.441 | 1.033 | 1.359 | 0.993 | 1.439 | 1.032 | 1.727 | 1.153 |
| AEN1 | 1.053 | 0.802 | 1.012 | 0.771 | 1.059 | 0.806 | 1.281 | 0.909 |
| AEN2 | 1.078 | 0.815 | 1.030 | 0.779 | 1.089 | 0.822 | 1.305 | 0.920 |
| AEN3 | 1.107 | 0.835 | 1.048 | 0.793 | 1.121 | 0.845 | 1.349 | 0.943 |
| AEN4 | 1.111 | 0.837 | 1.052 | 0.795 | 1.124 | 0.846 | 1.344 | 0.943 |

This table presents the monthly characteristics of the portfolio weight vectors. Panel A reports average turnover (TO) and maximum weight (MAX), whereas the standard deviation of the weights ($SD_\omega$) and mean absolute deviation from the equal-weighted benchmark ($MAD_{EW}$) can be found in Panel B. The average value of each weight characteristic over the out-of-sample period from January 1980 to December 2019 is reported. TO, $SD_\omega$ and $MAD_{EW}$ are reported as a percentage. The results are presented for the equal-weighted portfolio (EW) and minimum-variance portfolios based on the sample estimator (Sample) and four factor-implied covariance specifications: static factor covariance, dynamic beta covariance, dynamic factor covariance and dynamic error covariance. The factor specifications are based on the single factor model (Market), the Fama-French 3-factor model (FF3), principal component analysis (PCA), partial least squares (PLS), sparse principal component analysis (SPCA), sparse partial least squares (SPLS) and autoencoders with 1, 2, 3 and 4 hidden layers (AEN).

**Table 3**
**Portfolio performance based on standard deviation and Sharpe ratio**

|        | SD       | SR     |
|--------|----------|--------|
| EW     | 4.159    | 0.183  |
| Sample | 3.469*** | 0.209  |

|       | Static Factor Covariance | | Dynamic Beta Covariance | | Dynamic Factor Covariance | | Dynamic Error Covariance | |
|-------|-----------|-----------|-----------|-----------|-----------|-----------|-----------|-----------|
|       | SD        | SR        | SD        | SR        | SD        | SR        | SD        | SR        |
| Market | 3.630*** | 0.209     | 3.347*** | 0.218     | 3.640*** | 0.210     | 3.597*** | 0.205     |
| FF3   | 3.522***  | 0.218     | 3.329*** | 0.241     | 3.538*** | 0.213     | 3.469*** | 0.223     |
| PCA   | 3.361***  | 0.235**   | 3.290*** | 0.239**   | 3.360*** | 0.235**   | 3.268*** | 0.241**   |
| PLS   | 3.330***  | 0.229     | 3.276*** | 0.232*    | 3.335*** | 0.227     | 3.217*** | 0.238*    |
| SPCA  | 3.373***  | 0.241***  | 3.307*** | 0.243***  | 3.378*** | 0.240***  | 3.263*** | 0.246***  |
| SPLS  | 3.339***  | 0.233*    | 3.261*** | 0.242**   | 3.347*** | 0.231*    | 3.225*** | 0.242*    |
| AEN1  | 3.318***  | 0.239***  | 3.285*** | 0.243***  | 3.315*** | 0.239***  | 3.216*** | 0.244**   |
| AEN2  | 3.352***  | 0.232**   | 3.297*** | 0.235**   | 3.350*** | 0.232**   | 3.259*** | 0.237**   |
| AEN3  | 3.338***  | 0.238**   | 3.283*** | 0.240**   | 3.334*** | 0.238**   | 3.235*** | 0.246**   |
| AEN4  | 3.357***  | 0.225*    | 3.301*** | 0.227**   | 3.356*** | 0.224*    | 3.256*** | 0.227*    |

This table documents monthly portfolio performance measured using the standard deviation (SD) and Sharpe ratio (SR), over the out-of-sample period from January 1980 to December 2019. The results are presented for the equal-weighted portfolio (EW) and minimum-variance portfolios based on the sample estimator (Sample) and four factor-implied covariance specifications: static factor covariance, dynamic beta covariance, dynamic factor covariance and dynamic error covariance. The factor specifications are based on the single factor model (Market), the Fama-French 3-factor model (FF3), principal component analysis (PCA), partial least squares (PLS), sparse principal component analysis (SPCA), sparse partial least squares (SPLS) and autoencoders with 1, 2, 3 and 4 hidden layers (AEN). The significant outperformance of the alternative strategies from the equal-weighted strategy is denoted by: *, **, and *** for significance at the 10%, 5%, and 1% level, respectively.

**Table 4**
**Portfolio performance based on alternative risk measures**

|  | MAD | VaR | CVaR |
|---|---|---|---|
| EW | 3.693 | 6.079 | 7.817 |
| Sample | 3.111 | 4.981 | 6.431 |

|  | Static Factor Covariance | | | Dynamic Beta Covariance | | | Dynamic Factor Covariance | | | Dynamic Error Covariance | | |
|---|---|---|---|---|---|---|---|---|---|---|---|---|
|  | MAD | VaR | CVaR | MAD | VaR | CVaR | MAD | VaR | CVaR | MAD | VaR | CVaR |
| Market | 3.220 | 5.215 | 6.732 | 3.187 | 4.775 | 6.173 | 3.106 | 5.223 | 6.745 | 3.370 | 5.181 | 6.684 |
| FF3 | 3.168 | 5.024 | 6.495 | 3.065 | 4.674 | 6.065 | 3.178 | 5.066 | 6.545 | 3.254 | 4.932 | 6.381 |
| PCA | 2.819 | 4.739 | 6.143 | 2.775 | 4.625 | 6.000 | 2.826 | 4.737 | 6.141 | 2.917 | 4.586 | 5.952 |
| PLS | 2.882 | 4.715 | 6.107 | 2.886 | 4.627 | 5.995 | 2.911 | 4.728 | 6.121 | 2.798 | 4.525 | 5.869 |
| SPCA | 2.773 | 4.736 | 6.146 | 2.696 | 4.637 | 6.019 | 2.788 | 4.745 | 6.157 | 2.767 | 4.565 | 5.929 |
| SPLS | 2.963 | 4.715 | 6.111 | 2.897 | 4.577 | 5.940 | 2.952 | 4.734 | 6.132 | 2.883 | 4.526 | 5.874 |
| AEN1 | 2.774 | 4.664 | 6.050 | 2.699 | 4.605 | 5.978 | 2.780 | 4.661 | 6.046 | 2.785 | 4.504 | 5.848 |
| AEN2 | 2.841 | 4.737 | 6.137 | 2.867 | 4.648 | 6.025 | 2.847 | 4.735 | 6.135 | 2.874 | 4.590 | 5.952 |
| AEN3 | 2.772 | 4.697 | 6.092 | 2.818 | 4.613 | 5.984 | 2.806 | 4.691 | 6.084 | 2.831 | 4.526 | 5.878 |
| AEN4 | 2.893 | 4.767 | 6.169 | 2.853 | 4.681 | 6.061 | 2.894 | 4.768 | 6.171 | 2.878 | 4.618 | 5.978 |

In this table, we present the monthly portfolio performance measured using the mean absolute deviation (MAD), value-at-risk (VaR) and conditional value-at-risk (CVaR), over the out-of-sample period from January 1980 to December 2019. The VaR and CVaR are calculated at the 95% confidence level. The results are presented for the equal-weighted portfolio (EW) and minimum-variance portfolios based on the sample estimator (Sample) and four factor-implied covariance specifications: static factor covariance, dynamic beta covariance, dynamic factor covariance and dynamic error covariance. The factor specifications are based on the single factor model (Market), the Fama-French 3-factor model (FF3), principal component analysis (PCA), partial least squares (PLS), sparse principal component analysis (SPCA), sparse partial least squares (SPLS) and autoencoders with 1, 2, 3 and 4 hidden layers (AEN).

**Table 5**
**Portfolio performance based on certainty equivalent return**

|        | $\gamma = 2$ | $\gamma = 5$ | $\gamma = 10$ |
|--------|--------------|--------------|---------------|
| EW     | 0.589        | 0.330        | -0.103        |
| Sample | 0.016        | 0.095        | 0.226*        |

|        | Static Factor Covariance | | | Dynamic Beta Covariance | | | Dynamic Factor Covariance | | | Dynamic Error Covariance | | |
|--------|-------------|-------------|---------------|-------------|-------------|---------------|-------------|-------------|---------------|-------------|-------------|---------------|
|        | $\gamma = 2$ | $\gamma = 5$ | $\gamma = 10$ | $\gamma = 2$ | $\gamma = 5$ | $\gamma = 10$ | $\gamma = 2$ | $\gamma = 5$ | $\gamma = 10$ | $\gamma = 2$ | $\gamma = 5$ | $\gamma = 10$ |
| Market | 0.036 | 0.098 | 0.201 | 0.029 | 0.120 | 0.273* | 0.043 | 0.104 | 0.205 | 0.017 | 0.082 | 0.191 |
| FF3    | 0.056 | 0.129 | 0.252 | 0.101 | 0.194 | 0.350*** | 0.040 | 0.111 | 0.231 | 0.065 | 0.144 | 0.276* |
| PCA    | 0.087 | 0.177** | 0.327*** | 0.089 | 0.186** | 0.348*** | 0.087 | 0.177** | 0.327*** | 0.093 | 0.192* | 0.358*** |
| PLS    | 0.061 | 0.155 | 0.310*** | 0.065 | 0.163 | 0.327*** | 0.057 | 0.150 | 0.304*** | 0.074 | 0.178 | 0.352*** |
| SPCA   | 0.109 | 0.198** | 0.346*** | 0.104 | 0.200** | 0.359*** | 0.108 | 0.196** | 0.343*** | 0.106 | 0.206** | 0.372*** |
| SPLS   | 0.077 | 0.169 | 0.323*** | 0.092 | 0.192* | 0.359*** | 0.071 | 0.162 | 0.315*** | 0.086 | 0.190 | 0.362*** |
| AEN1   | 0.095 | 0.189** | 0.346*** | 0.101 | 0.199** | 0.362*** | 0.093 | 0.188** | 0.346*** | 0.092 | 0.197** | 0.371*** |
| AEN2   | 0.075 | 0.166* | 0.317*** | 0.077 | 0.173* | 0.334*** | 0.074 | 0.165* | 0.317*** | 0.075 | 0.176* | 0.343*** |
| AEN3   | 0.092 | 0.185** | 0.339*** | 0.090 | 0.188** | 0.351*** | 0.093 | 0.185** | 0.340*** | 0.102 | 0.204** | 0.375*** |
| AEN4   | 0.052 | 0.143 | 0.294*** | 0.051 | 0.147 | 0.307*** | 0.050 | 0.140 | 0.291*** | 0.043 | 0.143 | 0.311*** |

This table reports monthly portfolio performance measured using the difference in certainty equivalent return (ΔCER) for various levels of risk aversion, $\gamma$, over the out-of-sample period from January 1980 to December 2019. For the EW portfolio the monthly CER is reported as a percentage, while for the remaining portfolios the percentage ΔCER is provided, which is calculated as the difference in monthly CER between the alternative strategies and the equal-weighted strategy. The results are presented for the equal-weighted portfolio (EW) and minimum-variance portfolios based on the sample estimator (Sample) and four factor-implied covariance specifications: static factor covariance, dynamic beta covariance, dynamic factor covariance and dynamic error covariance. The factor specifications are based on the single factor model (Market), the Fama-French 3-factor model (FF3), principal component analysis (PCA), partial least squares (PLS), sparse principal component analysis (SPCA), sparse partial least squares (SPLS) and autoencoders with 1, 2, 3 and 4 hidden layers (AEN). The significant outperformance of the alternative strategies from the equal-weighted strategy is denoted by: *, **, and *** for significance at the 10%, 5%, and 1% level, respectively.

**Table 6**
**Portfolio performance during different volatility regimes**

*A. High volatility regime*

|  | SD | SR | | | | | | |
|---|---|---|---|---|---|---|---|---|
| EW | 5.037 | 0.128 | | | | | | |
| Sample | 3.994*** | 0.179 | | | | | | |

|  | Static Factor Covariance | | Dynamic Beta Covariance | | Dynamic Factor Covariance | | Dynamic Error Covariance | |
|---|---|---|---|---|---|---|---|---|
|  | SD | SR | SD | SR | SD | SR | SD | SR |
| Market | 4.075*** | 0.177 | 3.732*** | 0.197 | 4.153*** | 0.173 | 3.958*** | 0.180 |
| FF3 | 3.978*** | 0.192 | 3.792*** | 0.213** | 4.048*** | 0.175 | 3.846*** | 0.205 |
| PCA | 3.968*** | 0.181** | 3.874*** | 0.183** | 3.965*** | 0.181** | 3.820*** | 0.187** |
| PLS | 3.882*** | 0.180 | 3.820*** | 0.179 | 3.888*** | 0.178 | 3.700*** | 0.196* |
| SPCA | 3.988*** | 0.188** | 3.893*** | 0.188** | 3.994*** | 0.187** | 3.832*** | 0.192** |
| SPLS | 3.896*** | 0.186* | 3.798*** | 0.191* | 3.907*** | 0.183 | 3.714*** | 0.200* |
| AEN1 | 3.904*** | 0.198*** | 3.859*** | 0.198*** | 3.899*** | 0.197*** | 3.759*** | 0.205*** |
| AEN2 | 3.952*** | 0.178* | 3.878*** | 0.179* | 3.949*** | 0.178* | 3.819*** | 0.184* |
| AEN3 | 3.930*** | 0.189** | 3.857*** | 0.188** | 3.923*** | 0.189** | 3.776*** | 0.197** |
| AEN4 | 3.956*** | 0.179** | 3.882*** | 0.178* | 3.953*** | 0.179* | 3.801*** | 0.182* |

*B. Low volatility regime*

|  | SD | SR | | | | | | |
|---|---|---|---|---|---|---|---|---|
| EW | 2.283 | 0.411 | | | | | | |
| Sample | 2.493* | 0.298* | | | | | | |

|  | Static Factor Covariance | | Dynamic Beta Covariance | | Dynamic Factor Covariance | | Dynamic Error Covariance | |
|---|---|---|---|---|---|---|---|---|
|  | SD | SR | SD | SR | SD | SR | SD | SR |
| Market | 2.846*** | 0.285 | 2.677*** | 0.269** | 2.705*** | 0.307 | 2.986*** | 0.258* |
| FF3 | 2.705*** | 0.288 | 2.485* | 0.319 | 2.602** | 0.315 | 2.820*** | 0.267* |
| PCA | 2.157 | 0.414 | 2.14* | 0.423 | 2.160 | 0.414 | 2.199 | 0.410 |
| PLS | 2.268 | 0.377 | 2.229 | 0.393 | 2.271 | 0.377 | 2.319 | 0.357 |
| SPCA | 2.150* | 0.423 | 2.151* | 0.423 | 2.151* | 0.423 | 2.148 | 0.420 |
| SPLS | 2.269 | 0.378 | 2.230 | 0.396 | 2.269 | 0.378 | 2.316 | 0.361 |
| AEN1 | 2.171 | 0.381 | 2.163* | 0.392 | 2.173 | 0.381 | 2.167 | 0.372 |
| AEN2 | 2.163 | 0.409 | 2.154* | 0.415 | 2.165 | 0.409 | 2.169 | 0.402 |
| AEN3 | 2.174 | 0.399 | 2.158 | 0.408 | 2.177 | 0.399 | 2.195 | 0.398 |
| AEN4 | 2.176 | 0.377 | 2.160 | 0.388 | 2.181 | 0.376 | 2.207 | 0.366 |

In this table, we document the monthly portfolio performance measured using the standard deviation (SD) and Sharpe ratio (SR), during high (Panel A) and low (Panel B) volatility periods based on the filtered probabilities of a Markov-switching model estimated using the market factor. Observations where the filtered probability of the low volatility regime is above 0.5 are considered low-volatility periods, and observations where the filtered probability of the low volatility regime is below 0.5 are considered high-volatility periods. The results are presented for the equal-weighted portfolio (EW) and minimum-variance portfolios based on the sample estimator (Sample) and four factor-implied covariance specifications: static factor covariance, dynamic beta covariance, dynamic factor covariance and dynamic error covariance. The factor specifications are based on the single factor model (Market), the Fama-French 3-factor model (FF3), principal component analysis (PCA), partial least squares (PLS), sparse principal component analysis (SPCA), sparse partial least squares (SPLS) and autoencoders with 1, 2, 3 and 4 hidden layers (AEN). The significant outperformance of the alternative strategies from the equal-weighted strategy is denoted by: *, **, and *** for significance at the 10%, 5%, and 1% level, respectively.

**Table 7**
**Portfolio performance during different inflation regimes**

*A. High inflation regime*

|        | SD       | SR       |
|--------|----------|----------|
| EW     | 4.515    | 0.072    |
| Sample | 3.800*** | 0.178**  |

|        | Static Factor Covariance | | Dynamic Beta Covariance | | Dynamic Factor Covariance | | Dynamic Error Covariance | |
|--------|----------|----------|----------|----------|----------|----------|----------|----------|
|        | SD       | SR       | SD       | SR       | SD       | SR       | SD       | SR       |
| Market | 3.696**  | 0.224**  | 3.385*** | 0.261*** | 3.730**  | 0.217*   | 3.521*** | 0.258**  |
| FF3    | 3.657*** | 0.222**  | 3.412*** | 0.261*** | 3.693**  | 0.209**  | 3.548*** | 0.258*** |
| PCA    | 3.529*** | 0.186*** | 3.488*** | 0.200*** | 3.519*** | 0.187*** | 3.420*** | 0.206*** |
| PLS    | 3.552*** | 0.193*** | 3.510*** | 0.203*** | 3.558*** | 0.191*** | 3.335*** | 0.237*** |
| SPCA   | 3.570*** | 0.184*** | 3.502*** | 0.201*** | 3.574*** | 0.182*** | 3.450*** | 0.206*** |
| SPLS   | 3.564*** | 0.199*** | 3.488*** | 0.223*** | 3.572*** | 0.195*** | 3.347*** | 0.240*** |
| AEN1   | 3.552*** | 0.185*** | 3.513*** | 0.198*** | 3.547*** | 0.186*** | 3.417*** | 0.210*** |
| AEN2   | 3.539*** | 0.180*** | 3.510*** | 0.193*** | 3.530*** | 0.181*** | 3.438*** | 0.204*** |
| AEN3   | 3.513*** | 0.190*** | 3.498*** | 0.201*** | 3.500*** | 0.191*** | 3.404*** | 0.212*** |
| AEN4   | 3.547*** | 0.175*** | 3.518*** | 0.183*** | 3.543*** | 0.175*** | 3.417*** | 0.195*** |

*B. Low inflation regime*

|        | SD       | SR    |
|--------|----------|-------|
| EW     | 3.926    | 0.259 |
| Sample | 3.271*** | 0.231 |

|        | Static Factor Covariance | | Dynamic Beta Covariance | | Dynamic Factor Covariance | | Dynamic Error Covariance | |
|--------|----------|-------|----------|-------|----------|-------|----------|-------|
|        | SD       | SR    | SD       | SR    | SD       | SR    | SD       | SR    |
| Market | 3.598*   | 0.199 | 3.327*** | 0.193 | 3.594*   | 0.206 | 3.642    | 0.175 |
| FF3    | 3.448**  | 0.216 | 3.285*** | 0.228 | 3.453**  | 0.215 | 3.426*** | 0.202 |
| PCA    | 3.264*** | 0.265 | 3.176*** | 0.264 | 3.268*** | 0.264 | 3.182*** | 0.263 |
| PLS    | 3.200*** | 0.252 | 3.140*** | 0.252 | 3.205*** | 0.250 | 3.153*** | 0.239 |
| SPCA   | 3.258*** | 0.277 | 3.193*** | 0.270 | 3.263*** | 0.277 | 3.155*** | 0.271 |
| SPLS   | 3.209*** | 0.255 | 3.129*** | 0.254 | 3.216*** | 0.253 | 3.159*** | 0.243 |
| AEN1   | 3.179*** | 0.275 | 3.151*** | 0.272 | 3.178*** | 0.273 | 3.099*** | 0.266 |
| AEN2   | 3.243*** | 0.264 | 3.173*** | 0.262 | 3.246*** | 0.263 | 3.157*** | 0.257 |
| AEN3   | 3.237*** | 0.267 | 3.157*** | 0.264 | 3.239*** | 0.267 | 3.140*** | 0.267 |
| AEN4   | 3.246*** | 0.256 | 3.174*** | 0.255 | 3.248*** | 0.255 | 3.165*** | 0.246 |

In this table, we document the monthly portfolio performance measured using the standard deviation (SD) and Sharpe ratio (SR), during periods of high (Panel A) and low (Panel B) inflation. Periods of high (low) inflation are those when inflation for the specific month is higher (lower) than the median over the out-of-sample period. The results are presented for the equal-weighted portfolio (EW) and minimum-variance portfolios based on the sample estimator (Sample) and four factor-implied covariance specifications: static factor covariance, dynamic beta covariance, dynamic factor covariance and dynamic error covariance. The factor specifications are based on the single factor model (Market), the Fama-French 3-factor model (FF3), principal component analysis (PCA), partial least squares (PLS), sparse principal component analysis (SPCA), sparse partial least squares (SPLS) and autoencoders with 1, 2, 3 and 4 hidden layers (AEN). The significant outperformance of the alternative strategies from the equal-weighted strategy is denoted by: *, **, and *** for significance at the 10%, 5%, and 1% level, respectively.

**Table 8**
**Portfolio performance during different credit spread regimes**

*A. High credit spread regime*

|       | SD       | SR    |
|-------|----------|-------|
| EW    | 5.343    | 0.127 |
| Sample| 3.843*** | 0.176 |

|        | Static Factor Covariance | | Dynamic Beta Covariance | | Dynamic Factor Covariance | | Dynamic Error Covariance | |
|--------|------|-------|------|--------|------|-------|------|--------|
|        | SD   | SR    | SD   | SR     | SD   | SR    | SD   | SR     |
| Market | 3.966*** | 0.178 | 3.580*** | 0.223 | 4.112*** | 0.167 | 3.850*** | 0.195 |
| FF3    | 3.782*** | 0.192 | 3.555*** | 0.247** | 3.964*** | 0.168 | 3.662*** | 0.232 |
| PCA    | 4.050*** | 0.177 | 3.773*** | 0.192 | 4.053*** | 0.176 | 3.774*** | 0.201* |
| PLS    | 3.868*** | 0.185 | 3.701*** | 0.198 | 3.883*** | 0.181 | 3.624*** | 0.224* |
| SPCA   | 4.090*** | 0.171 | 3.844*** | 0.188 | 4.106*** | 0.169 | 3.769*** | 0.204* |
| SPLS   | 3.901*** | 0.186 | 3.674*** | 0.211 | 3.926*** | 0.181 | 3.621*** | 0.228* |
| AEN1   | 3.931*** | 0.188* | 3.781*** | 0.195 | 3.923*** | 0.186 | 3.702*** | 0.214** |
| AEN2   | 3.988*** | 0.170 | 3.736*** | 0.184 | 3.989*** | 0.169 | 3.759*** | 0.196 |
| AEN3   | 4.009*** | 0.180 | 3.770*** | 0.189 | 4.004*** | 0.181 | 3.737*** | 0.206* |
| AEN4   | 4.002*** | 0.184 | 3.764*** | 0.193 | 4.002*** | 0.183 | 3.755*** | 0.205* |

*B. Low credit spread regime*

|       | SD       | SR    |
|-------|----------|-------|
| EW    | 3.701    | 0.213 |
| Sample| 3.344**  | 0.222 |

|        | Static Factor Covariance | | Dynamic Beta Covariance | | Dynamic Factor Covariance | | Dynamic Error Covariance | |
|--------|------|-------|------|--------|------|-------|------|--------|
|        | SD   | SR    | SD   | SR     | SD   | SR    | SD   | SR     |
| Market | 3.520 | 0.220 | 3.272** | 0.216 | 3.479 | 0.227 | 3.516 | 0.208 |
| FF3    | 3.438 | 0.228 | 3.256*** | 0.238 | 3.394 | 0.230 | 3.409 | 0.220 |
| PCA    | 3.109*** | 0.261* | 3.122*** | 0.258 | 3.106*** | 0.262* | 3.091*** | 0.259 |
| PLS    | 3.140*** | 0.247 | 3.131*** | 0.246 | 3.141*** | 0.247 | 3.078*** | 0.244 |
| SPCA   | 3.110*** | 0.273** | 3.117*** | 0.266* | 3.110*** | 0.273** | 3.085*** | 0.263 |
| SPLS   | 3.141*** | 0.253 | 3.120*** | 0.254 | 3.141*** | 0.252 | 3.091*** | 0.247 |
| AEN1   | 3.098*** | 0.262* | 3.112*** | 0.263** | 3.098*** | 0.262* | 3.046*** | 0.257 |
| AEN2   | 3.122*** | 0.259* | 3.146*** | 0.256 | 3.119*** | 0.259* | 3.084*** | 0.254 |
| AEN3   | 3.094*** | 0.264* | 3.113*** | 0.261* | 3.090*** | 0.264* | 3.060*** | 0.263 |
| AEN4   | 3.124*** | 0.243 | 3.141*** | 0.241 | 3.123*** | 0.243 | 3.082*** | 0.236 |

In this table, we document the monthly portfolio performance measured using the standard deviation (SD) and Sharpe ratio (SR), during periods of high (Panel A) and low (Panel B) credit spread. Periods of high (low) credit spread are those when the spread for the specific month is higher (lower) than the median over the out-of-sample period. The results are presented for the equal-weighted portfolio (EW) and minimum-variance portfolios based on the sample estimator (Sample) and four factor-implied covariance specifications: static factor covariance, dynamic beta covariance, dynamic factor covariance and dynamic error covariance. The factor specifications are based on the single factor model (Market), the Fama-French 3-factor model (FF3), principal component analysis (PCA), partial least squares (PLS), sparse principal component analysis (SPCA), sparse partial least squares (SPLS) and autoencoders with 1, 2, 3 and 4 hidden layers (AEN). The significant outperformance of the alternative strategies from the equal-weighted strategy is denoted by: *, **, and *** for significance at the 10%, 5%, and 1% level, respectively.

**Table 9**
**Portfolio performance based on breakeven transaction costs**

| | Static Factor Covariance | Dynamic Beta Covariance | Dynamic Factor Covariance | Dynamic Error Covariance |
|---|---|---|---|---|
| Sample | 6.470 | | | |
| Market | 6.465 | 5.866 | 6.297 | 4.003 |
| FF3 | 8.392 | 8.208 | 6.474 | 7.148 |
| PCA | 18.146 | 16.479 | 17.971 | 12.884 |
| PLS | 14.230 | 12.690 | 13.574 | 11.414 |
| SPCA | 20.104 | 17.043 | 19.729 | 13.983 |
| SPLS | 15.484 | 14.948 | 14.883 | 12.261 |
| AEN1 | 18.595 | 16.739 | 18.473 | 13.397 |
| AEN2 | 16.345 | 15.023 | 16.179 | 11.797 |
| AEN3 | 17.158 | 15.668 | 16.925 | 13.187 |
| AEN4 | 13.017 | 12.156 | 12.498 | 9.177 |

This table presents monthly portfolio performance measured breakeven transaction costs ($c_{ew}$) with respect to the equal-weighted portfolio (EW), over the out-of-sample period from January 1980 to December 2019. The results are presented for the minimum-variance portfolios based on the sample estimator (Sample) and four factor-implied covariance specifications: static factor covariance, dynamic beta covariance, dynamic factor covariance and dynamic error covariance. The factor specifications are based on the single factor model (Market), the Fama-French 3-factor model (FF3), principal component analysis (PCA), partial least squares (PLS), sparse principal component analysis (SPCA), sparse partial least squares (SPLS) and autoencoders with 1, 2, 3 and 4 hidden layers (AEN). The breakeven transaction costs are reported in basis points and a positive value indicates that the alternative portfolio outperforms the EW.

**Table 10**
**Portfolio performance when short-selling is allowed**
*A. Standard deviation and Sharpe ratio*

|        | SD    | SR    |
|--------|-------|-------|
| EW     | 4.159 | 0.183 |
| Sample | 4.380 | 0.163 |

|        | Static Factor Covariance | | Dynamic Beta Covariance | | Dynamic Factor Covariance | | Dynamic Error Covariance | |
|--------|---------|---------|----------|---------|---------|---------|---------|---------|
|        | SD | SR | SD | SR | SD | SR | SD | SR |
| Market | 4.109 | 0.211 | 3.639** | 0.231 | 4.136 | 0.207 | 4.332 | 0.214 |
| FF3    | 4.120 | 0.214 | 3.515*** | 0.242 | 4.163 | 0.202 | 4.441 | 0.206 |
| PCA    | 3.387*** | 0.231** | 3.305*** | 0.238** | 3.386*** | 0.231** | 3.292*** | 0.240** |
| PLS    | 3.41*** | 0.224 | 3.289*** | 0.234* | 3.411*** | 0.223 | 3.308*** | 0.237 |
| SPCA   | 3.425*** | 0.235** | 3.342*** | 0.241*** | 3.426*** | 0.235** | 3.305*** | 0.241** |
| SPLS   | 3.432*** | 0.223 | 3.279*** | 0.239* | 3.435*** | 0.222 | 3.326*** | 0.234 |
| AEN1   | 3.328*** | 0.239*** | 3.283*** | 0.244*** | 3.324*** | 0.239*** | 3.226*** | 0.245** |
| AEN2   | 3.375*** | 0.229** | 3.309*** | 0.235** | 3.373*** | 0.229** | 3.28*** | 0.234** |
| AEN3   | 3.362*** | 0.237** | 3.298*** | 0.241** | 3.359*** | 0.237** | 3.258*** | 0.246** |
| AEN4   | 3.374*** | 0.225* | 3.316*** | 0.226* | 3.373*** | 0.225* | 3.278*** | 0.228* |

*B. Average turnover and breakeven transaction costs*

|        | TO    | $c_{ew}$ |
|--------|-------|------|
| EW     | 1.081 | NA   |
| Sample | 220.992 | -0.909 |

|        | Static Factor Covariance | | Dynamic Beta Covariance | | Dynamic Factor Covariance | | Dynamic Error Covariance | |
|--------|---------|---------|---------|---------|---------|---------|---------|---------|
|        | TO | $c_{ew}$ | TO | $c_{ew}$ | TO | $c_{ew}$ | TO | $c_{ew}$ |
| Market | 78.725 | 3.606 | 96.923 | 5.008 | 79.291 | 3.069 | 117.430 | 2.664 |
| FF3    | 93.821 | 3.343 | 121.381 | 4.904 | 96.111 | 1.999 | 144.917 | 1.599 |
| PCA    | 30.215 | 16.476 | 35.686 | 15.894 | 30.437 | 16.351 | 47.820 | 11.981 |
| PLS    | 35.856 | 11.790 | 42.400 | 12.343 | 35.929 | 11.478 | 55.090 | 9.998 |
| SPCA   | 29.738 | 18.146 | 36.120 | 16.553 | 29.798 | 18.108 | 47.116 | 12.599 |
| SPLS   | 35.123 | 11.750 | 42.530 | 13.511 | 35.075 | 11.473 | 54.284 | 9.586 |
| AEN1   | 31.497 | 18.411 | 37.252 | 16.864 | 31.683 | 18.299 | 47.918 | 13.237 |
| AEN2   | 31.597 | 15.074 | 36.269 | 14.778 | 31.860 | 14.945 | 48.511 | 10.964 |
| AEN3   | 33.660 | 16.575 | 38.092 | 15.671 | 33.954 | 16.427 | 50.529 | 12.741 |
| AEN4   | 34.001 | 12.758 | 37.986 | 11.652 | 34.410 | 12.602 | 50.839 | 9.044 |

This table reports monthly portfolio performance based on minimum-variance portfolios without short-selling constraints. Panel A reports the standard deviation (SD) and Sharpe ratio (SR), whereas the average turnover (TO) and breakeven transaction costs ($c_{ew}$) with respect to the EW portfolio can be found in Panel B. The out-of-sample period is from January 1980 to December 2019. The results are presented for the equal-weighted portfolio (EW) and minimum-variance portfolios based on the sample estimator (Sample) and four factor-implied covariance specifications: static factor covariance, dynamic beta covariance, dynamic factor covariance and dynamic error covariance. The factor specifications are based on the single factor model (Market), the Fama-French 3-factor model (FF3), principal component analysis (PCA), partial least squares (PLS), sparse principal component analysis (SPCA), sparse partial least squares (SPLS) and autoencoders with 1, 2, 3 and 4 hidden layers (AEN). NA indicates that the value is not available. The significant outperformance of the alternative strategies from the equal-weighted strategy is denoted by: *, **, and *** for significance at the 10%, 5%, and 1% level, respectively.

**Table 11**
**Portfolio performance using a penalized minimum-variance objective function**
*A. Standard deviation and Sharpe ratio*

|        | SD        | SR       |           |           |           |           |           |           |
|--------|-----------|----------|-----------|-----------|-----------|-----------|-----------|-----------|
| EW     | 4.159     | 0.183    |           |           |           |           |           |           |
| Sample | 3.626***  | 0.206    |           |           |           |           |           |           |

|        | Static Factor Covariance | | Dynamic Beta Covariance | | Dynamic Factor Covariance | | Dynamic Error Covariance | |
|--------|----------|-----------|----------|-----------|----------|-----------|----------|-----------|
|        | SD       | SR        | SD       | SR        | SD       | SR        | SD       | SR        |
| Market | 3.522*** | 0.228     | 3.331*** | 0.243*    | 3.59***  | 0.221     | 3.456*** | 0.225     |
| FF3    | 3.554*** | 0.232     | 3.33***  | 0.242*    | 3.591*** | 0.217     | 3.517*** | 0.229     |
| PCA    | 3.645*** | 0.207     | 3.527*** | 0.222**   | 3.623*** | 0.209     | 3.593*** | 0.214*    |
| PLS    | 3.547*** | 0.231**   | 3.444*** | 0.225**   | 3.553*** | 0.230**   | 3.458*** | 0.238***  |
| SPCA   | 3.648*** | 0.216**   | 3.539*** | 0.212*    | 3.656*** | 0.215**   | 3.606*** | 0.218**   |
| SPLS   | 3.551*** | 0.227**   | 3.426*** | 0.221*    | 3.563*** | 0.223**   | 3.484*** | 0.231**   |
| AEN1   | 3.647*** | 0.203     | 3.532*** | 0.215*    | 3.626*** | 0.206     | 3.607*** | 0.208     |
| AEN2   | 3.655*** | 0.204     | 3.538*** | 0.219*    | 3.633*** | 0.206     | 3.615*** | 0.213*    |
| AEN3   | 3.606*** | 0.210*    | 3.514*** | 0.224**   | 3.564*** | 0.213*    | 3.565*** | 0.219**   |
| AEN4   | 3.621*** | 0.206     | 3.524*** | 0.215*    | 3.599*** | 0.206     | 3.567*** | 0.218**   |

*B. Average turnover and breakeven transaction costs*

|        | TO     | $c_{ew}$ |
|--------|--------|----------|
| EW     | 1.081  | NA       |
| Sample | 19.863 | 12.246   |

|        | Static Factor Covariance | | Dynamic Beta Covariance | | Dynamic Factor Covariance | | Dynamic Error Covariance | |
|--------|--------|-----------|--------|-----------|--------|-----------|--------|-----------|
|        | TO     | $c_{ew}$  | TO     | $c_{ew}$  | TO     | $c_{ew}$  | TO     | $c_{ew}$  |
| Market | 24.111 | 19.540    | 30.242 | 20.575    | 25.009 | 15.881    | 25.955 | 16.885    |
| FF3    | 22.065 | 23.351    | 33.574 | 18.158    | 23.964 | 14.858    | 24.013 | 20.495    |
| PCA    | 13.251 | 19.721    | 16.390 | 25.475    | 14.117 | 19.945    | 14.634 | 22.873    |
| PLS    | 15.204 | 33.987    | 18.307 | 24.382    | 15.553 | 32.477    | 16.914 | 34.738    |
| SPCA   | 14.039 | 25.467    | 16.826 | 18.419    | 14.307 | 24.195    | 15.759 | 23.845    |
| SPLS   | 15.539 | 30.433    | 18.815 | 21.428    | 15.527 | 27.689    | 17.000 | 30.153    |
| AEN1   | 13.546 | 16.045    | 17.031 | 20.063    | 14.057 | 17.725    | 15.416 | 17.440    |
| AEN2   | 13.954 | 16.313    | 17.089 | 22.489    | 14.812 | 17.479    | 15.079 | 21.432    |
| AEN3   | 13.809 | 21.213    | 16.851 | 25.999    | 14.661 | 22.091    | 15.326 | 25.974    |
| AEN4   | 14.254 | 17.460    | 16.692 | 20.498    | 15.241 | 16.243    | 15.577 | 24.145    |

This table reports monthly portfolio performance based on minimum-variance portfolios with a turnover penalty for a parameter $c = 5$ bps. Panel A reports the standard deviation (SD) and Sharpe ratio (SR), whereas the average turnover (TO) and breakeven transaction costs ($c_{ew}$) with respect to the EW portfolio can be found in Panel B. The out-of-sample period is from January 1980 to December 2019. The results are presented for the equal-weighted portfolio (EW) and minimum-variance portfolios based on the sample estimator (Sample) and four factor-implied covariance specifications: static factor covariance, dynamic beta covariance, dynamic factor covariance and dynamic error covariance. The factor specifications are based on the single factor model (Market), the Fama-French 3-factor model (FF3), principal component analysis (PCA), partial least squares (PLS), sparse principal component analysis (SPCA), sparse partial least squares (SPLS) and autoencoders with 1, 2, 3 and 4 hidden layers (AEN). NA indicates that the value is not available. The significant outperformance of the alternative strategies from the equal-weighted strategy is denoted by: *, **, and *** for significance at the 10%, 5%, and 1% level, respectively.

**Table 12**
**Portfolio performance for the 49 Fama and French portfolios based on industry classification**

*A. Standard deviation and Sharpe ratio*

|  | SD | SR |  |  |  |  |  |  |
|---|---|---|---|---|---|---|---|---|
| EW | 4.606 | 0.157 |  |  |  |  |  |  |
| Sample | 3.481*** | 0.209* |  |  |  |  |  |  |

|  | Static Factor Covariance | | Dynamic Beta Covariance | | Dynamic Factor Covariance | | Dynamic Error Covariance | |
|---|---|---|---|---|---|---|---|---|
|  | SD | SR | SD | SR | SD | SR | SD | SR |
| Market | 3.443*** | 0.236** | 3.335*** | 0.224* | 3.47*** | 0.232** | 3.411*** | 0.239** |
| FF3 | 3.435*** | 0.225** | 3.371*** | 0.222** | 3.433*** | 0.221* | 3.416*** | 0.23** |
| PCA | 3.692*** | 0.219*** | 3.654*** | 0.215*** | 3.697*** | 0.22*** | 3.57*** | 0.226*** |
| PLS | 3.528*** | 0.23*** | 3.442*** | 0.23*** | 3.554*** | 0.227*** | 3.39*** | 0.237*** |
| SPCA | 3.655*** | 0.226*** | 3.598*** | 0.221*** | 3.667*** | 0.227*** | 3.539*** | 0.233*** |
| SPLS | 3.531*** | 0.229*** | 3.444*** | 0.23*** | 3.554*** | 0.225*** | 3.397*** | 0.238*** |
| AEN1 | 3.708*** | 0.218*** | 3.704*** | 0.212*** | 3.703*** | 0.218*** | 3.597*** | 0.231*** |
| AEN2 | 3.71*** | 0.215*** | 3.709*** | 0.206*** | 3.702*** | 0.217*** | 3.603*** | 0.221*** |
| AEN3 | 3.697*** | 0.218*** | 3.681*** | 0.211*** | 3.693*** | 0.22*** | 3.581*** | 0.226*** |
| AEN4 | 3.702*** | 0.216*** | 3.708*** | 0.204** | 3.695*** | 0.218*** | 3.583*** | 0.222*** |

*B. Average turnover and breakeven transaction costs*

|  | TO | $c_{ew}$ |  |  |  |  |  |  |
|---|---|---|---|---|---|---|---|---|
| EW | 0.053 | NA |  |  |  |  |  |  |
| Sample | 3.431 | 153.937 |  |  |  |  |  |  |

|  | Static Factor Covariance | | Dynamic Beta Covariance | | Dynamic Factor Covariance | | Dynamic Error Covariance | |
|---|---|---|---|---|---|---|---|---|
|  | TO | $c_{ew}$ | TO | $c_{ew}$ | TO | $c_{ew}$ | TO | $c_{ew}$ |
| Market | 2.094 | 391.965 | 25.793 | 26.030 | 6.201 | 121.991 | 15.446 | 53.271 |
| FF3 | 2.525 | 275.081 | 32.103 | 20.281 | 8.599 | 74.889 | 17.385 | 42.119 |
| PCA | 2.871 | 220.014 | 16.023 | 36.318 | 3.983 | 160.305 | 18.698 | 37.007 |
| PLS | 2.919 | 254.710 | 19.276 | 37.975 | 4.565 | 155.142 | 18.117 | 44.287 |
| SPCA | 9.092 | 76.336 | 20.967 | 30.602 | 10.231 | 68.776 | 24.621 | 30.935 |
| SPLS | 3.246 | 225.493 | 18.838 | 38.861 | 4.370 | 157.517 | 18.968 | 42.823 |
| AEN1 | 5.024 | 122.712 | 17.391 | 31.722 | 5.730 | 107.451 | 21.012 | 35.307 |
| AEN2 | 4.098 | 143.387 | 16.133 | 30.473 | 5.038 | 120.361 | 20.651 | 31.556 |
| AEN3 | 4.062 | 152.158 | 16.833 | 32.181 | 4.919 | 129.470 | 20.337 | 34.017 |
| AEN4 | 5.244 | 111.732 | 17.750 | 26.558 | 6.149 | 100.066 | 20.187 | 32.284 |

In this table, we report the monthly performance for portfolios constructed using the 49 Fama and French portfolios based on industry classification. Panel A reports the standard deviation (SD) and Sharpe ratio (SR), whereas the average turnover (TO) and breakeven transaction costs ($c_{ew}$) with respect to the EW portfolio can be found in Panel B. The out-of-sample period is from January 1980 to December 2019. The results are presented for the equal-weighted portfolio (EW) and minimum-variance portfolios based on the sample estimator (Sample) and four factor-implied covariance specifications: static factor covariance, dynamic beta covariance, dynamic factor covariance and dynamic error covariance. The factor specifications are based on the single factor model (Market), the Fama-French 3-factor model (FF3), principal component analysis (PCA), partial least squares (PLS), sparse principal component analysis (SPCA), sparse partial least squares (SPLS) and autoencoders with 1, 2, 3 and 4 hidden layers (AEN). NA indicates that the value is not available. The significant outperformance of the alternative strategies from the equal-weighted strategy is denoted by: *, **, and *** for significance at the 10%, 5%, and 1% level, respectively.

**Table 13**
**Portfolio performance for the S&P 500 and 20 Fama and French portfolios formed on size and book-to-market**

*A. Standard deviation and Sharpe ratio*

|  | SD | SR |
|---|---|---|
| EW | 5.106 | 0.156 |
| Sample | 4.305*** | 0.185* |

|  | Static Factor Covariance | | Dynamic Beta Covariance | | Dynamic Factor Covariance | | Dynamic Error Covariance | |
|---|---|---|---|---|---|---|---|---|
|  | SD | SR | SD | SR | SD | SR | SD | SR |
| Market | 4.534*** | 0.188** | 4.548*** | 0.184* | 4.553*** | 0.188** | 4.542*** | 0.19** |
| FF3 | 4.32*** | 0.183 | 4.454*** | 0.174 | 4.288*** | 0.178 | 4.307*** | 0.182 |
| PCA | 4.656*** | 0.181* | 4.676*** | 0.18* | 4.659*** | 0.183** | 4.599*** | 0.183* |
| PLS | 4.633*** | 0.182* | 4.639*** | 0.18* | 4.642*** | 0.183* | 4.576*** | 0.184* |
| SPCA | 4.65*** | 0.18* | 4.662*** | 0.181* | 4.655*** | 0.18* | 4.598*** | 0.18* |
| SPLS | 4.616*** | 0.181* | 4.626*** | 0.179 | 4.625*** | 0.182* | 4.561*** | 0.183* |
| AEN1 | 4.649*** | 0.181* | 4.663*** | 0.179* | 4.652*** | 0.182* | 4.59*** | 0.183** |
| AEN2 | 4.663*** | 0.181* | 4.684*** | 0.181** | 4.661*** | 0.182* | 4.602*** | 0.182* |
| AEN3 | 4.66*** | 0.18* | 4.69*** | 0.178* | 4.66*** | 0.181* | 4.601*** | 0.181* |
| AEN4 | 4.65*** | 0.181* | 4.664*** | 0.183** | 4.652*** | 0.182** | 4.598*** | 0.183* |

*B. Average turnover and breakeven transaction costs*

|  | TO | $c_{ew}$ |
|---|---|---|
| EW | 0.000 | NA |
| Sample | 2.835 | 102.293 |

|  | Static Factor Covariance | | Dynamic Beta Covariance | | Dynamic Factor Covariance | | Dynamic Error Covariance | |
|---|---|---|---|---|---|---|---|---|
|  | TO | $c_{ew}$ | TO | $c_{ew}$ | TO | $c_{ew}$ | TO | $c_{ew}$ |
| Market | 3.398 | 94.173 | 30.734 | 9.110 | 5.067 | 63.154 | 11.727 | 28.993 |
| FF3 | 2.762 | 97.755 | 35.149 | 5.121 | 11.270 | 19.521 | 6.109 | 40.923 |
| PCA | 2.216 | 112.816 | 17.862 | 13.436 | 4.109 | 65.709 | 19.594 | 13.780 |
| PLS | 1.514 | 171.731 | 16.169 | 14.843 | 2.575 | 104.854 | 18.302 | 15.299 |
| SPCA | 4.605 | 52.117 | 19.709 | 12.685 | 5.599 | 42.865 | 22.320 | 10.753 |
| SPLS | 1.411 | 177.179 | 15.964 | 14.407 | 2.550 | 101.961 | 18.924 | 14.268 |
| AEN1 | 1.618 | 154.512 | 16.887 | 13.620 | 2.695 | 96.475 | 19.181 | 14.076 |
| AEN2 | 2.128 | 112.782 | 18.770 | 13.319 | 3.598 | 69.483 | 20.025 | 12.984 |
| AEN3 | 2.882 | 83.276 | 21.064 | 10.444 | 4.217 | 59.284 | 20.615 | 12.127 |
| AEN4 | 2.421 | 103.263 | 18.233 | 14.808 | 3.571 | 72.809 | 20.186 | 13.376 |

In this table, we report the monthly performance for portfolios constructed using the S&P 500 and 20 Fama and French portfolios formed on size and book-to-market. Panel A reports the standard deviation (SD) and Sharpe ratio (SR), whereas the average turnover (TO) and breakeven transaction costs ($c_{ew}$) with respect to the EW portfolio can be found in Panel B. The out-of-sample period is from January 1980 to December 2019. The results are presented for the equal-weighted portfolio (EW) and minimum-variance portfolios based on the sample estimator (Sample) and four factor-implied covariance specifications: static factor covariance, dynamic beta covariance, dynamic factor covariance and dynamic error covariance. The factor specifications are based on the single factor model (Market), the Fama-French 3-factor model (FF3), principal component analysis (PCA), partial least squares (PLS), sparse principal component analysis (SPCA), sparse partial least squares (SPLS) and autoencoders with 1, 2, 3 and 4 hidden layers (AEN). NA indicates that the value is not available. The significant outperformance of the alternative strategies from the equal-weighted strategy is denoted by: *, **, and *** for significance at the 10%, 5%, and 1% level, respectively.

# Figures

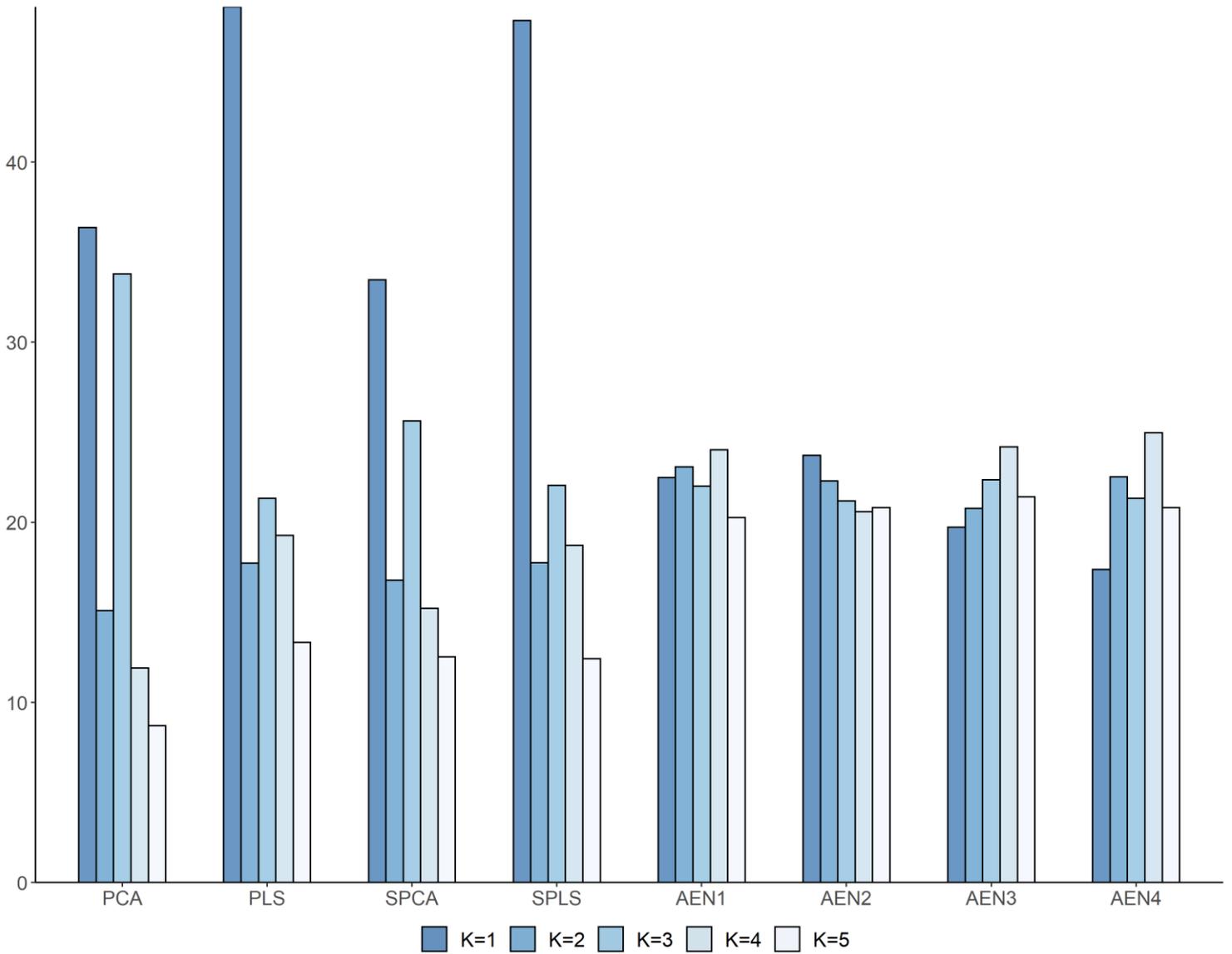

**Figure 1**

**Average $R^2_{adj}$ of the regressions of the latent factors on the Fama and French (2015) factors**

This figure shows the $R^2_{adj}$ as a percentage based on OLS estimation results for regressions of the latent factors on factors from the Fama and French (2015) five-factor model. The average over the out-of-sample period from July 1983 to December 2019 is given. The latent factor specifications are based on principal component analysis (PCA), partial least squares (PLS), sparse principal component analysis (SPCA), sparse partial least squares (SPLS) and autoencoders with 1, 2, 3 and 4 hidden layers (AEN).

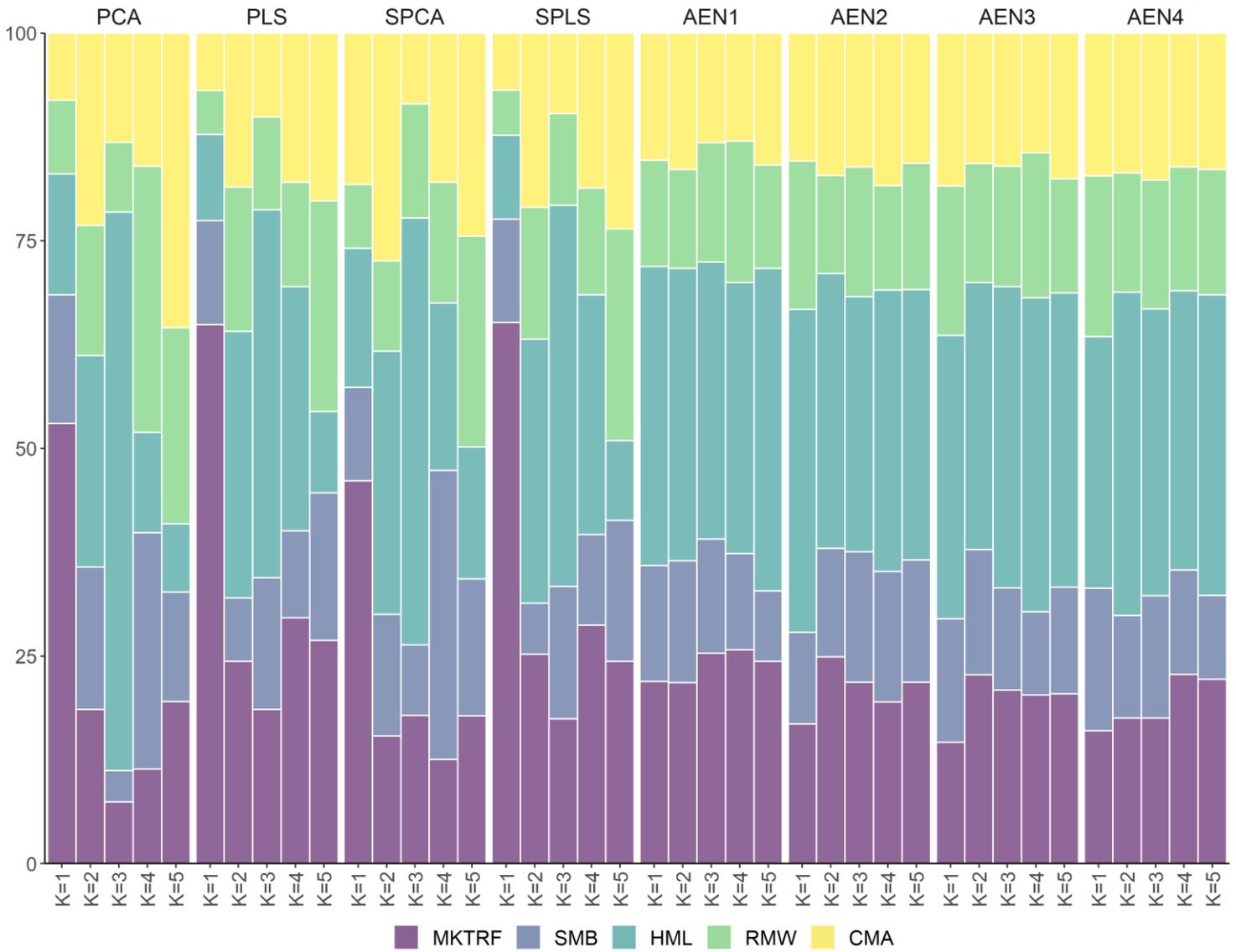

**Figure 2**
**Variable importance of the Fama and French (2015) factors**
This figure shows the variable importance based on OLS estimation results for regressions of the latent factors on factors from the Fama and French (2015) five-factor model. The measure of variable importance is calculated as the change in $R^2$ from setting the observations of a factor proxy to zero within each estimation window. The average over the out-of-sample period from July 1983 to December 2019 is given. The variable importance measures for each latent factor are scaled to sum to 100. The latent factor specifications are based on principal component analysis (PCA), partial least squares (PLS), sparse principal component analysis (SPCA), sparse partial least squares (SPLS) and autoencoders with 1, 2, 3 and 4 hidden layers (AEN). MKTRF, SMB, HML, RMW, and CMA are the Fama and French (2015) excess returns of the market from the risk-free rate, size, value, profitability, and investment factors, respectively.

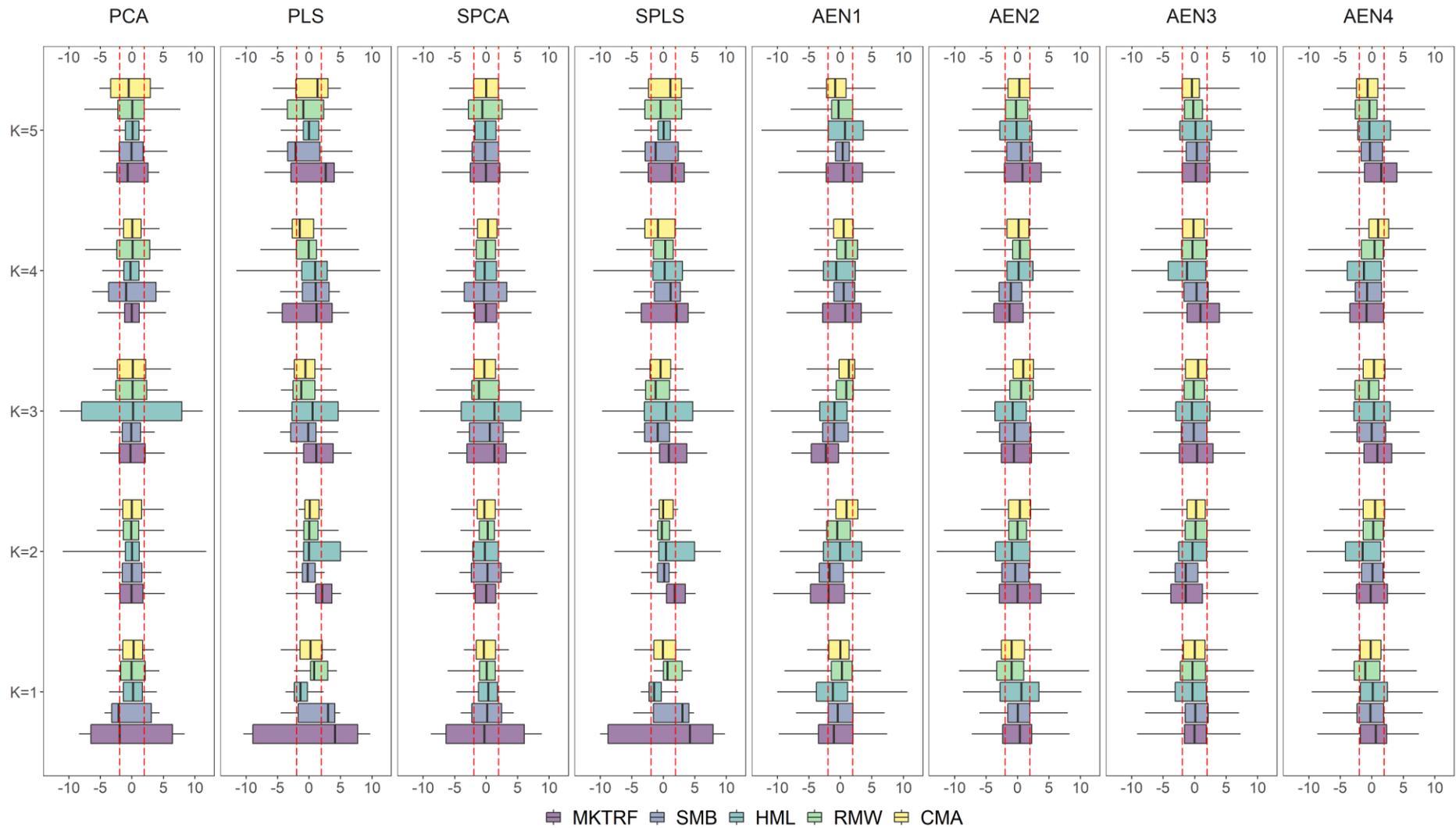

**Figure 3**
**Explaining the latent factors based on the Fama and French (2015) five-factor model**
This figure shows boxplots of the $t$-statistics based on OLS regressions of each of the five latent factors on factors from the Fama and French (2015) five-factor model. The horizontal axis reports $t$-statistics values ranging from -10 to 10, whereas the vertical axis reports the latent factors, $K = 1, ..., 5$. The sample period is from July 1963 to December 2019. The median is marked by the line within the box, the edges of the box denote the first and third quartiles, while the minimum and maximum $t$-statistics are depicted by the end of the lines outside the box. The latent factor specifications are based on principal component analysis (PCA), partial least squares (PLS), sparse principal component analysis (SPCA), sparse partial least squares (SPLS) and autoencoders with 1, 2, 3 and 4 hidden layers (AEN). MKTRF, SMB, HML, RMW, and CMA are the Fama and French (2015) excess returns of the market from the risk-free rate, size, value, profitability, and investment factors, respectively. The $t$-statistics are computed using heteroskedasticity and autocorrelation-robust standard errors (Newey and West, 1987). The red lines depict the Student's t critical values at the 5% level.

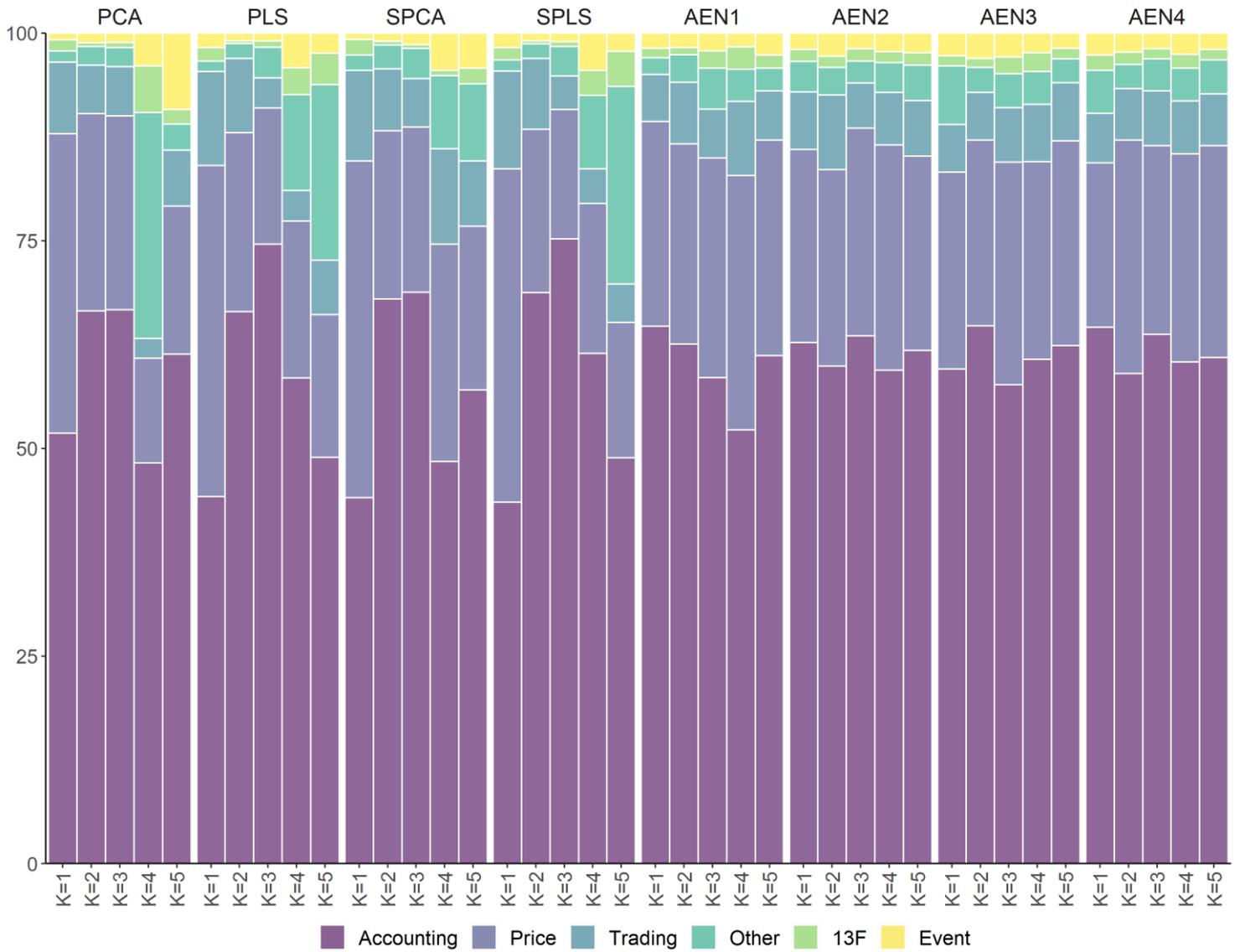

**Figure 4**
**Variable importance of the long-short anomaly portfolio returns from the Chen and Zimmermann (2020) dataset**
This figure shows the variable importance based on lasso regressions of the latent factors on long-short anomaly portfolio returns. The measure of variable importance is calculated as the change in $R^2$ from setting the observations of a feature to zero within each estimation window. The results are aggregated by summing the variable importance of the characteristics-based portfolios belonging in the same group. Details on the portfolios within each group can be found in Table A1 in the Appendix. The average over the out-of-sample period from January 1980 to December 2019 is given. The variable importance measures for each group are scaled to sum to 100. The latent factor specifications are based on principal component analysis (PCA), partial least squares (PLS), sparse principal component analysis (SPCA), sparse partial least squares (SPLS) and autoencoders with 1, 2, 3 and 4 hidden layers (AEN). The explanatory variables are 110 anomaly portfolios from the open-source asset pricing dataset by Chen and Zimmermann (2020), that have no missing values over the full sample period, from January 1960 to December 2019.

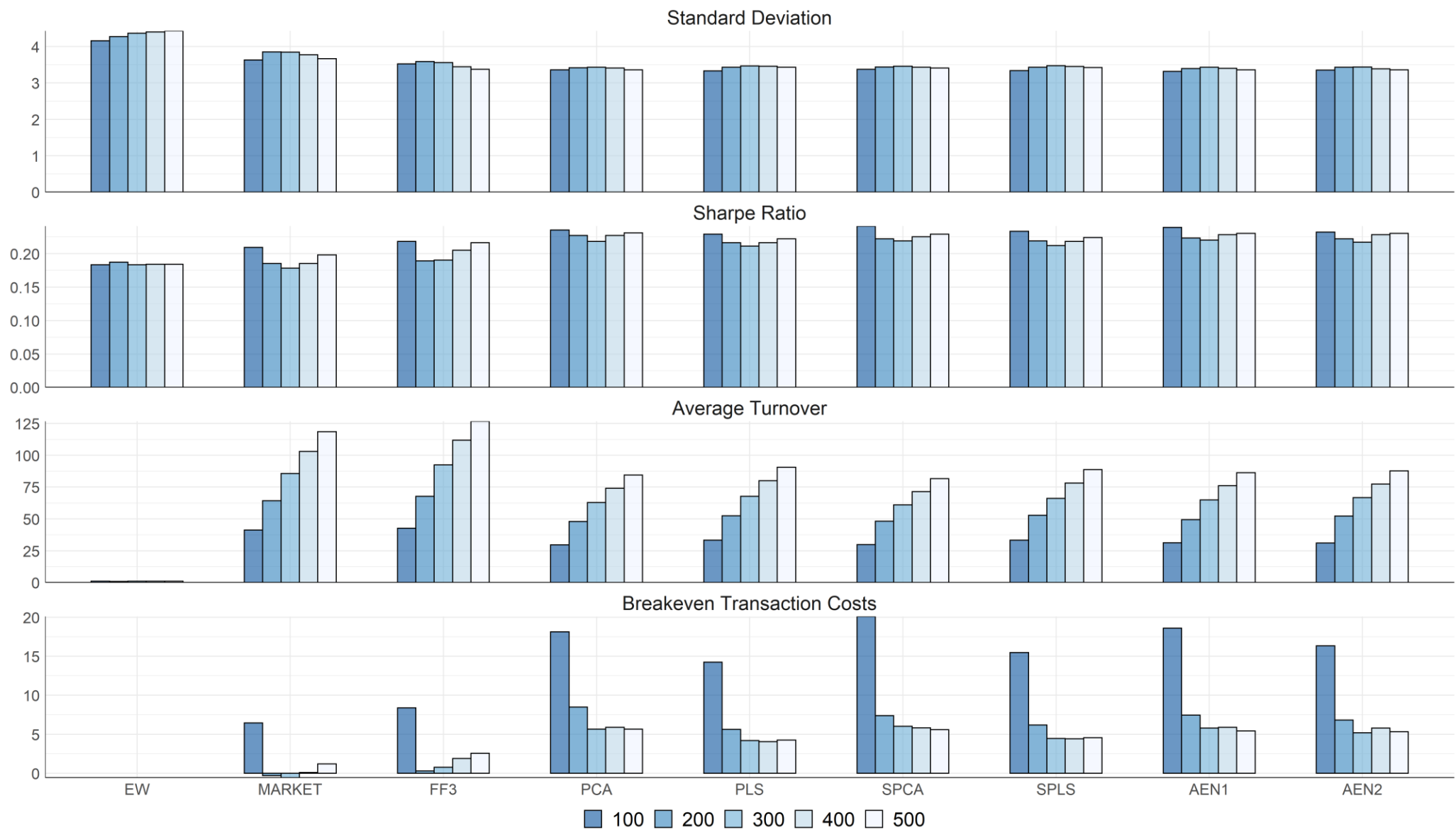

**Figure 5**
**Portfolio performance for a different number of stocks: Static Factor Covariance**
This figure shows the monthly portfolio performance for a varying number of assets. Performance is based on the standard deviation, Sharpe ratio, average turnover and breakeven transaction costs with respect to the EW portfolio. The out-of-sample period is from January 1980 to December 2019. The results are presented for the equal-weighted portfolio (EW) and for the static factor covariance. The factor specifications are based on the single factor model (Market), the Fama-French 3-factor model (FF3), principal component analysis (PCA), partial least squares (PLS), sparse principal component analysis (SPCA), sparse partial least squares (SPLS) and autoencoders with 1 and 2 hidden layers (AEN). The standard deviation and average turnover are reported as a percentage. The breakeven transaction costs are reported in basis points and a positive value indicates that the alternative portfolio outperforms the EW.

# Appendix

## Tables

**Table A1**
**Constituents of each group for the Chen and Zimmermann (2020) dataset**

| Group | # of Variables | Variables |
|---|---|---|
| 13F | 2 | RIO_Turnover, RIO_Volatility |
| Accounting | 46 | Accruals, AM, AssetGrowth, BMdec, BookLeverage, CashProd, CF, ChAssetTurnover, ChInv, ChInvIA, ChNNCOA, ChNWC, CompEquIss, CompositeDebtIssuance, DelCOA, DelCOL, DelFINL, DelLTI, DelNetFin, DivYieldST, EarningsConsistency, EntMult, GP, grcapx, grcapx3y, GrLTNOA, GrSaleToGrInv, GrSaleToGrOverhead, IntanCFP, IntanEP, IntanSP, Investment, InvestPPEInv, InvGrowth, MeanRankRevGrowth, OPLeverage, OrgCap, RD, RDAbility, ShareIss1Y, ShareIss5Y, SP, tang, Tax, TotalAccruals, VarCF |
| Event | 5 | ConvDebt, DivInit, DivSeason, Spinoff, SurpriseRD |
| Other | 5 | FirmAge, Herf, HerfAsset, HerfBE, sinAlgo |
| Price | 41 | Beta, BetaFP, BetaTailRisk, Coskewness, EP, High52, IdioRisk, IdioVol3F, IdioVolAHT, IndMom, IndRetBig, IntMom, Leverage, LRreversal, MaxRet, Mom12m, Mom12mOffSeason, Mom6m, MomOffSeason, MomOffSeason06YrPlus, MomOffSeason11YrPlus, MomOffSeason16YrPlus, MomRev, MomSeason, MomSeason06YrPlus, MomSeason11YrPlus, MomSeason16YrPlus, MomSeasonShort, MomVol, MRreversal, NetPayoutYield, PayoutYield, Price, PriceDelayRsq, PriceDelaySlope, PriceDelayTstat, ResidualMomentum, ReturnSkew, ReturnSkew3F, Size, STreversal |
| Trading | 11 | BidAskSpread, DolVol, Illiquidity, ShareVol, std_turn, VolMkt, VolSD, VolumeTrend, zerotrade, zerotradeAlt1, zerotradeAlt12 |

This table presents information on the six groups of variables for the Chen and Zimmermann (2020) dataset.

**Table A2**
**Constituents of each group for the McCracken and Ng (2015) dataset**

| Group | # of Variables | Variables |
|---|---|---|
| Consumption, orders, and inventories | 7 | DPCERA3M086SBEA, CMRMTSPLx, RETAILx, AMDMNOx, AMDMUOx, BUSINVx, ISRATIOx |
| Housing | 5 | HOUST, HOUSTNE, HOUSTMW, HOUSTS, HOUSTW |
| Interest and exchange rates | 19 | FEDFUNDS, TB3MS, TB6MS, GS1, GS5, GS10, AAA, BAA, TB3SMFFM, TB6SMFFM, T1YFFM, T5YFFM, T10YFFM, AAAFFM, BAAFFM, EXSZUSx, EXJPUSx, EXUSUKx, EXCAUSx |
| Labor market | 31 | HWI, HWIURATIO, CLF16OV, CE16OV, UNRATE, UEMPMEAN, UEMPLT5, UEMP5TO14, UEMP15OV, UEMP15T26, UEMP27OV, CLAIMSx, PAYEMS, USGOOD, CES1021000001, USCONS, MANEMP, DMANEMP, NDMANEMP, SRVPRD, USTPU, USWTRADE, USTRADE, USFIRE, USGOVT, CES0600000007, AWOTMAN, AWHMAN, CES0600000008, CES2000000008, CES3000000008 |
| Money and credit | 14 | M1SL, M2SL, M2REAL, BOGMBASE, TOTRESNS, NONBORRES, BUSLOANS, REALLN, NONREVSL, CONSPI, MZMSL, DTCOLNVHFNM, DTCTHFNM, INVEST |
| Output and income | 16 | RPI, W875RX1, INDPRO, IPFPNSS, IPFINAL, IPCONGD, IPDCONGD, IPNCONGD, IPBUSEQ, IPMAT, IPDMAT, IPNMAT, IPMANSICS, IPB51222S, IPFUELS, CUMFNS |
| Prices | 20 | WPSFD49207, WPSFD49502, WPSID61, WPSID62, OILPRICEx, PPICMM, CPIAUCSL, CPIAPPSL, CPITRNSL, CPIMEDSL, CUSR0000SAC, CUSR0000SAD, CUSR0000SAS, CPIULFSL, CUSR0000SA0L2, CUSR0000SA0L5, PCEPI, DDURRG3M086SBEA, DNDGRG3M086SBEA, DSERRG3M086SBEA |
| Stock market | 4 | SP500, SPindust, SPdivyield, SPPEratio |

This table presents information on the eight groups of variables for the McCracken and Ng (2015) dataset.

**Table A3**
**Portfolio performance for different validation window size**

*A. Validation subsample set to 10%*

|  | Static Factor Covariance | | Dynamic Beta Covariance | | Dynamic Factor Covariance | | Dynamic Error Covariance | |
|---|---|---|---|---|---|---|---|---|
|  | SD | SR | SD | SR | SD | SR | SD | SR |
| PCA | 3.361*** | 0.235** | 3.29*** | 0.239** | 3.36*** | 0.235** | 3.268*** | 0.241** |
| PLS | 3.33*** | 0.229 | 3.276*** | 0.233* | 3.335*** | 0.227 | 3.217*** | 0.238* |
| SPCA | 3.387*** | 0.235** | 3.322*** | 0.236** | 3.392*** | 0.234** | 3.273*** | 0.244*** |
| SPLS | 3.328*** | 0.232* | 3.264*** | 0.236* | 3.337*** | 0.23* | 3.21*** | 0.241* |
| AEN1 | 3.333*** | 0.232** | 3.288*** | 0.237** | 3.331*** | 0.232** | 3.235*** | 0.239** |
| AEN2 | 3.356*** | 0.23** | 3.294*** | 0.236** | 3.357*** | 0.229** | 3.271*** | 0.235** |
| AEN3 | 3.353*** | 0.234** | 3.305*** | 0.238** | 3.351*** | 0.234** | 3.245*** | 0.244** |
| AEN4 | 3.361*** | 0.225* | 3.304*** | 0.228** | 3.361*** | 0.225* | 3.265*** | 0.234** |

*B. Validation subsample set to 30%*

|  | Static Factor Covariance | | Dynamic Beta Covariance | | Dynamic Factor Covariance | | Dynamic Error Covariance | |
|---|---|---|---|---|---|---|---|---|
|  | SD | SR | SD | SR | SD | SR | SD | SR |
| PCA | 3.367*** | 0.236** | 3.312*** | 0.237** | 3.368*** | 0.235** | 3.265*** | 0.242** |
| PLS | 3.349*** | 0.226 | 3.296*** | 0.231* | 3.354*** | 0.225 | 3.231*** | 0.237* |
| SPCA | 3.386*** | 0.237** | 3.309*** | 0.238** | 3.395*** | 0.235** | 3.292*** | 0.243** |
| SPLS | 3.338*** | 0.232* | 3.262*** | 0.24** | 3.346*** | 0.230 | 3.218*** | 0.242* |
| AEN1 | 3.34*** | 0.238*** | 3.297*** | 0.24*** | 3.338*** | 0.239*** | 3.23*** | 0.243** |
| AEN2 | 3.36*** | 0.23** | 3.316*** | 0.229** | 3.36*** | 0.23** | 3.268*** | 0.234** |
| AEN3 | 3.338*** | 0.226* | 3.302*** | 0.231** | 3.336*** | 0.226* | 3.241*** | 0.236** |
| AEN4 | 3.363*** | 0.228** | 3.322*** | 0.229** | 3.366*** | 0.229* | 3.257*** | 0.236** |

This table documents monthly portfolio performance measured using the standard deviation (SD) and Sharpe ratio (SR), over the out-of-sample period from January 1980 to December 2019. The hyperparameters are selected using two alternative validation window sizes to the 20% of the baseline results: 10% (Panel A) and 30% (Panel B). The results are presented for the four factor-implied covariance specifications: static factor covariance, dynamic beta covariance, dynamic factor covariance and dynamic error covariance. The factor specifications are based on principal component analysis (PCA), partial least squares (PLS), sparse principal component analysis (SPCA), sparse partial least squares (SPLS) and autoencoders with 1, 2, 3 and 4 hidden layers (AEN). The significant outperformance of the alternative strategies from the equal-weighted strategy is denoted by: *, **, and *** for significance at the 10%, 5%, and 1% level, respectively.

**Table A4**
**Portfolio performance for different levels of transaction costs**

*A. Transaction costs of 5 bps*

|  | SD | SR |  |  |  |  |  |  |
|---|---|---|---|---|---|---|---|---|
| EW | 4.159 | 0.183 |  |  |  |  |  |  |
| Sample | 3.470*** | 0.203 |  |  |  |  |  |  |

|  | Static Factor Covariance | | Dynamic Beta Covariance | | Dynamic Factor Covariance | | Dynamic Error Covariance | |
|---|---|---|---|---|---|---|---|---|
|  | SD | SR | SD | SR | SD | SR | SD | SR |
| Market | 3.630*** | 0.203 | 3.346*** | 0.209 | 3.640*** | 0.204 | 3.596*** | 0.197 |
| FF3 | 3.521*** | 0.212 | 3.327*** | 0.23 | 3.538*** | 0.206 | 3.468*** | 0.215 |
| PCA | 3.361*** | 0.230** | 3.290*** | 0.234** | 3.359*** | 0.230** | 3.267*** | 0.234** |
| PLS | 3.329*** | 0.224 | 3.276*** | 0.226 | 3.334*** | 0.222 | 3.216*** | 0.231 |
| SPCA | 3.373*** | 0.236** | 3.306*** | 0.237** | 3.377*** | 0.236** | 3.262*** | 0.239** |
| SPLS | 3.339*** | 0.228 | 3.260*** | 0.236* | 3.346*** | 0.226 | 3.224*** | 0.234 |
| AEN1 | 3.317*** | 0.235** | 3.284*** | 0.238*** | 3.314*** | 0.234** | 3.215*** | 0.237** |
| AEN2 | 3.351*** | 0.227** | 3.296*** | 0.230** | 3.349*** | 0.227** | 3.258*** | 0.229* |
| AEN3 | 3.337*** | 0.233** | 3.282*** | 0.234** | 3.333*** | 0.233** | 3.234*** | 0.238** |
| AEN4 | 3.356*** | 0.22 | 3.300*** | 0.221* | 3.355*** | 0.219 | 3.255*** | 0.219 |

*B. Transaction costs of 20 bps*

|  | SD | SR |  |  |  |  |  |  |
|---|---|---|---|---|---|---|---|---|
| EW | 4.159 | 0.183 |  |  |  |  |  |  |
| Sample | 3.471*** | 0.185 |  |  |  |  |  |  |

|  | Static Factor Covariance | | Dynamic Beta Covariance | | Dynamic Factor Covariance | | Dynamic Error Covariance | |
|---|---|---|---|---|---|---|---|---|
|  | SD | SR | SD | SR | SD | SR | SD | SR |
| Market | 3.629*** | 0.186 | 3.344*** | 0.182 | 3.641*** | 0.186 | 3.595*** | 0.173 |
| FF3 | 3.520*** | 0.194 | 3.323*** | 0.197 | 3.537*** | 0.186 | 3.464*** | 0.19 |
| PCA | 3.361*** | 0.217 | 3.291*** | 0.218 | 3.359*** | 0.217 | 3.266*** | 0.213 |
| PLS | 3.328*** | 0.209 | 3.276*** | 0.208 | 3.333*** | 0.207 | 3.213*** | 0.208 |
| SPCA | 3.371*** | 0.223* | 3.305*** | 0.221* | 3.376*** | 0.222* | 3.26*** | 0.217 |
| SPLS | 3.337*** | 0.213 | 3.260*** | 0.217 | 3.345*** | 0.211 | 3.222*** | 0.211 |
| AEN1 | 3.315*** | 0.220* | 3.282*** | 0.221* | 3.312*** | 0.220* | 3.213*** | 0.215 |
| AEN2 | 3.349*** | 0.213 | 3.295*** | 0.213 | 3.347*** | 0.213 | 3.256*** | 0.208 |
| AEN3 | 3.335*** | 0.218 | 3.281*** | 0.217 | 3.331*** | 0.218 | 3.232*** | 0.216 |
| AEN4 | 3.355*** | 0.205 | 3.299*** | 0.204 | 3.354*** | 0.204 | 3.253*** | 0.196 |

This table presents monthly portfolio performance measured using the standard deviation (SD) and Sharpe ratio (SR), after transaction costs are taken into account. Panel A reports the results for transaction costs of 5 bps, while Panel B presents the results for transaction costs of 20 bps. The out-of-sample period is from January 1980 to December 2019. The results are presented for the equal-weighted portfolio (EW) and minimum-variance portfolios based on the sample estimator (Sample) and four factor-implied covariance specifications: static factor covariance, dynamic beta covariance, dynamic factor covariance and dynamic error covariance. The factor specifications are based on the single factor model (Market), the Fama-French 3-factor model (FF3), principal component analysis (PCA), partial least squares (PLS), sparse principal component analysis (SPCA), sparse partial least squares (SPLS) and autoencoders with 1, 2, 3 and 4 hidden layers (AEN). The significant outperformance of the alternative strategies from the equal-weighted strategy is denoted by: *, **, and *** for significance at the 10%, 5%, and 1% level, respectively.

**Table A5**
**Characteristics of the portfolio weight vectors when short-selling is allowed**

|  | MAX | $SD_\omega$ | $MAD_{EW}$ | | | | | | | | | |
|---|---|---|---|---|---|---|---|---|---|---|---|---|
| EW | 0.010 | 0.000 | 0.000 | | | | | | | | | |
| Sample | 0.378 | 6.317 | 4.873 | | | | | | | | | |
| | Static Factor Covariance | | | Dynamic Beta Covariance | | | Dynamic Factor Covariance | | | Dynamic Error Covariance | | |
| | MAX | $SD_\omega$ | $MAD_{EW}$ | MAX | $SD_\omega$ | $MAD_{EW}$ | MAX | $SD_\omega$ | $MAD_{EW}$ | MAX | $SD_\omega$ | $MAD_{EW}$ |
| Market | 0.226 | 2.855 | 2.122 | 0.176 | 2.412 | 1.864 | 0.253 | 2.803 | 2.082 | 0.297 | 3.319 | 2.327 |
| FF3 | 0.237 | 3.242 | 2.383 | 0.201 | 2.628 | 2.001 | 0.278 | 3.132 | 2.313 | 0.309 | 3.929 | 2.765 |
| PCA | 0.064 | 1.051 | 0.785 | 0.065 | 1.014 | 0.763 | 0.063 | 1.056 | 0.789 | 0.132 | 1.262 | 0.894 |
| PLS | 0.078 | 1.320 | 0.979 | 0.072 | 1.255 | 0.940 | 0.078 | 1.322 | 0.981 | 0.140 | 1.556 | 1.095 |
| SPCA | 0.060 | 1.010 | 0.727 | 0.073 | 1.014 | 0.743 | 0.060 | 1.009 | 0.726 | 0.128 | 1.216 | 0.839 |
| SPLS | 0.082 | 1.317 | 0.947 | 0.077 | 1.266 | 0.927 | 0.081 | 1.316 | 0.947 | 0.163 | 1.554 | 1.069 |
| AEN1 | 0.062 | 1.024 | 0.776 | 0.064 | 0.984 | 0.748 | 0.062 | 1.028 | 0.780 | 0.114 | 1.234 | 0.883 |
| AEN2 | 0.058 | 1.042 | 0.783 | 0.068 | 0.997 | 0.752 | 0.059 | 1.049 | 0.788 | 0.112 | 1.251 | 0.888 |
| AEN3 | 0.063 | 1.070 | 0.804 | 0.066 | 1.016 | 0.765 | 0.063 | 1.078 | 0.810 | 0.130 | 1.288 | 0.911 |
| AEN4 | 0.067 | 1.071 | 0.804 | 0.069 | 1.018 | 0.766 | 0.067 | 1.080 | 0.810 | 0.128 | 1.283 | 0.911 |

This table presents monthly characteristics of the portfolio weight vectors based on the maximum weight (MAX), standard deviation of the weights ($SD_\omega$) and mean absolute deviation from the equal-weighted benchmark ($MAD_{EW}$). The results are based on minimum-variance portfolios without short-selling constraints. The average value of each weight characteristic over the out-of-sample period from January 1980 to December 2019 is reported. The results are presented for the equal-weighted portfolio (EW) and minimum-variance portfolios based on the sample estimator (Sample) and four factor-implied covariance specifications: static factor covariance, dynamic beta covariance, dynamic factor covariance and dynamic error covariance. The factor specifications are based on the single factor model (Market), the Fama-French 3-factor model (FF3), principal component analysis (PCA), partial least squares (PLS), sparse principal component analysis (SPCA), sparse partial least squares (SPLS) and autoencoders with 1, 2, 3 and 4 hidden layers (AEN).

**Table A6**
**Characteristics of the portfolio weight vectors when using a penalized minimum-variance objective function**

|        | MAX   | $SD_\omega$ | $MAD_{EW}$ |
|--------|-------|-------|-------|
| EW     | 0.010 | 0.000 | 0.000 |
| Sample | 0.330 | 2.912 | 1.514 |

|        | Static Factor Covariance ||| Dynamic Beta Covariance ||| Dynamic Factor Covariance ||| Dynamic Error Covariance |||
|--------|-------|-------|-------|-------|-------|-------|-------|-------|-------|-------|-------|-------|
|        | MAX   | $SD_\omega$ | $MAD_{EW}$ | MAX   | $SD_\omega$ | $MAD_{EW}$ | MAX   | $SD_\omega$ | $MAD_{EW}$ | MAX   | $SD_\omega$ | $MAD_{EW}$ |
| Market | 0.258 | 2.647 | 1.547 | 0.297 | 2.466 | 1.522 | 0.392 | 2.873 | 1.572 | 0.351 | 3.305 | 1.624 |
| FF3    | 0.271 | 2.659 | 1.513 | 0.289 | 2.456 | 1.501 | 0.354 | 2.819 | 1.539 | 0.336 | 3.228 | 1.602 |
| PCA    | 0.073 | 1.209 | 0.982 | 0.085 | 1.298 | 1.042 | 0.079 | 1.259 | 1.017 | 0.145 | 1.446 | 1.071 |
| PLS    | 0.081 | 1.447 | 1.148 | 0.092 | 1.453 | 1.144 | 0.101 | 1.487 | 1.168 | 0.150 | 1.726 | 1.238 |
| SPCA   | 0.072 | 1.188 | 0.984 | 0.087 | 1.290 | 1.051 | 0.076 | 1.212 | 0.990 | 0.168 | 1.447 | 1.054 |
| SPLS   | 0.085 | 1.465 | 1.169 | 0.107 | 1.503 | 1.173 | 0.095 | 1.476 | 1.172 | 0.194 | 1.726 | 1.227 |
| AEN1   | 0.071 | 1.144 | 0.931 | 0.090 | 1.259 | 1.005 | 0.076 | 1.183 | 0.950 | 0.128 | 1.390 | 1.014 |
| AEN2   | 0.070 | 1.202 | 0.981 | 0.085 | 1.304 | 1.050 | 0.083 | 1.262 | 1.020 | 0.137 | 1.456 | 1.077 |
| AEN3   | 0.074 | 1.225 | 0.999 | 0.080 | 1.299 | 1.048 | 0.085 | 1.286 | 1.039 | 0.139 | 1.478 | 1.097 |
| AEN4   | 0.075 | 1.246 | 1.001 | 0.083 | 1.309 | 1.048 | 0.082 | 1.299 | 1.037 | 0.147 | 1.476 | 1.072 |

This table reports monthly characteristics of the portfolio weight vectors based on the maximum weight (MAX), standard deviation of the weights ($SD_\omega$) and mean absolute deviation from the equal-weighted benchmark ($MAD_{EW}$). The results are based on minimum-variance portfolios with a turnover penalty for a parameter $c = 5$ bps. The average value of each weight characteristic over the out-of-sample period from January 1980 to December 2019 is reported. The results are presented for the equal-weighted portfolio (EW) and minimum-variance portfolios based on the sample estimator (Sample) and four factor-implied covariance specifications: static factor covariance, dynamic beta covariance, dynamic factor covariance and dynamic error covariance. The factor specifications are based on the single factor model (Market), the Fama-French 3-factor model (FF3), principal component analysis (PCA), partial least squares (PLS), sparse principal component analysis (SPCA), sparse partial least squares (SPLS) and autoencoders with 1, 2, 3 and 4 hidden layers (AEN).

# Figures

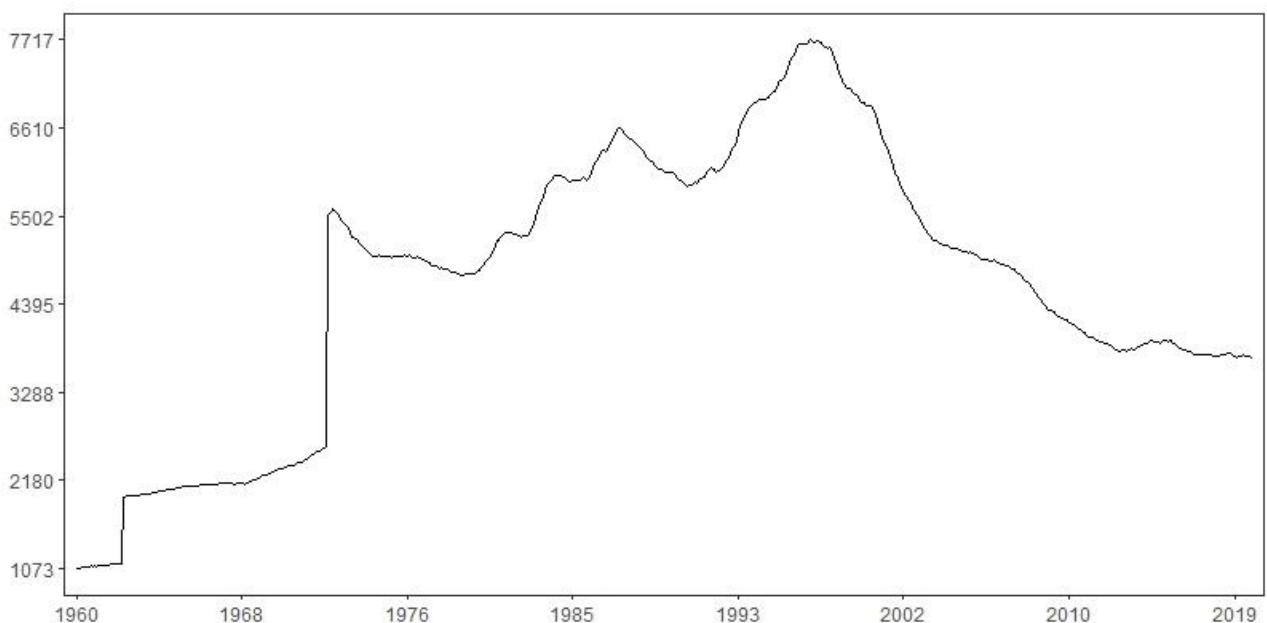

**Figure A1**
**Number of stocks per month after the filters have been applied to the CRSP dataset**
This figure shows the monthly number of stocks that are listed to the NYSE, AMEX, and NASDAQ stock exchanges (exchange codes 1, 2 or 3) and are ordinary common shares (share codes 10 or 11), over the full sample period from January 1960 to December 2019 (720 observations).

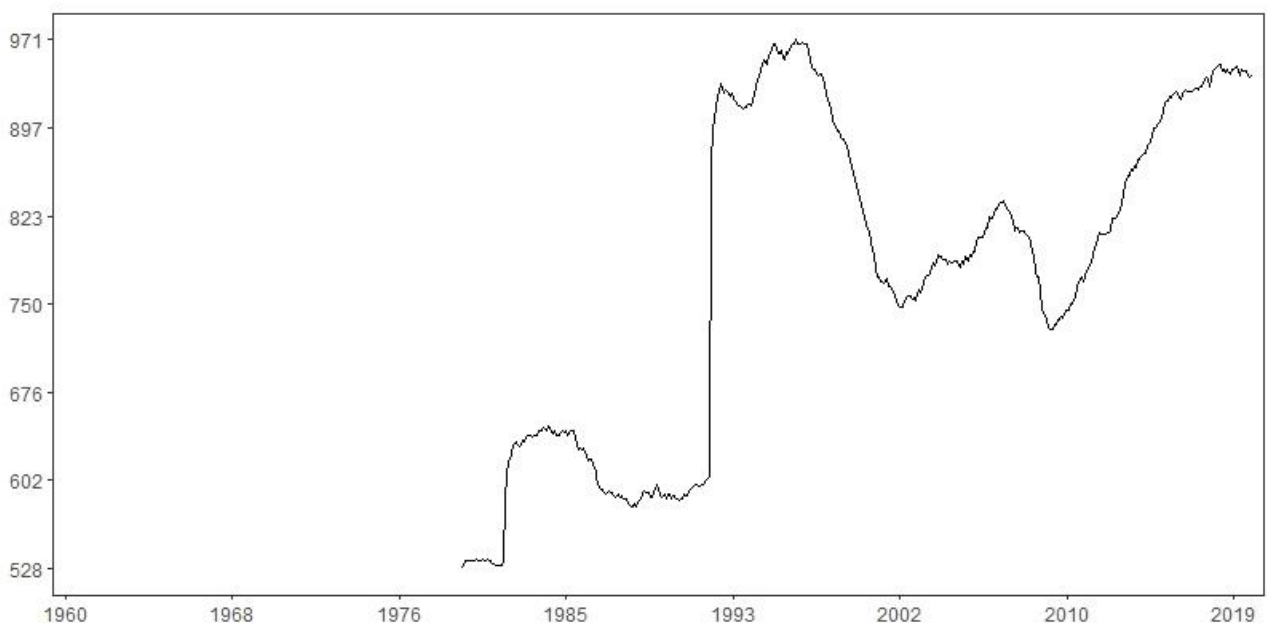

**Figure A2**
**Number of stocks per month that fulfil the conditions of the rolling window**
This figure shows the number of stocks in each iteration of the rolling window. Stocks are considered if they have at least 97.5% of the in-sample observations available, if they are not missing a return observation for the next month after the end of the rolling window and have a price greater than $5. The out-of-sample period is from December 1980 to December 2019 (480 observations).

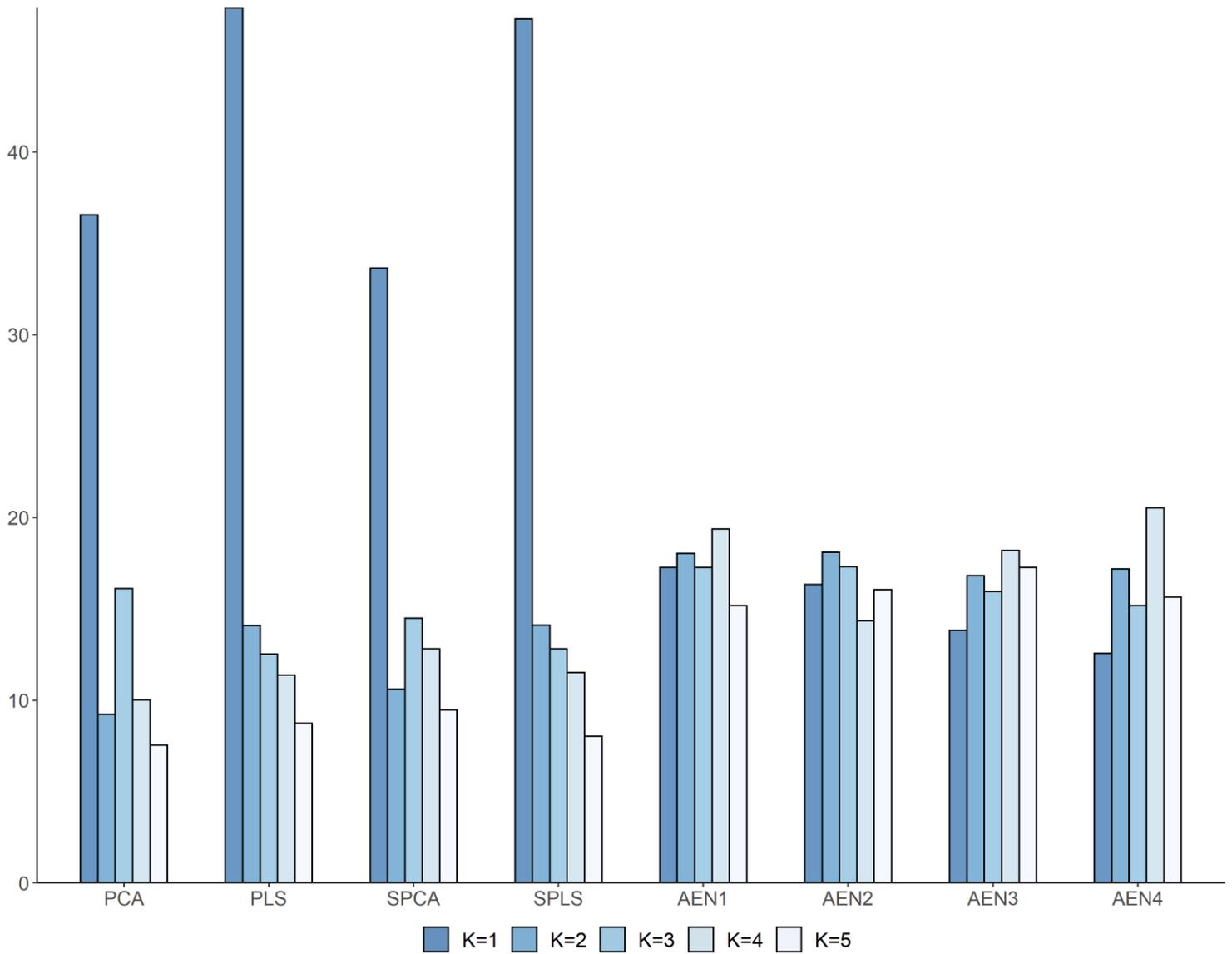

**Figure A3**
**Average $R^2_{adj}$ of the regressions of the latent factors on factors from the augmented q-factor model**
This figure shows the $R^2_{adj}$ as a percentage based on OLS estimation results for regressions of the latent factors on factors from the augmented q-factor model. The average over the out-of-sample period from January 1987 to December 2019 is given. The latent factor specifications are based on principal component analysis (PCA), partial least squares (PLS), sparse principal component analysis (SPCA), sparse partial least squares (SPLS) and autoencoders with 1, 2, 3 and 4 hidden layers (AEN).

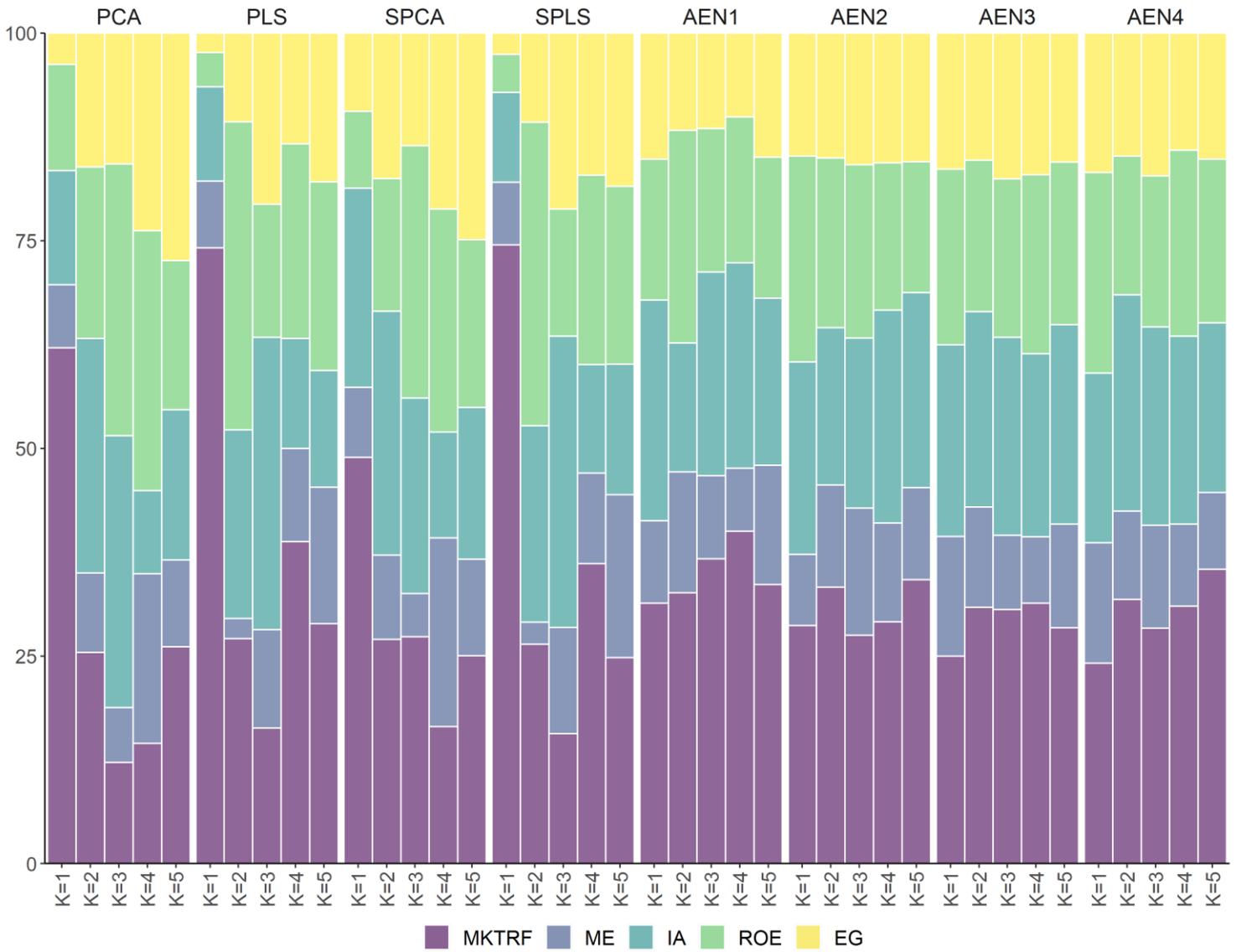

**Figure A4**
**Variable importance based on the augmented q-factor model**
This figure shows the variable importance based on OLS estimation results for regressions of the latent factors on factors from the augmented q-factor model. The measure of variable importance is calculated as the change in $R^2$ from setting the observations of a factor proxy to zero within each estimation window. The average over the out-of-sample period from January 1987 to December 2019 is given. The variable importance measures for each latent factor are scaled to sum to 100. The latent factor specifications are based on principal component analysis (PCA), partial least squares (PLS), sparse principal component analysis (SPCA), sparse partial least squares (SPLS) and autoencoders with 1, 2, 3 and 4 hidden layers (AEN). MKTRF, ME, IA, ROE and EG are the Fama and French excess returns of the market from the risk-free rate, the Hou, Xue and Zhang (2015) size, investment, return-on-equity factors and the Hou, Mo, Xue and Zhang (2021) expected growth factor, respectively.

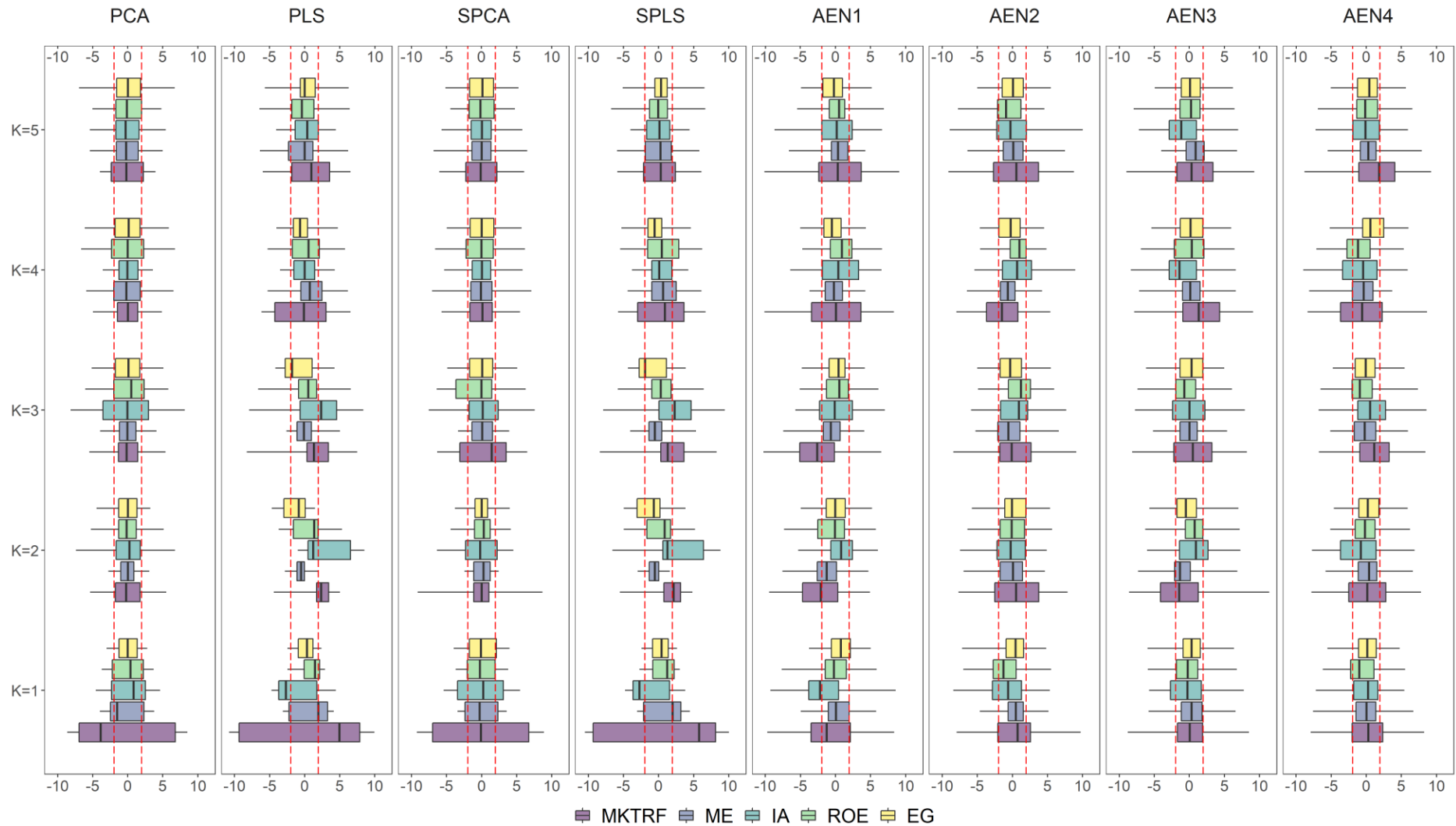

**Figure A5**
**Explaining the latent factors based on the augmented q-factor model**
This figure shows boxplots of the $t$-statistics based on OLS regressions of each of the five latent factors on factors from the augmented q-factor model. The horizontal axis reports $t$-statistics values ranging from -10 to 10 whereas, the vertical axis reports the latent factors, $K = 1, ..., 5$. The sample period is from January 1967 to December 2019. The median is marked by the line within the box, the edges of the box denote the first and third quartiles, while the minimum and maximum $t$-statistics are depicted by the end of the lines outside the box. The latent factor specifications are based on principal component analysis (PCA), partial least squares (PLS), sparse principal component analysis (SPCA), sparse partial least squares (SPLS) and autoencoders with 1, 2, 3 and 4 hidden layers (AEN). MKTRF, ME, IA, ROE and EG are the Fama and French excess returns of the market from the risk-free rate, the Hou, Xue and Zhang (2015) size, investment, return-on-equity factors and the Hou, Mo, Xue and Zhang (2021) expected growth factor, respectively. The $t$-statistics are computed using heteroskedasticity and autocorrelation-robust standard errors (Newey and West, 1987). The red lines depict the Student's t critical values at the 5% level.

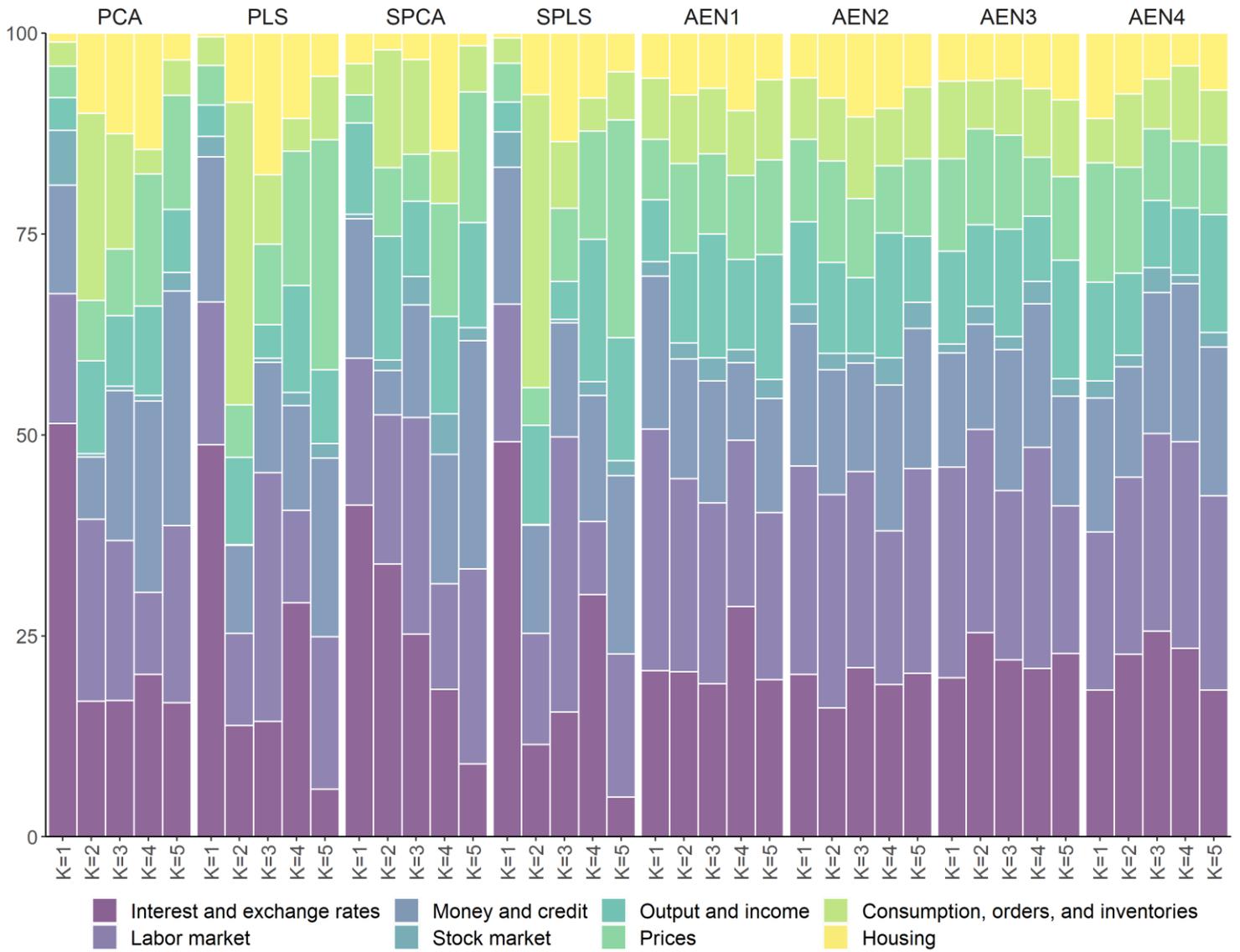

**Figure A6**
**Variable importance of the economic indicators from the McCracken and Ng (2015) dataset**
This figure shows the variable importance based on lasso regressions of the latent factors on economic indicators. The measure of variable importance is calculated as the change in $R^2$ from setting the observations of a feature to zero within each estimation window. The results are aggregated by summing the variable importance of the economic indicators belonging in the same group. Details on the variables within each group can be found in Table A2 in the Appendix. The average over the out-of-sample period from January 1980 to December 2019 is given. The variable importance measures for each group are scaled to sum to 100. The latent factor specifications are based on principal component analysis (PCA), partial least squares (PLS), sparse principal component analysis (SPCA), sparse partial least squares (SPLS) and autoencoders with 1, 2, 3 and 4 hidden layers (AEN). The explanatory variables are the 116 lagged economic indicators from the FRED-MD dataset by McCracken and Ng (2015) that have no missing values over the full sample period, from January 1960 to December 2019.

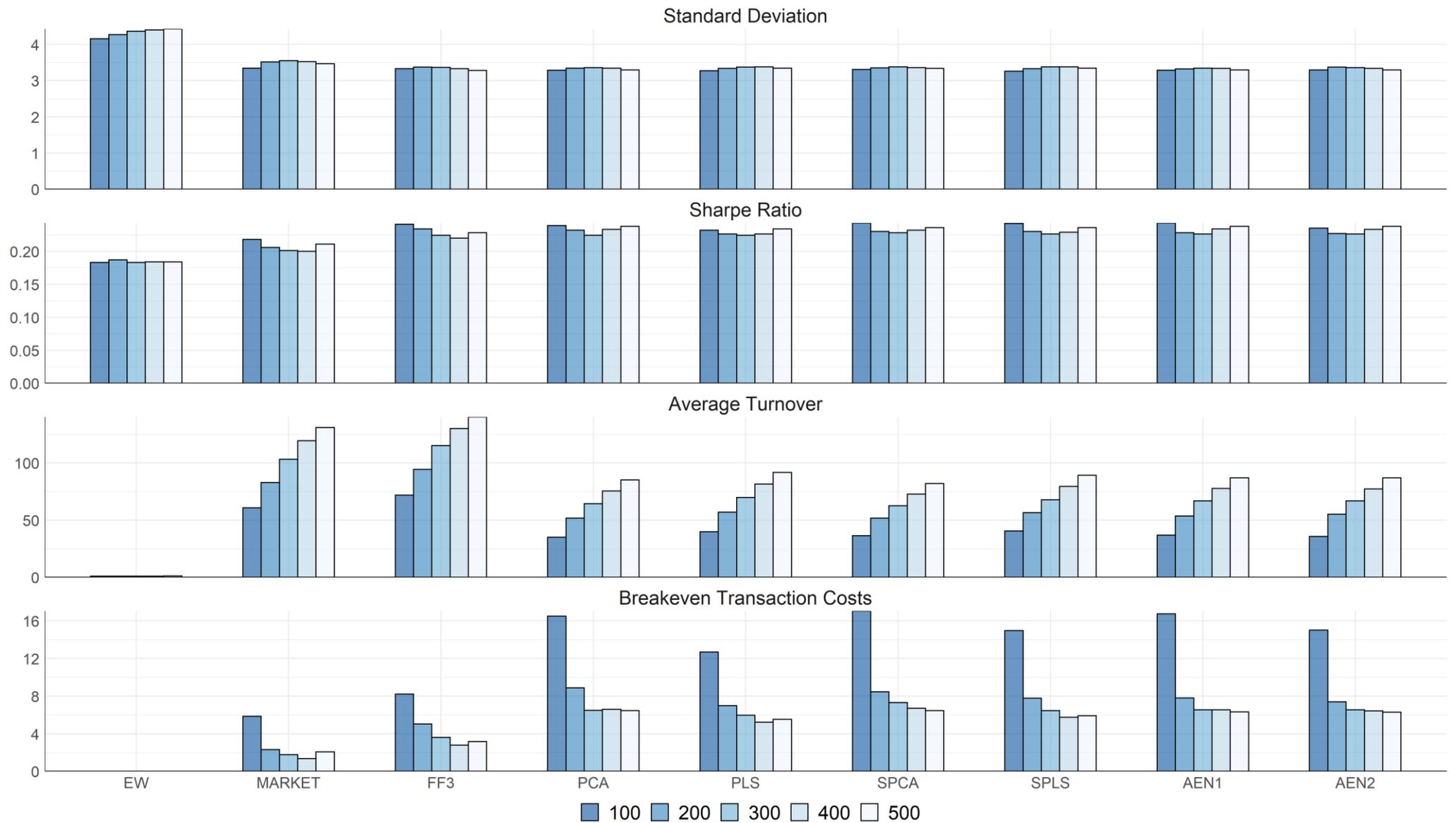

**Figure A7**
**Portfolio performance for a different number of stocks: Dynamic Beta Covariance**
This figure shows the monthly portfolio performance for a varying number of assets. Performance is based on the standard deviation, Sharpe ratio, average turnover and breakeven transaction costs with respect to the EW portfolio. The out-of-sample period is from January 1980 to December 2019. The results are presented for the equal-weighted portfolio (EW) and for the dynamic beta covariance. The factor specifications are based on the single factor model (Market), the Fama-French 3-factor model (FF3), principal component analysis (PCA), partial least squares (PLS), sparse principal component analysis (SPCA), sparse partial least squares (SPLS) and autoencoders with 1 and 2 hidden layers (AEN). The standard deviation and average turnover are reported as a percentage. The breakeven transaction costs are reported in basis points and a positive value indicates that the alternative portfolio outperforms the EW.

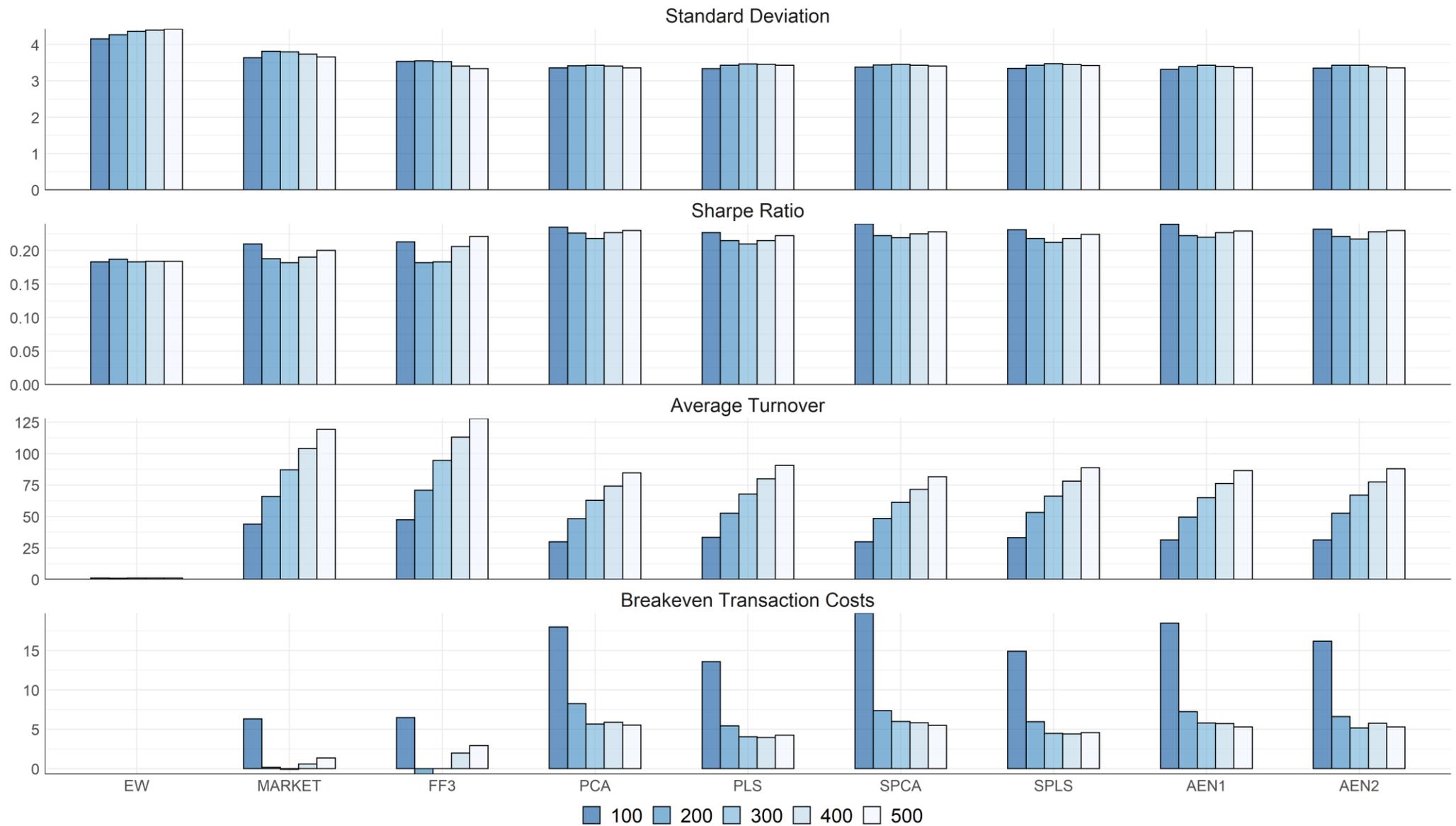

**Figure A8**
**Portfolio performance for a different number of stocks: Dynamic Factor Covariance**
This figure shows the monthly performance for a varying number of assets. Performance is based on the standard deviation, Sharpe ratio, average turnover and breakeven transaction costs with respect to the EW portfolio. The out-of-sample period is from January 1980 to December 2019. The results are presented for the equal-weighted portfolio (EW) and for the dynamic factor covariance. The factor specifications are based on the single factor model (Market), the Fama-French 3-factor model (FF3), principal component analysis (PCA), partial least squares (PLS), sparse principal component analysis (SPCA), sparse partial least squares (SPLS) and autoencoders with 1 and 2 hidden layers (AEN). The standard deviation and average turnover are reported as a percentage. The breakeven transaction costs are reported in basis points and a positive value indicates that the alternative portfolio outperforms the EW.

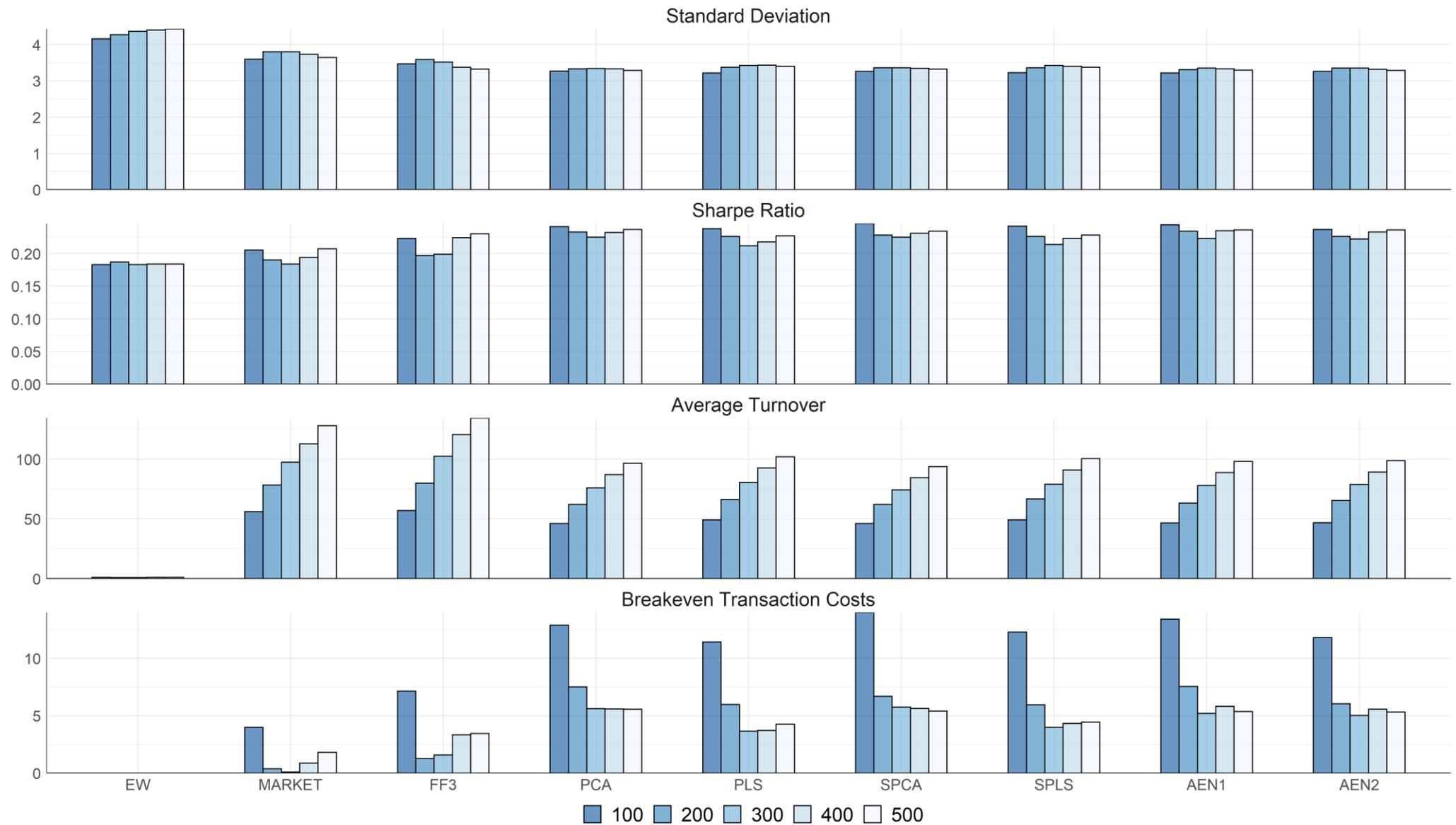

**Figure A9**
**Portfolio performance for a different number of stocks: Dynamic Error Covariance**
This figure shows the monthly performance for a varying number of assets. Performance is based on the standard deviation, Sharpe ratio, average turnover and breakeven transaction costs with respect to the EW portfolio. The out-of-sample period is from January 1980 to December 2019. The results are presented for the equal-weighted portfolio (EW) and for the dynamic error covariance. The factor specifications are based on the single factor model (Market), the Fama-French 3-factor model (FF3), principal component analysis (PCA), partial least squares (PLS), sparse principal component analysis (SPCA), sparse partial least squares (SPLS) and autoencoders with 1 and 2 hidden layers (AEN). The standard deviation and average turnover are reported as a percentage. The breakeven transaction costs are reported in basis points and a positive value indicates that the alternative portfolio outperforms the EW.